# Computational Design of Finite Strain Auxetic Metamaterials via Topology Optimization and Nonlinear Homogenization

Guodong Zhang[*] and Kapil Khandelwal[§]

[*]Graduate Student, *Dept. of Civil & Env. Engg. & Earth Sci., University of Notre Dame*
[§]Associate Professor, *Dept. of Civil & Env. Engg. & Earth Sci., 156 Fitzpatrick Hall, University of Notre Dame, Notre Dame, IN 46556, United States, Email:kapil.khandelwal@nd.edu,*
ORCID: 0000-0002-5748-6019, (Corresponding Author).

*Preprint Submitted*

## Abstract

A novel computational framework for designing metamaterials with negative Poisson's ratio over a large strain range is presented in this work by combining the density-based topology optimization together with a mixed stress/deformation driven nonlinear homogenization method. A measure of Poisson's ratio based on the macro deformations is proposed, which is further validated through direct numerical simulations. With the consistent optimization formulations based on nonlinear homogenization, auxetic metamaterial designs with respect to different loading orientations and with different unit cell domains are systematically explored. In addition, the extension to multimaterial auxetic metamaterial designs is also considered, and stable optimization formulations are presented to obtain discrete metamaterial topologies under finite strains. Various new auxetic designs are obtained based on the proposed framework. To validate the performance of optimized designs, a multiscale stability analysis is carried out using the Bloch wave method and rank-one convexity check. As demonstrated, short and/or long wavelength instabilities can occur during the loading process, leading to a change of periodicity of the microstructure, which can affect the performance of an optimized design.

**Keywords:** Negative Poisson's ratio; Nonlinear auxetic metamaterials; Topology optimization; Nonlinear homogenization; Multiscale stability.

## 1 Introduction

Design of mechanical metamaterials to achieve extreme properties has been an active research area in the past decades [1-3], wherein the overall macroscopic properties of these metamaterials are tailored by carefully designing the geometries of the underlying microstructures. Thus, the properties of these metamaterials are largely determined by their microstructural designs rather than by the chemical compositions of the constituent materials. Among these metamaterials, materials with negative Poisson's ratio (NPR), also known as *auxetic* materials [4], have received considerable attention ever since the pioneering works by Robert [5] and Lakes [6]. The Poisson's ratio measures the extent by which a material contracts transversally relative to the axial stretch. As opposed to the common engineering materials with positive Poisson's ratio, the auxetic materials expand transversally when stretched. Due to this unusual behavior, auxetic materials have been shown to possess many desirable mechanical properties such as high shear resistance, synclastic curvature and indentation resistance, among others [7]. The auxetic material behavior arises due to the purposeful design of material microstructural geometries, and among the microstructural geometric features that lead to auxetic behavior, three main types have been proposed, i.e., *re-entrant* type [8], *chiral* type [9] and *rotating units* type [10]. By manipulating and combining these three features, a variety of auxetic designs has been heuristically obtained [7]. Though promising, most of the auxetic material designs are valid only under a relatively small deformation region. Due to the high geometric nonlinearity during the deformation process, the design of auxetic metamaterials with NPR over a large deformation range is still a challenging issue. More information regarding the auxetic materials can be found in the recent review articles [7, 11].

Aside from the experimental or heuristic driven design approaches, novel metamaterial designs can be also discovered by mechanical modeling and optimization based approaches. Among other design approaches, the topology optimization method has shown great potential in structural and material designs since its initiation by Bendsøe and Kikuchi [12]. Combining with asymptotic homogenization under small deformations, Sigmund [13] first demonstrated the use of topology optimization for designing materials with desirable effective properties, including NPR and other thermoelastic properties [14]. Since then topology optimization has been successfully applied for discovering metamaterials with various extreme properties, e.g. extreme thermal expansion [15], extremal bulk and shear modulus [16, 17], desirable band-gaps [18], optimal damping characteristics [19] and optimal energy absorbing capability [20], among others. However, most of these design studies are still confined to small deformation range and there are only limited studies that considered finite deformations in topology optimization [21-23].

The difficulties in incorporating the nonlinear elastic behavior at finite deformations in designing nonlinear metamaterials are, in part, due to the characterization of overall homogenized metamaterial's properties. Starting with the work of Hill [24], a series of studies have been carried out to predict the effective properties of nonlinear composites based on homogenization theories, see [25] and references therein. In contrast to these analytical approaches, where it is difficult to incorporate the full geometric information of the underlying microstructures, the computational homogenization approaches which are developed using the exact finite element (FE) models of representative volume element (RVE) are capable of predicting the overall macroscopic metamaterial response [26], and are therefore, more appropriate for metamaterials analysis and design.

Another critical challenge, while considering metamaterials design under finite deformations, is related to the potential instabilities that can happen at both the *micro* and *macro* scales. Previous studies have shown that a nonlinear composite material with quasiconvex, and thus strictly rank-one convex constituents, can lose strict rank-one convexity at adequate loads, which allows for the emergence of discontinuous deformation gradient field, and this response is termed as *macroscale* instability [27]. Depending on the geometries of the microstructure as well as the loading conditions, short wavelength buckling or *microscale* instability can occur before the macroscale instability [28], which leads to a change of the periodicity of the microstructure. As a result, the RVE which includes the smallest periodic cell (one unit cell) cannot serve as RVE after the onset of instability. This presents a barrier for nonlinear material characterization, since the wavelength of the buckling mode is not a priori known. Therefore, the nonlinear metamaterial designs should be investigated w.r.t. stability issues, as a design might lose its validity when structural micro or macro stability is lost.

In this study, a nonlinear homogenization based topology optimization framework is presented that can be used for design of auxetic metamaterials under finite deformations. Compared to the previous work on nonlinear auxetic metamaterials design via topology optimization by Wang et al. [23], which is based on numerical tensile experiments instead of homogenization, the main contributions of this work are: (a) Consistent mixed stress/deformation driven nonlinear homogenization analysis is incorporated in the topology optimization that enables a clear transition from the microstructural behavior to macroscale properties, and an overall measure of Poisson's ratio in terms of the macro deformations is proposed; (b) Design space is greatly expanded in the proposed design framework, wherein different unit cell domains and loading orientations can be consistently considered, due to the underlying homogenization framework; (c) Design of auxetic materials is extended from single to multiple materials phases and many novel auxetic metamaterials designs are discovered; (d) Optimized designs are further validated by a direct numerical simulation of extended periodic solids with the optimized topologies; (e) Multiscale stability investigation is carried out for validating the structural behavior of the optimized designs and the stability issues are discussed in the design of auxetic metamaterials. The rest of the paper is organized as follows: in Section 2, finite deformation homogenization method is reviewed. In Section 3, the density based multimaterial topology optimization formulations are presented. The sensitivity analysis is given in Section 4. Section 5 shows the illustrative numerical examples, which is followed by numerical validation studies of homogenization analysis and multiscale stability investigations in Section 6 and Section 7, respectively. Final remarks and conclusions are given in Section 8.

## 2 Finite Deformation Homogenization

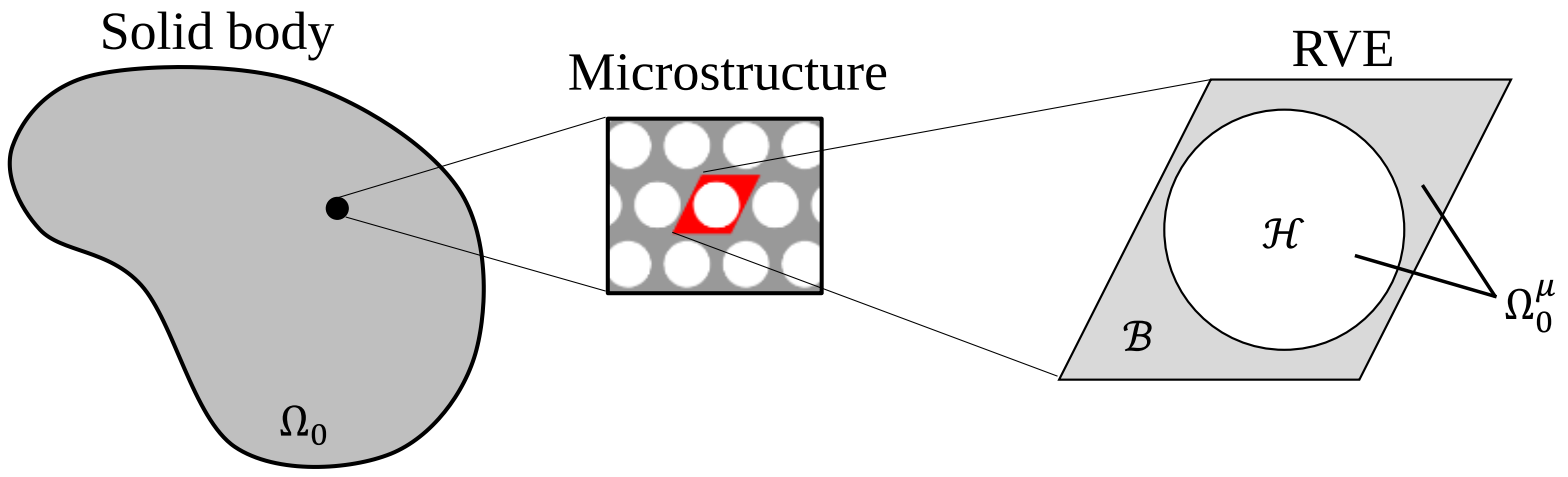

Figure 1. Solid body with periodic microstructure and corresponding RVE.

Homogenization theory is used for the prediction of overall properties of a metamaterial with underlying periodic microstructure. For example, Figure 1 shows a solid body made of material that has a specific microstructure that can be characterized by a periodic arrangement of a unit cell. This unit cell can be taken as a representative volume element (RVE), which can be used to determine the overall material properties. The domain that RVE occupies is denoted by $\Omega_0^\mu$, which includes solid part $\mathcal{B}$ and void part $\mathcal{H}$, i.e., $\Omega_0^\mu = \mathcal{B} \cup \mathcal{H}$, as shown in Figure 1, and $\partial\mathcal{B} = \partial\Omega_0^\mu \cup \partial\mathcal{H}$ where $\partial(\blacksquare)$ denotes the boundary of $\blacksquare$. To fulfill the scale separation assumption, the characteristic length of RVE should be much smaller than the dimension of the macroscale solid body. In this section, a brief review of the first-order finite deformation homogenization theory is presented that is used to obtain the effective properties of the periodic unit cells.

Let $\nabla\bar{\boldsymbol{u}}(\bar{\boldsymbol{X}})$ denote the macroscopic displacement gradient at the material point $\bar{\boldsymbol{X}}$ of the solid body which occupies the domain $\Omega_0$. For the remainder of this paper, a bar will be used to denote macroscopic quantities. Without loss of generality, the microscale displacement field $\boldsymbol{u}(\boldsymbol{X})$ over the RVE domain $\Omega_0^\mu$, which is associated with the material point $\bar{\boldsymbol{X}} \in \Omega_0$, is linked to the macroscale kinematical variables by

$$\boldsymbol{u}(\boldsymbol{X}) = \nabla\bar{\boldsymbol{u}}.\boldsymbol{X} + \tilde{\boldsymbol{u}}(\boldsymbol{X}) \tag{1}$$

where $\tilde{\boldsymbol{u}}(\boldsymbol{X})$ is defined as the displacement fluctuation field. The microscale deformation gradient can be thus expressed as

$$\boldsymbol{F}(\boldsymbol{X}) = \bar{\boldsymbol{F}} + \nabla_{\boldsymbol{X}}\tilde{\boldsymbol{u}}(\boldsymbol{X}) \tag{2}$$

where $\bar{\boldsymbol{F}} = \boldsymbol{I} + \nabla\bar{\boldsymbol{u}}$ represents the macroscale deformation gradient, $\nabla_{\boldsymbol{X}}(\cdot)$ denotes the gradient operator with respect to the microscale coordinates $\boldsymbol{X}$. Following [29], the microscale displacement field has to satisfy the kinematical admissibility constraints, which are postulated as

$$\int_{\mathcal{B}} \boldsymbol{u}(\boldsymbol{X})dV = \boldsymbol{0} \qquad \text{and} \qquad \bar{\boldsymbol{F}} = \boldsymbol{I} + \frac{1}{V}\int_{\partial\Omega_0^\mu} \boldsymbol{u}(\boldsymbol{X}) \otimes \boldsymbol{N}dS \tag{3}$$

in which $V$ is the volume of the domain $\Omega_0^\mu$ and $\boldsymbol{N}$ is the unit normal vector on the boundary $\partial\Omega_0^\mu$. Using the divergence theorem, it can be further shown that Eqns. $(3)_1$ and $(3)_2$ are equivalent to

$$\int_{\mathcal{B}} \tilde{\boldsymbol{u}}(\boldsymbol{X})dV = \boldsymbol{0} \qquad \text{and} \qquad \int_{\partial\Omega_0^\mu} \tilde{\boldsymbol{u}}(\boldsymbol{X}) \otimes \boldsymbol{N}dS = \boldsymbol{0} \tag{4}$$

where it is assumed that the microscale coordinate system is chosen such that $\int_{\mathcal{B}} \boldsymbol{X}dV = \boldsymbol{0}$. Moreover, applying the divergence theorem to Eq. $(3)_2$ gives

$$\bar{\boldsymbol{F}} = \boldsymbol{I} + \frac{1}{V}\int_{\mathcal{B}} \nabla_{\boldsymbol{X}}\boldsymbol{u}(\boldsymbol{X})dV - \frac{1}{V}\int_{\partial\mathcal{H}} \boldsymbol{u}(\boldsymbol{X}) \otimes \boldsymbol{N}dS \tag{5}$$

which means that the macroscopic deformation gradient $\bar{\boldsymbol{F}}$ is not equal to the volume average of the microscopic deformation gradient $\boldsymbol{F}$, in general, due to the presence of the voids [30]. Finally, the kinematically admissible displacement fluctuation field $\tilde{\boldsymbol{u}}(\boldsymbol{X})$ is defined in a functional space $\mathcal{V}_{min}$

$$\mathcal{V}_{min} = \left\{\tilde{\boldsymbol{u}}(\boldsymbol{X}) \middle| \tilde{\boldsymbol{u}}(\boldsymbol{X}) \in H^1(\mathcal{B}), \int_{\mathcal{B}} \tilde{\boldsymbol{u}}(\boldsymbol{X})dV = \boldsymbol{0},\ \int_{\partial\Omega_0^\mu} \tilde{\boldsymbol{u}}(\boldsymbol{X}) \otimes \boldsymbol{N}dS = \boldsymbol{0}\right\} \tag{6}$$

where $H^1(\mathcal{B}) = \{\boldsymbol{v} | v_i \in L^2(\mathcal{B}), \partial v_i/\partial X_j \in L^2(\mathcal{B}), i,j = 1,2,\dots,d\}$ and $L^2(\mathcal{B})$ represents the space of square integrable functions defined on $\mathcal{B}$ and $d$ is the number of space dimensions. The subscript *min* means that this set of constraints is the minimal set required for kinematical admissibility. It has been shown in [29] that this set of constraints corresponds to the constant traction boundary conditions. Additional constraints can be introduced in a consistent way that may lead to periodic boundary conditions or linear displacement boundary conditions [29]. It is noted that the constraint in Eq. $(3)_1$ is equivalent to removing rigid-body translation, while the constraint in Eq. $(3)_2$ implicitly removes rigid-body rotation.

The micro-to-macro transition is governed by the principle of multiscale virtual power [29], which reads

$$\overline{\boldsymbol{P}}:\delta\overline{\boldsymbol{F}} = \frac{1}{V}\int_{\mathcal{B}} \boldsymbol{P}:\delta\boldsymbol{F}dV \qquad \forall\, \delta\overline{\boldsymbol{F}} \in \mathrm{Lin}, \delta\widetilde{\boldsymbol{u}} \in \mathcal{V} \subseteq \mathcal{V}_{min} \tag{7}$$

where $\overline{\boldsymbol{P}}$ and $\boldsymbol{P}$ are the macro and micro 1st Piola-Kirchhoff stress tensors, respectively, and the space of virtual kinematical admissible fluctuation field $\delta\widetilde{\boldsymbol{u}}$ is identical to that of the kinematical admissible fluctuation field $\widetilde{\boldsymbol{u}}$, i.e. $\mathcal{V}$ which is a subspace of $\mathcal{V}_{min}$. Equation (7) can be seen as the variational form of the Hill-Mandel condition [24, 31] that states the incremental internal energy equivalence between the macroscale and microscale. In Eq. (7), inertia and body forces have been assumed zero and the interested readers are referred to the Refs [29, 32] for further extensions to dynamic cases.

The stress homogenization relation

$$\overline{\boldsymbol{P}} = \frac{1}{V}\int_{\mathcal{B}} \boldsymbol{P}dV \tag{8}$$

and the microscale equilibrium equation

$$\int_{\mathcal{B}} \boldsymbol{P}:\boldsymbol{\nabla_X}\delta\widetilde{\boldsymbol{u}}\, dV = 0 \qquad \forall\, \delta\widetilde{\boldsymbol{u}} \in \mathcal{V} \tag{9}$$

can be obtained from Eq. (7) by choosing $\delta\widetilde{\boldsymbol{u}} = \boldsymbol{0}$ and $\delta\overline{\boldsymbol{F}} = \boldsymbol{0}$, respectively.

### *2.1 Deformation-driven homogenization – periodic boundary conditions*

In this section, a deformation-driven homogenization formulation for computing the homogenized stresses and tangent moduli of a given discretized microstructure is presented. Since in the metamaterial topological design, the underlying microstructure is always assumed to be periodic with repeating unit cells (Figure 1), it motivates the use of periodic boundary conditions. To be compatible with the assumption of periodicity, the RVE must satisfy certain geometrical constraints. For example, for 2D problems, the most general RVE shape is the parallelogram (Figure 1 or Figure 2a), and square or rectangular shapes are special cases of parallelogram. The hexagonal unit cell (Figure 2b) can also be equivalently recast into a parallelogram. For a 2D RVE, the boundary can be divided into a pair of negative and positive sides, denoting as $\partial\Omega_0^{\mu-}$ and $\partial\Omega_0^{\mu+}$, respectively, see Figure 2, where points on the positive side can be reached by translating the corresponding points on the negative side using a periodic lattice vector $\boldsymbol{a}_1$ or $\boldsymbol{a}_2$ or $\pm(\boldsymbol{a}_1 \pm \boldsymbol{a}_2)$. For the periodic boundary conditions, the displacement fluctuations on the negative side equal the corresponding ones on the positive side, i.e.,

$$\tilde{\boldsymbol{u}}^{+} = \tilde{\boldsymbol{u}}^{-} \quad \text{on } \partial\Omega_0^{\mu} \tag{10}$$

which can be proved to automatically satisfy the constraints in Eqns. $(4)_2$ or $(3)_2$. As a result, the kinematically admissible displacement fluctuation field considering periodic boundary condition is defined in a functional space $\mathcal{V}$

$$\mathcal{V} = \left\{ \tilde{\boldsymbol{u}}(\boldsymbol{X}) \middle| \tilde{\boldsymbol{u}}(\boldsymbol{X}) \in H^1(\mathcal{B}), \int_{\mathcal{B}} \tilde{\boldsymbol{u}}(\boldsymbol{X}) dV = \boldsymbol{0},\ \tilde{\boldsymbol{u}}^{+} = \tilde{\boldsymbol{u}}^{-} \text{ on } \partial\Omega_0^{\mu} \right\} \tag{11}$$

and the multiscale homogenization problem is completely defined by Eq. (7).

Consider now for a given discretized RVE (Figure 2), the constraints in Eq. (10) is discretized as

$$\tilde{\boldsymbol{u}}_q^{+} = \tilde{\boldsymbol{u}}_q^{-}, \quad q = 1,2,\dots,m \tag{12}$$

where $m$ pairs of nodes lying on the negative and positive boundary sides are identified. The rigid-body translation constraint (Eq. $(4)_1$) can be equivalently replaced by fixing one arbitrary point, e.g. $\tilde{\boldsymbol{u}}_o = \boldsymbol{0}$ in $\mathcal{B}$. Thus, the discretized functional space $\mathcal{V}^h$ is defined by

$$\mathcal{V}^h = \left\{ \tilde{\boldsymbol{u}}(\boldsymbol{X}) \middle| \tilde{\boldsymbol{u}}(\boldsymbol{X}) \in H^1(\mathcal{B}), \tilde{\boldsymbol{u}}_o = \boldsymbol{0},\ \tilde{\boldsymbol{u}}_q^{+} = \tilde{\boldsymbol{u}}_q^{-}\ (q = 1, 2, \dots, m) \right\} \tag{13}$$

*2.1.1 Principle of multiscale virtual power with Lagrange multiplier – discrete form*

Applying discrete Lagrange multipliers to enforce the constraints in Eq. (13), the discretized version of the principle of multiscale virtual power reads

$$-V(\bar{\boldsymbol{P}}:\delta\bar{\boldsymbol{F}}) + \int_{\mathcal{B}} \boldsymbol{P}:\delta\boldsymbol{F}dV - \delta\boldsymbol{\lambda}^T\boldsymbol{u}_o - \boldsymbol{\lambda}^T\delta\boldsymbol{u}_o - \sum_{q=1}^{m} \delta\boldsymbol{\mu}_q^T\left[\boldsymbol{u}_q^+ - \boldsymbol{u}_q^- - (\bar{\boldsymbol{F}} - \boldsymbol{I}).\boldsymbol{L}_q\right]$$
$$- \sum_{q=1}^{m} \boldsymbol{\mu}_q^T\left[\delta\boldsymbol{u}_q^+ - \delta\boldsymbol{u}_q^- - \delta\bar{\boldsymbol{F}}.\boldsymbol{L}_q\right] = 0 \qquad \forall\, \delta\bar{\boldsymbol{F}} \in \mathrm{Lin},\ \delta\boldsymbol{u},\ \delta\boldsymbol{\lambda},\ \delta\boldsymbol{\mu} \tag{14}$$

where $\boldsymbol{\lambda}$ and $\boldsymbol{\mu} = [\boldsymbol{\mu}_1, \dots, \boldsymbol{\mu}_m]^T$ are the Lagrange multipliers, and the constraints are restated in terms of the displacement field $\boldsymbol{u}(\boldsymbol{X})$ instead of fluctuation field $\widetilde{\boldsymbol{u}}(\boldsymbol{X})$. Note that $\boldsymbol{u}_o = \boldsymbol{0}$ is equivalent to $\widetilde{\boldsymbol{u}}_o = \boldsymbol{0}$ in the sense of removing rigid-body translation. The vector $\boldsymbol{L}_q$ in Eq. (14) is defined as $\boldsymbol{L}_q = \sum_{i=1}^{d} c_i\boldsymbol{a}_i$ with $c_i$ any integers that satisfies $\boldsymbol{X}_q^+ = \boldsymbol{X}_q^- + \boldsymbol{L}_q$ where $\boldsymbol{X}_q^+$ and $\boldsymbol{X}_q^-$ are the coordinates of the nodes on a pair of positive and negative sides, see Figure 2a.

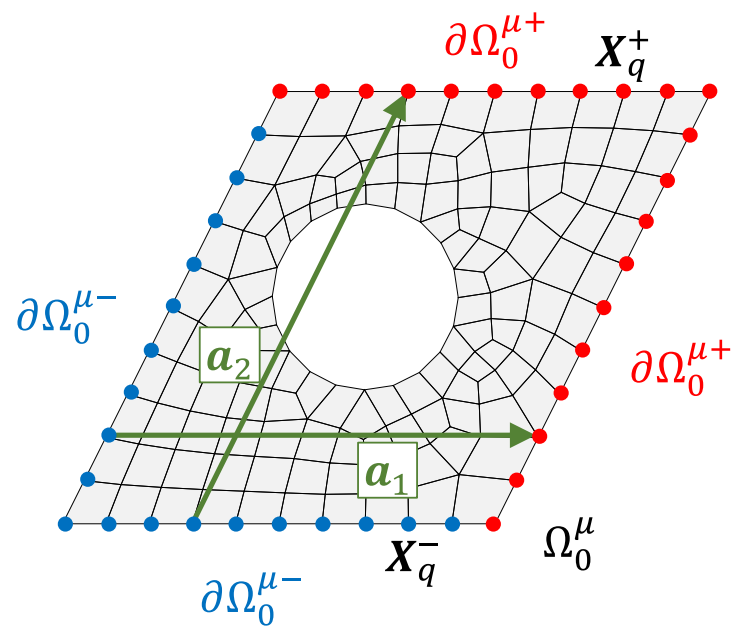

(a) Parallelogram RVE

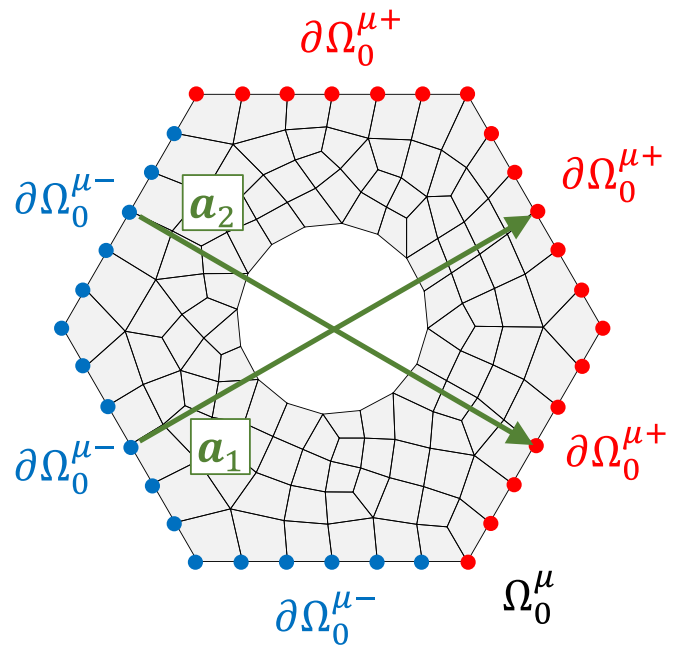

(b) Hexagonal RVE

Figure 2. Geometries and partitioning of boundary nodes of discretized microstructures of RVE (blue color denotes the negative sides and red color denotes the positive sides).

*2.1.2 Homogenized stress*

The homogenized stress can be determined by the discrete forces on the boundary, which are related to the Lagrange multipliers. To this end, the interpretation of the Lagrange multipliers in Eq. (14) is first carried out. Assuming $\delta\bar{\boldsymbol{F}} = \boldsymbol{0}$, $\delta\boldsymbol{\lambda} = \boldsymbol{0}$ and $\delta\boldsymbol{\mu} = \boldsymbol{0}$ and $\delta\boldsymbol{u} = \boldsymbol{c}_0$ (with $\boldsymbol{c}_0$ constant in $\mathcal{B}$) in Eq. (14) gives $\delta\boldsymbol{u}_q^+ - \delta\boldsymbol{u}_q^- = \boldsymbol{0}$ ($q = 1, \dots, m$), which leads to $\boldsymbol{\lambda}^T\boldsymbol{c}_0 = 0$. Therefore,

$$\boldsymbol{\lambda} = \boldsymbol{0} \tag{15}$$

has to be satisfied, which means that for a self-equilibrated system, fixing one arbitrary point for removing rigid-body translation does not create any reaction force.

Now taking $\delta\bar{\boldsymbol{F}} = \boldsymbol{0}$, $\delta\boldsymbol{\lambda} = \boldsymbol{0}$ and $\delta\boldsymbol{\mu} = \boldsymbol{0}$ in Eq. (14) but with $\delta\boldsymbol{u}(\boldsymbol{X}) = \boldsymbol{A}_0.\boldsymbol{X}$ in $\mathcal{B}$ where $\boldsymbol{A}_0 \in$ Lin (a constant 2nd-order tensor), gives $\delta\boldsymbol{F} = \boldsymbol{A}_0$ and $\delta\boldsymbol{u}_q^+ - \delta\boldsymbol{u}_q^- = \boldsymbol{A}_0.\boldsymbol{L}_q$, which leads to

$$\left(\int_{\mathcal{B}} \boldsymbol{P} dV - \sum_{q=1}^{m} \boldsymbol{\mu}_q \otimes \boldsymbol{L}_q\right) : \boldsymbol{A}_0 = 0 \quad \forall \boldsymbol{A}_0 \in \mathrm{Lin} \tag{16}$$

where $\boldsymbol{\mu}_q$ and $\boldsymbol{L}_q$ $(q = 1, \ldots, m)$ are both vectors (or 1st-order tensors) while $\boldsymbol{P}$ and $\boldsymbol{A}_0$ are 2nd-order tensors in the space of dimension $d = 2$ or 3 for 2D and 3D cases, respectively. Since $\boldsymbol{A}_0$ can be chosen arbitrarily, it follows from Eqns. (8) and (16) that

$$\overline{\boldsymbol{P}} = \frac{1}{V}\int_{\mathcal{B}} \boldsymbol{P} dV = \frac{1}{V}\sum_{q=1}^{m} \boldsymbol{\mu}_q \otimes \boldsymbol{L}_q \tag{17}$$

Therefore, it can be seen that the homogenized stress can be computed from the Lagrange multipliers $\boldsymbol{\mu}$.

### *2.1.3 Finite element formulation*

Considering the unknown variables to be solved as $\boldsymbol{u}$, $\boldsymbol{\lambda}$ and $\boldsymbol{\mu}$, the resulting set of nonlinear constrained equilibrium equations, from Eq. (14), reads

$$\boldsymbol{R}(\boldsymbol{u}, \boldsymbol{\lambda}, \boldsymbol{\mu}) = \begin{bmatrix} \boldsymbol{R}_1(\boldsymbol{u}, \boldsymbol{\lambda}, \boldsymbol{\mu}) \\ \boldsymbol{R}_2(\boldsymbol{u}, \boldsymbol{\lambda}, \boldsymbol{\mu}) \\ \boldsymbol{R}_3(\boldsymbol{u}, \boldsymbol{\lambda}, \boldsymbol{\mu}) \end{bmatrix} = \begin{bmatrix} \boldsymbol{F}_{int}(\boldsymbol{u}) - \boldsymbol{M}_1^T \boldsymbol{\lambda} - \boldsymbol{M}_2^T \boldsymbol{\mu} \\ -\boldsymbol{M}_1 \boldsymbol{u} \\ -\boldsymbol{M}_2 \boldsymbol{u} \end{bmatrix} + \begin{bmatrix} \boldsymbol{0} \\ \boldsymbol{0} \\ \boldsymbol{b} \end{bmatrix} = \boldsymbol{0} \tag{18}$$

where $\boldsymbol{F}_{int}$ represents the global internal force vector defined by

$$\boldsymbol{F}_{int}(\boldsymbol{u}) = \mathop{\mathcal{A}}_{e=1}^{n_{ele}} \boldsymbol{F}_{int}^e \quad \text{with} \quad \boldsymbol{F}_{int}^e = \int_{\Omega^e} \boldsymbol{B}^T \boldsymbol{P} dV \tag{19}$$

where $\boldsymbol{B}$ is the shape function derivative matrix, $\Omega^e$ represents the $e^{th}$ element integration domain satisfying $\mathcal{B} = \bigcup_{e=1}^{n_{ele}} \Omega^e$ and $n_{ele}$ are the total number of elements in the RVE. In the topology optimization, with the design domain defined as the RVE, fictitious domain approach is adopted in which void area $\mathcal{H}$ is also included in the finite element analysis (FEA) and is assigned vanishing material properties, i.e. $\Omega_0^\mu = \bigcup_{e=1}^{n_{ele}} \Omega^e$.

The matrices $\boldsymbol{M}_1$ and $\boldsymbol{M}_2$, and vector $\boldsymbol{b}$ are constructed such that

$$\begin{aligned} \boldsymbol{u}_o &= \boldsymbol{M}_1 \boldsymbol{u} \\ \boldsymbol{u}^+ - \boldsymbol{u}^- &= \boldsymbol{M}_2 \boldsymbol{u} \end{aligned} \tag{20}$$

$$\boldsymbol{b}=\begin{bmatrix}(\overline{\boldsymbol{F}}-\boldsymbol{I}).\boldsymbol{L}_1\\ \vdots\\ (\overline{\boldsymbol{F}}-\boldsymbol{I}).\boldsymbol{L}_m\end{bmatrix}=[\boldsymbol{L}_M]([\overline{\boldsymbol{F}}]-[\boldsymbol{I}])=\begin{bmatrix}\tilde{X}_1 & 0 & \tilde{Y}_1 & 0\\ 0 & \tilde{X}_1 & 0 & \tilde{Y}_1\\ \vdots & \vdots & \vdots & \vdots\\ \tilde{X}_m & 0 & \tilde{Y}_m & 0\\ 0 & \tilde{X}_m & 0 & \tilde{Y}_m\end{bmatrix}_{2m\times 4}\left(\begin{bmatrix}\bar{F}_{11}\\ \bar{F}_{21}\\ \bar{F}_{12}\\ \bar{F}_{22}\end{bmatrix}-\begin{bmatrix}1\\0\\0\\1\end{bmatrix}\right)$$

where $\boldsymbol{u}$ is the global nodal displacement vector, $\boldsymbol{u}^+=[\boldsymbol{u}_1^+,\dots,\boldsymbol{u}_m^+]^T$ and $\boldsymbol{u}^-=[\boldsymbol{u}_1^-,\dots,\boldsymbol{u}_m^-]^T$ includes $m$ nodal displacements defined on the positive and negative boundary sides, respectively. $\boldsymbol{L}_q=\left[\tilde{X}_q,\tilde{Y}_q\right]^T$ is the translational vector from the $q^{th}$ node on the negative side to the $q^{th}$ node on the positive side. The expression of $\boldsymbol{b}$ vector is written for 2D case in a matrix vector form in Eq. $(20)_3$.

The nonlinear system in Eq. (18) is solved using the Newton-Raphson method and the Jacobian matrix, which is needed for Newton-Raphson solver, can be calculated as

$$\mathbf{J}_T=\begin{bmatrix}\partial\boldsymbol{R}_1/\partial\boldsymbol{u} & \partial\boldsymbol{R}_1/\partial\boldsymbol{\lambda} & \partial\boldsymbol{R}_1/\partial\boldsymbol{\mu}\\ \partial\boldsymbol{R}_2/\partial\boldsymbol{u} & \partial\boldsymbol{R}_2/\partial\boldsymbol{\lambda} & \partial\boldsymbol{R}_2/\partial\boldsymbol{\mu}\\ \partial\boldsymbol{R}_3/\partial\boldsymbol{u} & \partial\boldsymbol{R}_3/\partial\boldsymbol{\lambda} & \partial\boldsymbol{R}_3/\partial\boldsymbol{\mu}\end{bmatrix}=\begin{bmatrix}\boldsymbol{K}_T & -\boldsymbol{M}_1^T & -\boldsymbol{M}_2^T\\ -\boldsymbol{M}_1 & \boldsymbol{0} & \boldsymbol{0}\\ -\boldsymbol{M}_2 & \boldsymbol{0} & \boldsymbol{0}\end{bmatrix} \tag{21}$$

where the term $\boldsymbol{K}_T=\partial\boldsymbol{F}_{int}/\partial\boldsymbol{u}$ is the usual tangent structural stiffness matrix calculated by

$$\boldsymbol{K}_T=\frac{\partial\boldsymbol{F}_{int}}{\partial\boldsymbol{u}}=\mathop{\mathcal{A}}_{e=1}^{n_{ele}}\boldsymbol{k}_T^e \quad \text{with} \quad \boldsymbol{k}_T^e=\int_{\Omega^e}\boldsymbol{B}^T\left[\frac{\partial\boldsymbol{P}}{\partial\boldsymbol{F}}\right]\boldsymbol{B}\,dV \tag{22}$$

in which the tangent moduli $\partial\boldsymbol{P}/\partial\boldsymbol{F}$ is obtained from material subroutine.

### *2.1.4 Homogenized tangent moduli*

Using Eq. (17) and the definition of matrix $[\boldsymbol{L}_M]$ given in Eq. $(20)_3$, it can be shown that the homogenized stress $\overline{\boldsymbol{P}}$ is given by

$$[\overline{\boldsymbol{P}}]=\frac{1}{V}[\boldsymbol{L}_M]^T\boldsymbol{\mu} \tag{23}$$

where the bracket outside $\overline{\boldsymbol{P}}$ means that it is arranged in a $4\times1$ vector form, similarly as $[\overline{\boldsymbol{F}}]$ used in Eq. $(20)_3$.

The 4$^{th}$-order tensor homogenized tangent moduli $\overline{\mathbb{A}}$ is defined by

$$\overline{\mathbb{A}}=\frac{\partial\overline{\boldsymbol{P}}}{\partial\overline{\boldsymbol{F}}} \tag{24}$$

and can be rephrased in a matrix form as $[\overline{\mathbb{A}}]=\partial[\overline{\boldsymbol{P}}]/\partial[\overline{\boldsymbol{F}}]$, which is of size $4\times4$ for 2D case. From Eq. (23), it is clear that $[\overline{\mathbb{A}}]$ is determined by the derivative of Lagrange multiplier $\boldsymbol{\mu}$ with respect to $\bar{\boldsymbol{F}}$. To this end, the set of global equilibrium equations (18) is perturbed at the equilibrium state by a perturbation $\Delta\bar{\boldsymbol{F}}$, i.e.,

$$\begin{bmatrix} \boldsymbol{K}_T & -\boldsymbol{M}_1^T & -\boldsymbol{M}_2^T \\ -\boldsymbol{M}_1 & \boldsymbol{0} & \boldsymbol{0} \\ -\boldsymbol{M}_2 & \boldsymbol{0} & \boldsymbol{0} \end{bmatrix} \begin{bmatrix} \Delta\boldsymbol{u} \\ \Delta\boldsymbol{\lambda} \\ \Delta\boldsymbol{\mu} \end{bmatrix} + \begin{bmatrix} \boldsymbol{0} \\ \boldsymbol{0} \\ \boldsymbol{L}_M \end{bmatrix} [\Delta\bar{\boldsymbol{F}}] = \boldsymbol{0} \tag{25}$$

which results in

$$\begin{bmatrix} \Delta\boldsymbol{u} \\ \Delta\boldsymbol{\lambda} \\ \Delta\boldsymbol{\mu} \end{bmatrix} = -\mathbf{J}_T^{-1} \begin{bmatrix} \boldsymbol{0} \\ \boldsymbol{0} \\ \boldsymbol{L}_M \end{bmatrix} [\Delta\bar{\boldsymbol{F}}] \tag{26}$$

Combining Eq. (26) with Eqns. (23) and (24), it can be shown that

$$[\bar{\mathbb{A}}] = -\frac{1}{V}\left[\hat{\boldsymbol{L}}_M\right]^T \mathbf{J}_T^{-1}\left[\hat{\boldsymbol{L}}_M\right] \tag{27}$$

where the matrix $\left[\hat{\boldsymbol{L}}_M\right]$ is of size $(N + 2 + 2m) \times 4$ and is defined by

$$\left[\hat{\boldsymbol{L}}_M\right] = \begin{bmatrix} \boldsymbol{0}_{N\times 4} \\ \boldsymbol{0}_{2\times 4} \\ [\boldsymbol{L}_M]_{2m\times 4} \end{bmatrix} \tag{28}$$

where $N$ is the number of total DOFs in the displacement field, i.e. the size of $\boldsymbol{u}$ vector.

### *2.2 Uniaxial loading case – mixed stress/deformation driven formulation*

In the deformation driven homogenization analysis, the macroscopic deformation gradient $\bar{\boldsymbol{F}}$ is prescribed, serving as the macroscopic loading imposed on the RVE. Without loss of generality, the macroscopic rigid-body rotation is ignored ($\bar{\boldsymbol{R}} = \boldsymbol{I}$) and the principal macro stretch ratios $\bar{\lambda}_a$ are at a fixed angle $\theta$ with respect to the standard Euclidean bases $\{\boldsymbol{e}_a\}$, i.e.,

$$\bar{\boldsymbol{F}} = \bar{\boldsymbol{U}} = \boldsymbol{Q}\bar{\boldsymbol{F}}^Q\boldsymbol{Q}^T = \begin{bmatrix} \left(\bar{\lambda}_1\cos^2\theta + \bar{\lambda}_2\sin^2\theta\right) & \sin\theta\cos\theta\left(\bar{\lambda}_1 - \bar{\lambda}_2\right) \\ \sin\theta\cos\theta\left(\bar{\lambda}_1 - \bar{\lambda}_2\right) & \left(\bar{\lambda}_1\sin^2\theta + \bar{\lambda}_2\cos^2\theta\right) \end{bmatrix} \tag{29}$$

where $\bar{\boldsymbol{F}}^Q = \operatorname{diag}\left(\bar{\lambda}_1, \bar{\lambda}_2\right)$ and $\boldsymbol{Q}$ the bases transformation matrix expressed as

$$\boldsymbol{Q}(\theta) = \begin{bmatrix} \cos\theta & -\sin\theta \\ \sin\theta & \cos\theta \end{bmatrix} \tag{30}$$

or written equivalently in matrix-vector form as $[\bar{\boldsymbol{F}}] = [\boldsymbol{Q}_M][\bar{\boldsymbol{F}}^Q]$ with

$$[\boldsymbol{Q}_M] = \begin{bmatrix} \cos^2\theta & -\sin\theta\cos\theta & -\sin\theta\cos\theta & \sin^2\theta \\ \sin\theta\cos\theta & \cos^2\theta & -\sin^2\theta & -\sin\theta\cos\theta \\ \sin\theta\cos\theta & -\sin^2\theta & \cos^2\theta & -\sin\theta\cos\theta \\ \sin^2\theta & \sin\theta\cos\theta & \sin\theta\cos\theta & \cos^2\theta \end{bmatrix} \tag{31}$$

where $[\bar{\boldsymbol{F}}] = [\bar{F}_{11} \quad \bar{F}_{21} \quad \bar{F}_{12} \quad \bar{F}_{22}]^T$ and $[\bar{\boldsymbol{F}}^Q] = [\bar{\lambda}_1 \quad 0 \quad 0 \quad \bar{\lambda}_2]^T$ are the vector forms of $\bar{\boldsymbol{F}}$ and $\bar{\boldsymbol{F}}^Q$, respectively.

The axial stress along stretch $\bar{\lambda}_1$ direction can be calculated through a bases transformation as

$$\bar{P}_{11}^{Q} = \bar{P}_{11}\cos^2\theta + \bar{P}_{21}\cos\theta\sin\theta + \bar{P}_{12}\cos\theta\sin\theta + \bar{P}_{22}\sin^2\theta \tag{32}$$

where $\bar{P}_{ij}$ is the component of the homogenized stress tensor $\bar{\boldsymbol{P}}$ expressed in terms of the standard bases $\{\boldsymbol{e}_a\}$.

The uniaxial loading condition is achieved by a mixed stress/deformation driven homogenization formulation, where $\bar{\lambda}_2$ and $\theta$ is prescribed together with the stress condition $\bar{P}_{11}^{Q} = 0$. Thus, the macro stretch $\bar{\lambda}_1$ is unknown, and is obtained by the requirement that

$$H\left(\bar{P}_{ij}(\bar{\lambda}_1)\right) = \bar{P}_{11}^{Q} = 0 \tag{33}$$

As a result, the deformation driven homogenization presented in Section 2.1 can be seen as an inner loop where $\bar{\boldsymbol{F}}$ is given, and the outer loop is a scalar-valued nonlinear equation $H\left(\bar{P}_{ij}(\bar{\lambda}_1)\right) = 0$ that solves for $\bar{\lambda}_1$. The Jacobian for the outer loop is derived as

$$\begin{aligned}
\mathrm{J}_{oT} &\stackrel{\text{def}}{=} \frac{\partial H\left(\bar{P}_{ij}(\bar{\lambda}_1)\right)}{\partial\bar{\lambda}_1} = \frac{\partial H\left(\bar{P}_{ij}(\bar{\lambda}_1)\right)}{\partial\bar{F}_{pq}}\frac{\partial\bar{F}_{pq}}{\partial\bar{\lambda}_1} \\
&= \left(\bar{\mathbb{A}}_{1111}\frac{\partial\bar{F}_{11}}{\partial\bar{\lambda}_1} + \bar{\mathbb{A}}_{1112}\frac{\partial\bar{F}_{12}}{\partial\bar{\lambda}_1} + \bar{\mathbb{A}}_{1121}\frac{\partial\bar{F}_{21}}{\partial\bar{\lambda}_1} + \bar{\mathbb{A}}_{1122}\frac{\partial\bar{F}_{22}}{\partial\bar{\lambda}_1}\right)\cos^2\theta \\
&+ \left(\bar{\mathbb{A}}_{2111}\frac{\partial\bar{F}_{11}}{\partial\bar{\lambda}_1} + \bar{\mathbb{A}}_{2112}\frac{\partial\bar{F}_{12}}{\partial\bar{\lambda}_1} + \bar{\mathbb{A}}_{2121}\frac{\partial\bar{F}_{21}}{\partial\bar{\lambda}_1} + \bar{\mathbb{A}}_{2122}\frac{\partial\bar{F}_{22}}{\partial\bar{\lambda}_1}\right)\cos\theta\sin\theta \\
&+ \left(\bar{\mathbb{A}}_{1211}\frac{\partial\bar{F}_{11}}{\partial\bar{\lambda}_1} + \bar{\mathbb{A}}_{1212}\frac{\partial\bar{F}_{12}}{\partial\bar{\lambda}_1} + \bar{\mathbb{A}}_{1221}\frac{\partial\bar{F}_{21}}{\partial\bar{\lambda}_1} + \bar{\mathbb{A}}_{1222}\frac{\partial\bar{F}_{22}}{\partial\bar{\lambda}_1}\right)\cos\theta\sin\theta \\
&+ \left(\bar{\mathbb{A}}_{2211}\frac{\partial\bar{F}_{11}}{\partial\bar{\lambda}_1} + \bar{\mathbb{A}}_{2212}\frac{\partial\bar{F}_{12}}{\partial\bar{\lambda}_1} + \bar{\mathbb{A}}_{2221}\frac{\partial\bar{F}_{21}}{\partial\bar{\lambda}_1} + \bar{\mathbb{A}}_{2222}\frac{\partial\bar{F}_{22}}{\partial\bar{\lambda}_1}\right)\sin^2\theta
\end{aligned} \tag{34}$$

where $\bar{\mathbb{A}}_{ijkl} \stackrel{\text{def}}{=} \partial\bar{P}_{ij}/\partial\bar{F}_{kl}$ is computed in the inner loop (Eq. (27)) and

$$\frac{\partial\bar{F}_{11}}{\partial\bar{\lambda}_1} = \cos^2\theta, \quad \frac{\partial\bar{F}_{12}}{\partial\bar{\lambda}_1} = \frac{\partial\bar{F}_{21}}{\partial\bar{\lambda}_1} = \sin\theta\cos\theta, \quad \frac{\partial\bar{F}_{22}}{\partial\bar{\lambda}_1} = \sin^2\theta \tag{35}$$

## 3 Density Based Multimaterial Topology Optimization

In the density based multimaterial topology optimization, in particular with three phases including void phase, the design is described by two element-wise density fields $\rho_1(\boldsymbol{X})$ and $\rho_2(\boldsymbol{X})$, where $\rho_1$ indicates if the material, either material-1 or material-2, is present ($\rho_1 = 1$) or absent ($\rho_1 = 0$), and $\rho_2$ denotes the proportion of the material-1, i.e. $\rho_2 = 1$ means full of material-1 while $\rho_2 = 0$ means full of material-2. To accommodate gradient-based optimizers, the density variables are relaxed to continuous values, i.e. $\rho_1 \in [0,1]$ and $\rho_2 \in [0,1]$, where $0 < \rho_1 < 1$ represents the mixture of void phase and solid phase, either material-1 or material-2, while $0 < \rho_2 < 1$ represents the mixture of two material phases.

### *3.1 Material interpolation scheme*

In this section, a multi-incompressible-material interpolation is proposed. The considered three phases include a void phase and two nearly incompressible isotropic hyperelastic solid phases. The material interpolation scheme can be seen as an extension of the existing multimaterial interpolation scheme [15] that is tailored to include the incompressible case. The material interpolation is carried out on the Helmholtz free energy, where the void phase is also modeled as a hyperelastic material, but with diminishing properties. Thus, the free energy in the considered multimaterial interpolation scheme is expressed as

$$\psi(\rho_1, \rho_2, \boldsymbol{F}) = \psi_v(\rho_1, \boldsymbol{F}) + \psi_1(\rho_1, \rho_2, \boldsymbol{F}) + \psi_2(\rho_1, \rho_2, \boldsymbol{F}) \tag{36}$$

where $\psi_v$, $\psi_1$ and $\psi_2$ denote the interpolated free energy of the void ($v$) phase, the 1st hyperelastic phase and the 2nd hyperelastic phase, respectively. Free energy of the void phase is interpolated as

$$\psi_v(\rho_1, \boldsymbol{F}) = \left(1 - \rho_1^{p_e}\right)\left[\hat{\psi}_v(\boldsymbol{F}) + \tilde{\psi}_v(\boldsymbol{F})\right] \tag{37}$$

where $\hat{\psi}_v$ and $\tilde{\psi}_v$ are the volumetric and isochoric contributions of the free energy, respectively, and $p_e$ is the penalization power used for penalizing the density field $\rho_1$. Free energies of the 1st and 2nd hyperelastic phases are interpolated as

$$\psi_1(\rho_1, \rho_2, \boldsymbol{F}) = \rho_2^p\left[\zeta^\kappa(\rho_1)\hat{\psi}_1(\boldsymbol{F}) + \zeta^\mu(\rho_1)\tilde{\psi}_1(\boldsymbol{F})\right] \tag{38}$$

and

$$\psi_2(\rho_1, \rho_2, \boldsymbol{F}) = (1 - \rho_2)^p\left[\zeta^\kappa(\rho_1)\hat{\psi}_2(\boldsymbol{F}) + \zeta^\mu(\rho_1)\tilde{\psi}_2(\boldsymbol{F})\right] \tag{39}$$

where the volumetric ($\hat{\psi}_1$ or $\hat{\psi}_2$) and isochoric contributions ($\tilde{\psi}_1$ or $\tilde{\psi}_2$) are separately interpolated, and $p$ is the penalization power used for penalizing the density field $\rho_2$. The interpolation functions $\zeta^\kappa(\rho_1)$ and $\zeta^\mu(\rho_1)$ in the Eqns. (38) and (39) are evaluated using material parameters according to the phase to which it is attached and are defined based on the $E$-$\nu$ interpolation rule proposed in [33], which is used for relaxing the incompressibility in the intermediate-density elements. Following the work by Zhang et al. [33], the functions $\zeta^\kappa(\rho_1)$ and $\zeta^\mu(\rho_1)$ are determined by

$$\zeta^{\kappa}(\rho_1) = \frac{\kappa(\rho_1)}{\kappa_0} \qquad \text{and} \qquad \zeta^{\mu}(\rho_1) = \frac{\mu(\rho_1)}{\mu_0} \tag{40}$$

where the bulk modulus $\kappa(\rho_1)$ and shear modulus $\mu(\rho_1)$ are related to the Young's modulus $E(\rho_1)$ and Poisson's ratio $\nu(\rho_1)$ by

$$\kappa(\rho_1) = \frac{E(\rho_1)}{3\big(1 - 2\nu(\rho_1)\big)} \qquad \text{and} \qquad \mu(\rho_1) = \frac{E(\rho_1)}{2\big(1 + \nu(\rho_1)\big)} \tag{41}$$

with $\kappa_0$ and $\mu_0$ the initial bulk and shear modulus of the solid material phase, and $E(\rho_1)$ and $\nu(\rho_1)$ interpolated using the $E$-$\nu$ interpolation scheme as

$$\begin{aligned} E(\rho_1) &= \rho_1^{p_e} E_0 \\ \nu(\rho_1) &= [\epsilon_\nu + (1 - \epsilon_\nu)(1 - (1 - \rho_1)^{p_\nu})]\nu_0 \end{aligned} \tag{42}$$

where the initial Young's modulus $E_0$ and Poisson's ratio $\nu_0$ of the solid phase are determined from $\kappa_0$ and $\mu_0$; $\epsilon_\nu$ is the lower bound parameter for the Poisson's ratio and is chosen as $\epsilon_\nu = 0.4$ and $p_\nu$ is the penalization power. It is noted that the material interpolation can be recovered for single material (only 1st hyperelastic material) topology optimization by letting $\rho_2 \equiv 1$ in the entire design domain, which is similar to the interpolation scheme proposed in [33].

### *3.2 Topology optimization formulation*

To design a material with NPR it is considered that the material under uniaxial tension (or compression) along the $\bar{\lambda}_2$ direction ($\bar{\lambda}_2$ is specified and $\bar{P}_{11}^{Q} = 0$ is enforced, see Section 2.2) expands (or contract) along the $\bar{\lambda}_1$ direction. The Poisson's ratio $\bar{\nu}$ of the metamaterial is defined by

$$\bar{\nu} \stackrel{\text{def}}{=} -\frac{\bar{\lambda}_1 - 1}{\bar{\lambda}_2 - 1} \tag{43}$$

For a nonlinear hyperelastic material, the material's behavior depends on the applied loads. Thus, in order to design a material that has a constant negative Poisson's ratio over the considered load path with $n$ loading steps, $\bar{\nu}^1 = \bar{\nu}^2 = \cdots = \bar{\nu}^n$ is required, where $\bar{\nu}^k$ $(k = 1,2, \dots, n)$ denotes the Poisson's ratio at loading step $k$. To enforce this condition, the objective function $f_0$ to be minimized can be constructed as

$$f_0 = \sum_{k=1}^{n} \left( -\frac{\bar{\lambda}_1^k - 1}{\bar{\lambda}_2^k - 1} - \bar{\nu}_T \right)^2 \tag{44}$$

with the target value of Poisson's ratio is assumed to be $\bar{\nu}_T$, where $\bar{\nu}_T$ is prescribed based on the design requirements. The formulation in Eq. (44) is, however, numerically unstable due to the potential singularity when $\bar{\lambda}_2^k \to 1$. Thus, this objective function is modified to

$$f_0 = \sum_{k=1}^{n} \left(\bar{\lambda}_1^k + \bar{\nu}_T \bar{\lambda}_2^k - \bar{\nu}_T - 1\right)^2 \tag{45}$$

which gives as numerically stable measure of Poisson's ratio. In this study, both single and multimaterial topology optimizations are considered for the NPR metamaterials designs. For the NPR designs with single material phase, the optimization formulation (OF-1) is given by

$$\begin{aligned} &\min_{\boldsymbol{x}_1 \in \mathcal{D}} \ f_0(\boldsymbol{x}_1) + \alpha_1 V_f(\boldsymbol{x}_1) \\ &\text{s.t.} \ f_1(\boldsymbol{x}_1) = 1 - \left[\bar{\mathbb{A}}_0^Q\right]_{11} / \bar{k} \le 0 \\ &\qquad f_2(\boldsymbol{x}_1) = 1 - \left[\bar{\mathbb{A}}_0^Q\right]_{44} / \bar{k} \le 0 \\ &\qquad f_3(\boldsymbol{x}_1) = V_f(\boldsymbol{x}_1) - V_T \le 0 \end{aligned} \tag{46}$$

For the NPR designs with multiple material phases, the optimization formulation (OF-2) is given by

$$\begin{aligned} &\min_{\boldsymbol{x}_1, \boldsymbol{x}_2 \in \mathcal{D}} f_0(\boldsymbol{x}_1, \boldsymbol{x}_2) + \alpha_2 M_f(\boldsymbol{x}_1, \boldsymbol{x}_2) \\ &\text{s.t.} \ f_1(\boldsymbol{x}_1, \boldsymbol{x}_2) = 1 - \left[\bar{\mathbb{A}}_0^Q\right]_{11} / \bar{k} \le 0 \\ &\qquad f_2(\boldsymbol{x}_1, \boldsymbol{x}_2) = 1 - \left[\bar{\mathbb{A}}_0^Q\right]_{44} / \bar{k} \le 0 \\ &\qquad f_3(\boldsymbol{x}_1, \boldsymbol{x}_2) = M_f(\boldsymbol{x}_1, \boldsymbol{x}_2) - 1 \le 0 \end{aligned} \tag{47}$$

where $\alpha_1 \ge 0$ and $\alpha_2 \ge 0$.

$$V_f(\boldsymbol{x}_1) = \frac{1}{V} \sum_{e=1}^{n_{ele}} \rho_1^e(\boldsymbol{x}_1) v_e \tag{48}$$

is the total material volume fraction and

$$M_f(\boldsymbol{x}_1, \boldsymbol{x}_2) = \frac{1}{M^*} \sum_{e=1}^{n_{ele}} \left[\omega_1 \rho_1^e \rho_2^e v_e + \omega_2 \rho_1^e (1 - \rho_2^e) v_e\right] \tag{49}$$

is the total material weight ratio. $\left[\bar{\mathbb{A}}_0^Q\right]_{11}$ and $\left[\bar{\mathbb{A}}_0^Q\right]_{44}$ in Eqns. (46) and (47) are the macroscale initial stiffness along the loading axis and the direction orthogonal to the loading axis, respectively. The initial stiffness matrix $\left[\bar{\mathbb{A}}_0^Q\right]$, which is expressed in the eigenspace of $\bar{\boldsymbol{U}}$, is defined by $\left[\bar{\mathbb{A}}_0^Q\right] \stackrel{\text{def}}{=}$ $[\boldsymbol{Q}_M]^T[\bar{\mathbb{A}}_0][\boldsymbol{Q}_M]$, with $[\bar{\mathbb{A}}_0]$ given in Eq. (27) but with the tangent Jacobian matrix $\mathbf{J}_T$ replaced by the initial Jacobian matrix $\mathbf{J}_0$ and $[\boldsymbol{Q}_M]$ is given in Eq. (31). The scalar $\bar{k}$ represents a predefined macroscopic required initial stiffness value; $v_e$ denotes the $e^{th}$ element volume; $V_T$ defines the maximum allowable material volume fraction; $\omega_1$ and $\omega_2$ represent the physical densities of 1st and 2nd hyperelastic material phases, and $M^*$ is allowable upper limit on the total mass. Further box constraints are enforced on the design variables such that $\mathcal{D} = [0,1]^{n_{ele}}$.

***Remark:*** The addition of the terms $\alpha_1 V_f(\boldsymbol{x}_1)$ or $\alpha_2 M_f(\boldsymbol{x}_1, \boldsymbol{x}_2)$ on the objective function $f_0$ in Eqns. (46) and (47) can be seen as a multi-objective formulation, where beside the minimization of function $f_0$ the material usage is also minimized under the given constraints. These multi-objective formulations are needed in some cases, since it is difficult to predetermine an appropriate combination of the target Poisson's ratio ($\bar{\nu}_T$), stiffness ($\bar{k}$) and material volume ($V_T$) or weight ($M^*$) constraints that will lead to a discrete NPR design. For instance, under a given material volume constraint, if the stiffness constraint is set too high, the optimized topology will approach a stiffness design, which is different from a NPR design. On the other hand, if the stiffness constraint is set too low, even though the target NPR design is achievable, a discrete topology is not guaranteed. This is because of the fact that the penalty effect brought by the penalization in the material interpolation scheme is not completely reflected in the objective function, i.e. Poisson's ratio, since it is not related to the stiffness of material. This necessity of adding $\alpha_1 V_f(\boldsymbol{x}_1)$ or $\alpha_2 M_f(\boldsymbol{x}_1, \boldsymbol{x}_2)$ terms in the objective functions will be made clear through the numerical examples in the following sections.

### *3.3 Density filter – periodic formulation*

Density filter [34, 35] is utilized in this study to avoid the mesh dependency and checkerboard issues. The filter can be expressed in a matrix form as

$$\boldsymbol{\rho}_1 = \boldsymbol{W}\boldsymbol{x}_1 \quad \text{and} \quad \boldsymbol{\rho}_2 = \boldsymbol{W}\boldsymbol{x}_2 \tag{50}$$

where $\boldsymbol{\rho}_1$ and $\boldsymbol{\rho}_2$ are the vectors containing the filtered design variables; $\boldsymbol{W}$ is the filtering matrix that can be expressed in component form as

$$W_{ij} = \frac{w_{ij} v_j}{\sum_{j=1}^{n_{ele}} w_{ij} v_j} \quad \text{with} \quad w_{ij} = \max\left(r_{min} - d\left(\boldsymbol{X}_i, \boldsymbol{X}_j\right), 0\right) \tag{51}$$

where $r_{min}$ is the filter radius and $\boldsymbol{X}_i$ denotes the coordinates of the centroid of $i^{\text{th}}$ element. The distance between points $\boldsymbol{X}_i$ and $\boldsymbol{X}_j$ should take the spatial periodicity of the RVE into account, i.e.,

$$d(\boldsymbol{X}_i, \boldsymbol{X}_j) = \min_{\boldsymbol{L} \in \mathcal{Q}} \left\| \boldsymbol{X}_i - (\boldsymbol{X}_j + \boldsymbol{L}) \right\|_2 \quad \text{with } \mathcal{Q} \stackrel{\text{def}}{=} \{\boldsymbol{L} | \boldsymbol{L} = \textstyle\sum_{i=1}^{d} c_i \boldsymbol{a}_i \ , \ c_i \in \mathbb{Z}\} \tag{52}$$

where $\mathbb{Z}$ stands for the set of integers. See Figure 3 for an illustration filter in parallelogram and hexagonal RVE domains.

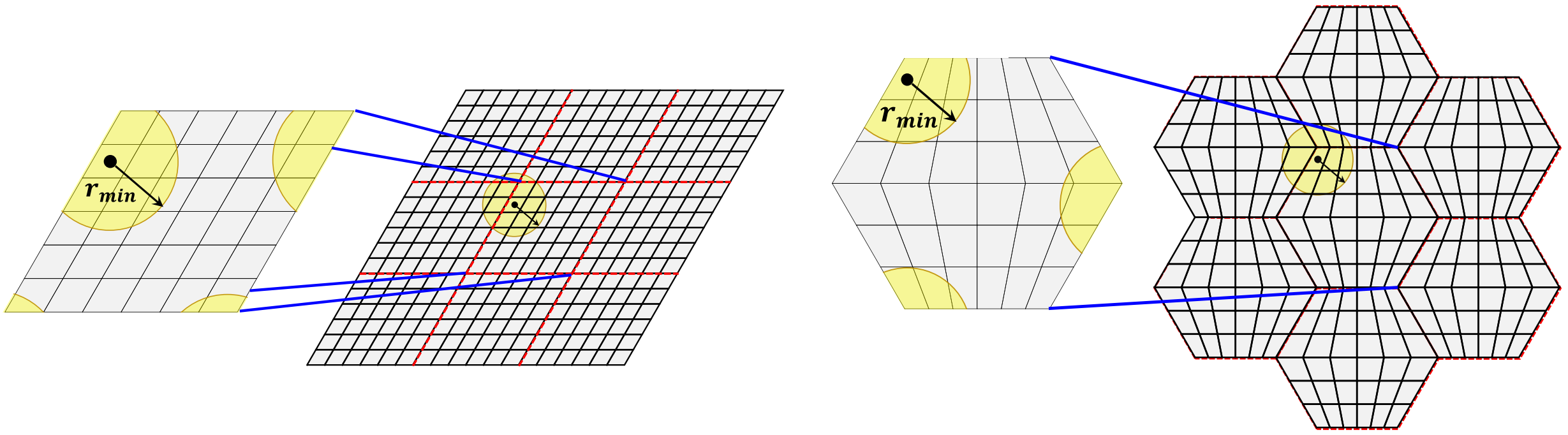

(a) Parallelogram RVE (b) Hexagonal RVE

Figure 3. Density filter considering periodicity.

### 3.4 FE mesh distortion

The finite element mesh distortion issue due to the use of fictitious domain approach is addressed by introducing the linear energy interpolation scheme, which was first proposed in [36] and later extended to an adaptive scheme in [33]. Following [33], the idea is to interpolate between the linear and nonlinear kinematics based on the solid/void density field $\rho_1$. Thus, the deformation gradient $\boldsymbol{F}$ is interpolated using the volume fraction of void phase as

$$\boldsymbol{F} = \boldsymbol{I} + \gamma \nabla_{\boldsymbol{X}} \boldsymbol{u} \quad \text{with} \quad \gamma(\rho_1) = \frac{\exp(\beta \rho_1)}{\exp(c\beta) + \exp(\beta \rho_1)} \tag{53}$$

where $c$ and $\beta$ are interpolation parameters. Following [33], the element internal force vector Eq. (19) is modified to

$$\boldsymbol{F}_{int}^{e} = \int_{\Omega^e} \gamma \boldsymbol{B}^T \boldsymbol{P} \, dV + \int_{\Omega^e} (1 - \gamma^2) \boldsymbol{B}_L^T [\mathbb{C} \colon \boldsymbol{\varepsilon}] \, dV \tag{54}$$

where $\boldsymbol{B}_L$ denotes the derivative of the shape functions of a regular 4-node element, $\boldsymbol{\varepsilon} = \nabla_{\boldsymbol{X}}^{s} \boldsymbol{u}$ represents the small strain measure and $\mathbb{C}$ the linear isotropic elastic moduli determined by the interpolated Young's modulus $E(\rho_1) = [\epsilon_E + (1 - \epsilon_E)\rho_1^{p_L}]E_0$ and constant Poisson's ratio $\nu = 0.2$, where $\epsilon_E = 10^{-8}$ and $E_0$ is chosen identical to that of the initial Young's modulus of the hyperelastic material phase (the soft one if two solid hyperelastic phases are considered), and $p_L$ is the penalization power. The remaining parameters are $\beta = 120$ and $c = 0.08$, where $c$ is adaptively updated (if needed) using the scheme proposed in [33], i.e. $c$ is updated to $c + \Delta c$ with $\Delta c$ = 0.05 until the convergence of FEA is achieved.

### 3.5 F-bar elements

Since incompressible material phases are considered, to avoid volumetric locking F-bar element formulation [37] is adopted, where the deformation gradient $\boldsymbol{F}$ is modified to

$$\boldsymbol{F}^b = r^{1/2}\boldsymbol{F} \quad \text{(in-plane part)} \qquad \text{with } r = \frac{\det \boldsymbol{F}_0}{\det \boldsymbol{F}} \tag{55}$$

where $\boldsymbol{F}_0$ is the deformation gradient evaluated at the centroid of the element. As a result, the 1st Piola-Kirchhoff stress tensor is modified to

$$\boldsymbol{P} = r^{-1/2}\boldsymbol{P}^b \quad \text{(in-plane part)} \tag{56}$$

where the deformation gradient-stress pair $(\boldsymbol{F}^b, \boldsymbol{P}^b)$ serves as the input-output of the regular material subroutine. Note that due to the introduction of the linear energy interpolation scheme (Section 3.4), both $\boldsymbol{F}$ and $\boldsymbol{F}_0$ in Eq. (55) are evaluated based on the interpolated displacement field. Further details on the implementation and performance of F-bar elements can be found in [37].

## 4 Sensitivity Analysis

The use of hyperelastic material leads to a path-independent finite element analysis. However, due to the path dependency of the objective function $f_0$ as defined in Eq. (45), the sensitivity analysis of $f_0$ still needs to include all the analysis steps. In contrast, the sensitivity analyses of the constraints ($f_1$ and $f_2$) take only the initial undeformed configuration into account. The sensitivity calculations of $f_3$, $V_f(\boldsymbol{x}_1)$ and $M_f(\boldsymbol{x}_1, \boldsymbol{x}_2)$ are straightforward and are therefore omitted. The sensitivity of important quantities with respect to the physical density variables, $\boldsymbol{\rho}_1$ and $\boldsymbol{\rho}_2$, are provided and the density filter (Eq. (50)) can be incorporated using a simple chain rule.

### 4.1 Sensitivity analysis of $f_0$

The path-dependent sensitivity analysis is carried out using the adjoint sensitivity framework presented in [38]. Due to the path dependency of the objective function described by the principal stretch $\bar{\lambda}_1$ along the loading process, the adjoint function is constructed as

$$\hat{f}_0 = f_0(v^1, \dots, v^n) + \sum_{k=1}^{n} \boldsymbol{\gamma}^{k^T} \boldsymbol{R}^k(\hat{\boldsymbol{u}}^k, v^k, \boldsymbol{\rho}_1, \boldsymbol{\rho}_2) + \sum_{k=1}^{n} \eta^k H^k(\hat{\boldsymbol{u}}^k) \tag{57}$$

where $\hat{\boldsymbol{u}}^k \equiv [\boldsymbol{u}^k, \boldsymbol{\lambda}^k, \boldsymbol{\mu}^k]^T$ and $v^k \equiv \bar{\lambda}_1^k$ are the solution and auxiliary variables at step $k$ and are determined by the corresponding global system $\boldsymbol{R}^k = \boldsymbol{0}$ given in Eq. (18) and a nonlinear equation $H^k = 0$ given in Eq. (33); $\boldsymbol{\gamma}^k$ and $\eta^k$ are the adjoint variables and $n$ is the total number of load steps. Clearly, $d\hat{f}_0/d\boldsymbol{\rho}_1 \equiv df_0/d\boldsymbol{\rho}_1$ and $d\hat{f}_0/d\boldsymbol{\rho}_2 \equiv df_0/d\boldsymbol{\rho}_2$ since the constraints $\boldsymbol{R}^k = \boldsymbol{0}$ and $H^k = 0$ are always satisfied irrespective of the density variables $\boldsymbol{\rho}_1$ and $\boldsymbol{\rho}_2$. Taking derivatives of $\hat{f}_0$ with respect to $\boldsymbol{\rho}_1$ (or $\boldsymbol{\rho}_2$) and eliminating all the terms that contain the implicit derivatives $d\hat{\boldsymbol{u}}^k/d\boldsymbol{\rho}_1$ and $dv^k/d\boldsymbol{\rho}_1$ $(k = 1, \dots, n)$ yield

$$\frac{d\hat{f}_0}{d\boldsymbol{\rho}_1} = \sum_{k=1}^{n} \boldsymbol{\gamma}^{k^T} \frac{\partial \boldsymbol{R}^k}{\partial \boldsymbol{\rho}_1} \quad \text{or} \quad \frac{d\hat{f}_0}{d\boldsymbol{\rho}_2} = \sum_{k=1}^{n} \boldsymbol{\gamma}^{k^T} \frac{\partial \boldsymbol{R}^k}{\partial \boldsymbol{\rho}_2} \tag{58}$$

since $\partial f_0/\partial \boldsymbol{\rho}_1 = \partial f_0/\partial \boldsymbol{\rho}_2 = \boldsymbol{0}$ and $\partial H^k/\partial \boldsymbol{\rho}_1 = \partial H^k/\partial \boldsymbol{\rho}_2 = \boldsymbol{0}$, and the adjoint variables $\boldsymbol{\gamma}^k$ and $\eta^k$ are calculated by $(k = 1, \dots, n)$

$$\begin{bmatrix} \boldsymbol{\gamma}^{k^T} & \eta^k \end{bmatrix} \begin{bmatrix} \dfrac{\partial \boldsymbol{R}^k}{\partial \hat{\boldsymbol{u}}^k} & \dfrac{\partial \boldsymbol{R}^k}{\partial v^k} \\ \dfrac{\partial H^k}{\partial \hat{\boldsymbol{u}}^k} & 0 \end{bmatrix} = \begin{bmatrix} \boldsymbol{0} & -\dfrac{\partial f_0}{\partial v^k} \end{bmatrix} \quad \left( \text{Noticing } \frac{\partial f_0}{\partial \hat{\boldsymbol{u}}^k} = \boldsymbol{0} \text{ and } \frac{\partial H^k}{\partial v^k} = 0 \right) \tag{59}$$

Hence, to complete the sensitivity analysis, the derivatives that need to be calculated are

$$\frac{\partial f_0}{\partial v^k}, \quad \frac{\partial \boldsymbol{R}^k}{\partial \hat{\boldsymbol{u}}^k}, \quad \frac{\partial \boldsymbol{R}^k}{\partial v^k}, \quad \frac{\partial \boldsymbol{R}^k}{\partial \boldsymbol{\rho}_1}, \quad \frac{\partial \boldsymbol{R}^k}{\partial \boldsymbol{\rho}_2}, \quad \frac{\partial H^k}{\partial \hat{\boldsymbol{u}}^k}$$

The detailed expressions for these derivatives are given in Appendix A.

### *4.2 Sensitivity analysis of $f_1$ and $f_2$*

Using the chain rule yields

$$\frac{df_1}{d\rho_A^e} = -\frac{1}{\bar{k}} \frac{d[\bar{\mathbb{A}}_0^Q]_{11}}{d\rho_A^e} \quad , \quad \frac{df_2}{d\rho_A^e} = -\frac{1}{\bar{k}} \frac{d[\bar{\mathbb{A}}_0^Q]_{44}}{d\rho_A^e} \quad , \quad A = 1, 2\,; \quad e = 1, 2, \dots, n_{ele} \tag{60}$$

On the other hand, since $[\bar{\mathbb{A}}_0^Q]$ is related to $[\bar{\mathbb{A}}_0]$ by a basis transformation, so are their derivatives, i.e.,

$$\frac{d[\bar{\mathbb{A}}_0^Q]}{d\rho_A^e} = [\boldsymbol{Q}_M]^T \frac{d[\bar{\mathbb{A}}_0]}{d\rho_A^e} [\boldsymbol{Q}_M] \quad , \quad A = 1,2\,; \quad e = 1,2, \dots, n_{ele} \tag{61}$$

As $[\bar{\mathbb{A}}_0]$ is determined by the initial Jacobian matrix evaluated at the undeformed configuration $\mathbf{J}_0$, i.e. $[\bar{\mathbb{A}}_0] = -\frac{1}{V}[\hat{\boldsymbol{L}}_M]^T \mathbf{J}_0^{-1} [\hat{\boldsymbol{L}}_M]$, the derivative $d[\bar{\mathbb{A}}_0]/d\rho_A^e$ can be computed by

$$\frac{d[\bar{\mathbb{A}}_0]}{d\rho_A^e} = -\frac{1}{V}\left[\hat{\boldsymbol{L}}_M\right]^T \frac{d\mathbf{J}_0^{-1}}{d\rho_A^e}\left[\hat{\boldsymbol{L}}_M\right] \tag{62}$$

with

$$\frac{d\mathbf{J}_0^{-1}}{d\rho_A^e} = -\mathbf{J}_0^{-1}\frac{d\mathbf{J}_0}{d\rho_A^e}\mathbf{J}_0^{-1} \quad \text{and} \quad \frac{d\mathbf{J}_0}{d\rho_A^e} = \begin{bmatrix} d\boldsymbol{K}_0/d\rho_A^e & -\boldsymbol{M}_1^T & -\boldsymbol{M}_2^T \\ -\boldsymbol{M}_1 & \mathbf{0} & \mathbf{0} \\ -\boldsymbol{M}_2 & \mathbf{0} & \mathbf{0} \end{bmatrix} \tag{63}$$

where the remaining term $d\boldsymbol{K}_0/d\rho_A^e$ can be easily computed.

## 5 Numerical Examples

In the following examples, the isotropic hyperelastic material phases are modeled using regularized neo-Hookean model for which the free energy is expressed as

$$\begin{aligned} &\psi(\boldsymbol{C}) = \hat{\psi}(J) + \tilde{\psi}(\widetilde{\boldsymbol{C}}) \quad \text{with} \\ &\hat{\psi}(J) = \frac{\kappa}{2}(J-1)^2 \quad \text{and} \quad \tilde{\psi}(\widetilde{\boldsymbol{C}}) = \frac{\mu}{2}(\tilde{I}_1 - 3) \end{aligned} \tag{64}$$

where $J = \det \boldsymbol{F}$, $\boldsymbol{C} = \boldsymbol{F}^T\boldsymbol{F}$ and $\tilde{I}_1 = \mathrm{tr}(\widetilde{\boldsymbol{C}})$ with $\widetilde{\boldsymbol{C}} = J^{-2/3}\boldsymbol{C}$, and $\kappa$ and $\mu$ are the bulk and shear modulus, which are related to Young's modulus $E$ and Poisson's ratio $\nu$ through Eq. (41). Except for the last example with multiple materials, the other examples consider NPR designs with a single hyperelastic material with properties: $E$ = 100 and $\nu$ = 0.49. The void phase in both the single and multimaterial cases is modeled with material properties $E = 10^{-6}$ and $\nu$ = 0.2.

When considering finite deformations, continuation is usually needed to avoid FE analysis failure during early optimization iterations when intermediate density design is present [33]. The employed continuation scheme incrementally updates $p_e$ and $p$ from 1 to 3, $p_L$ from 4 to 6 and $p_\nu$ from 3 to 1 with an increment/decrement of 0.1 every 20 iterations. This increased penalization of $p_L$ compared to $p_e$ is done so that the optimizer does not use low-density values to exploit small deformation kinematics [33]. It is noted that when evaluating the initial tangent stiffness constraints, i.e. $\left[\bar{\mathbb{A}}_0^Q\right]_{11}$ and $\left[\bar{\mathbb{A}}_0^Q\right]_{44}$ in Eqns. (46) and (47), the material penalization $p_e$ and $p$ are taken to be 3 for the first 50 iterations and increased to 5 with an increment of 0.1 every 20 iterations, while $p_\nu$ is taken to be 1 throughout the optimization process. Also, the linear energy interpolation (Eq. (53)) is not considered in the evaluation of $\left[\bar{\mathbb{A}}_0^Q\right]_{11}$ and $\left[\bar{\mathbb{A}}_0^Q\right]_{44}$, since there is no mesh distortion at the initial step.

All design domains – with square, parallelogram and hexagonal unit cells – are discretized by a 80×80 FE mesh. In nonlinear FEA, the tolerance criterion for convergence of the adaptive step size Newton-Raphson (NR) scheme utilized for analysis is $10^{-12}$ in terms of the energy residual. In the adaptive load-stepping NR scheme, the initial and maximum load ratios are set to 0.05, while the minimum load ratio is set to 0.001. Therefore, at least 20 steps are involved in the FEA, i.e. $n \geq 20$, resulting in an adequate control of the Poisson's ratio over the entire loading process. The Method of Moving Asymptotes (MMA) [39] is used as the optimizer with default parameter settings. The density filter radius is set to $r_{min}$ = 0.0375 for all problems. All the numerical computations are carried out in a Matlab based in-house finite element library *CPSSL-FEA* developed at the University of Notre Dame.

### *5.1 Stiffness constraint – a parametric study*

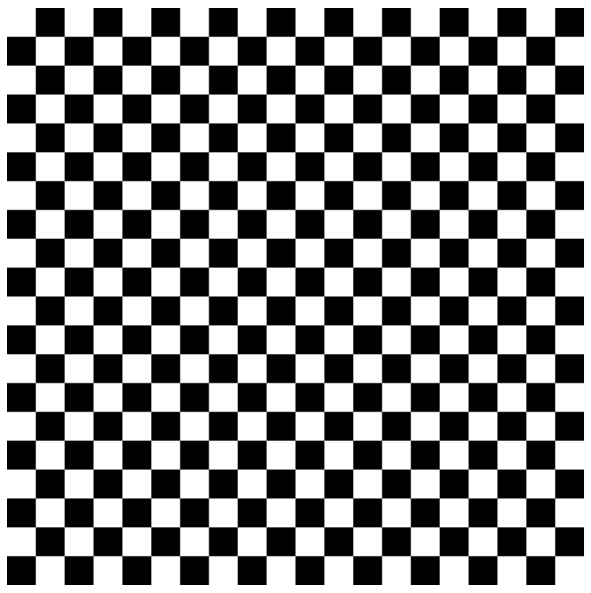

Figure 4. Initial design.

The first example is used to illustrate the importance of using an appropriate combination of stiffness and material volume constraints to achieve a discrete optimized topology with a target NPR. The design domain is a square unit cell of dimension 1×1 with a unit thickness. Uniaxial tension with stretch ratio of $\bar{\lambda}_2$ = 1.2 is considered together with OF-1 and $\bar{\nu}_T = -1$ . The initial design is chosen as a checkerboard design, see Figure 4. With a fixed material volume $V_T$ = 0.4, different stiffness constraints are selected with $\bar{k}$ = 2, 3 and 5. Without minimizing the material usage, i.e. $\alpha_1$ = 0 in Eq. (46), the final optimized topologies are shown in Figure 5, where it can be seen that the discreteness of a topology can be improved by increasing the stiffness constraint. However, the target negative Poisson's ratio is eventually deteriorated by increasing the initial stiffness constraint. Existence of the intermediate density elements in Figure 5a can be predicated on the fact that there is more material than that is required for designing the microstructure with $\bar{\nu}_T = -1$, while simultaneously fulfilling the stiffness constraints with $\bar{k}$ = 2. The penalization built in the material interpolation scheme is accordingly not reflected in the objective function $f_0$. From Figure 5a and Figure 5b, it is reasonable to conjecture that for the material volume constraint $V_T$ = 0.4, an appropriate value for $\bar{k}$ should be between 2 and 3 to obtain a discrete topology with no intermediate floating densities. However, it is not straightforward to predetermine $\bar{k}$ such that a discrete topology can be achieved with the prescribed $V_T$ and $\bar{v}_T$.

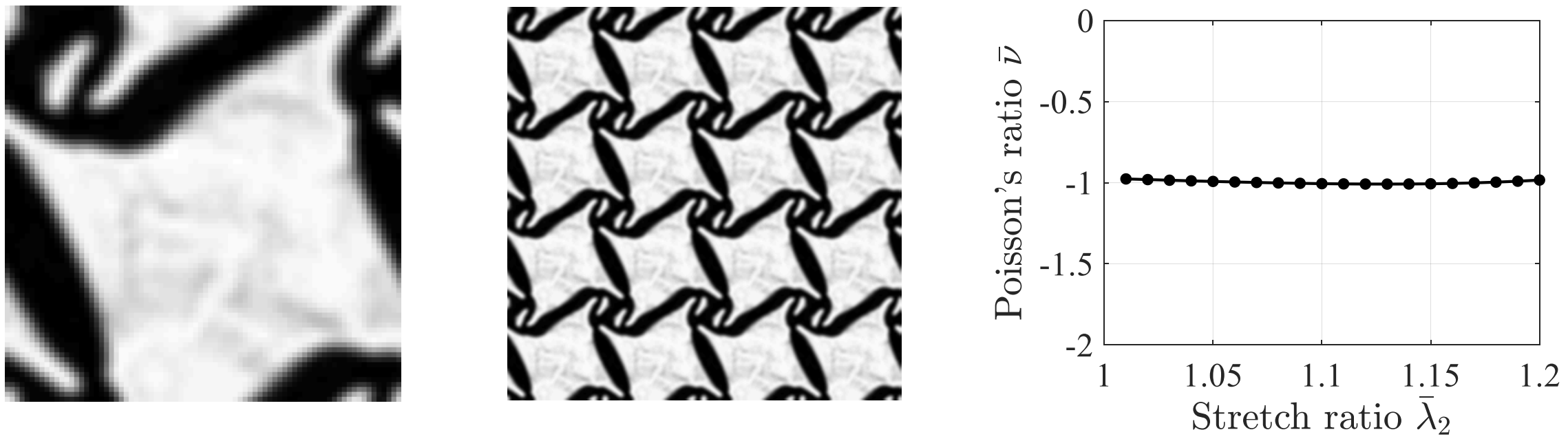

(a) $\bar{k}$ = 2 ($f_0$ = 2.11×10$^{-5}$, $V_f$ = 0.400)

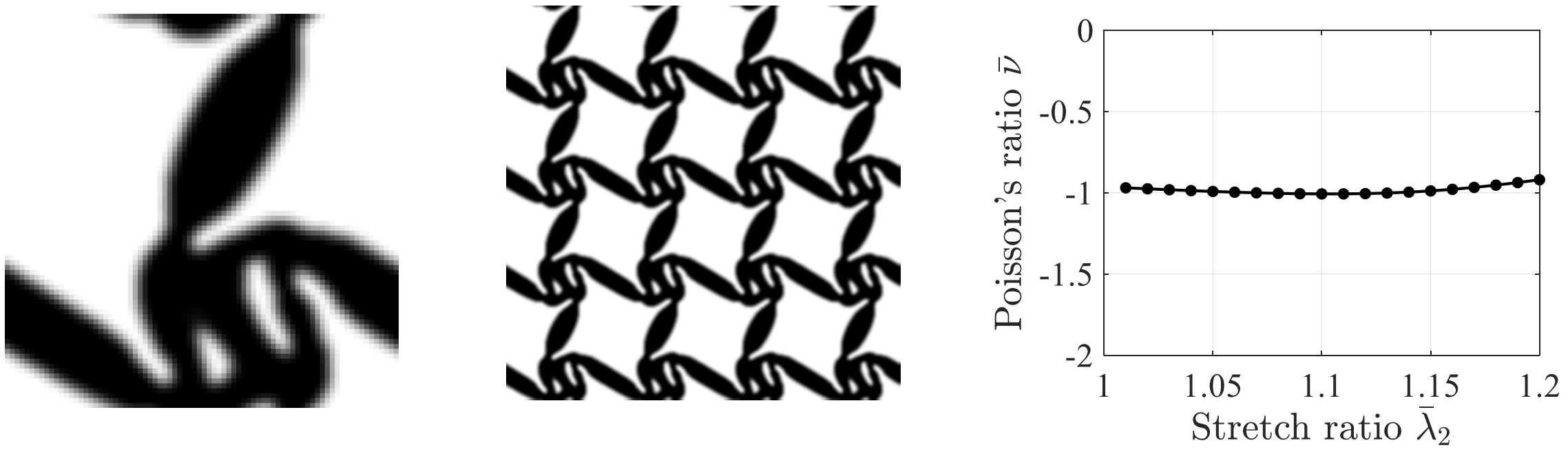

(b) $\bar{k}$ = 3 ($f_0$ = 5.20×10$^{-4}$, $V_f$ = 0.400)

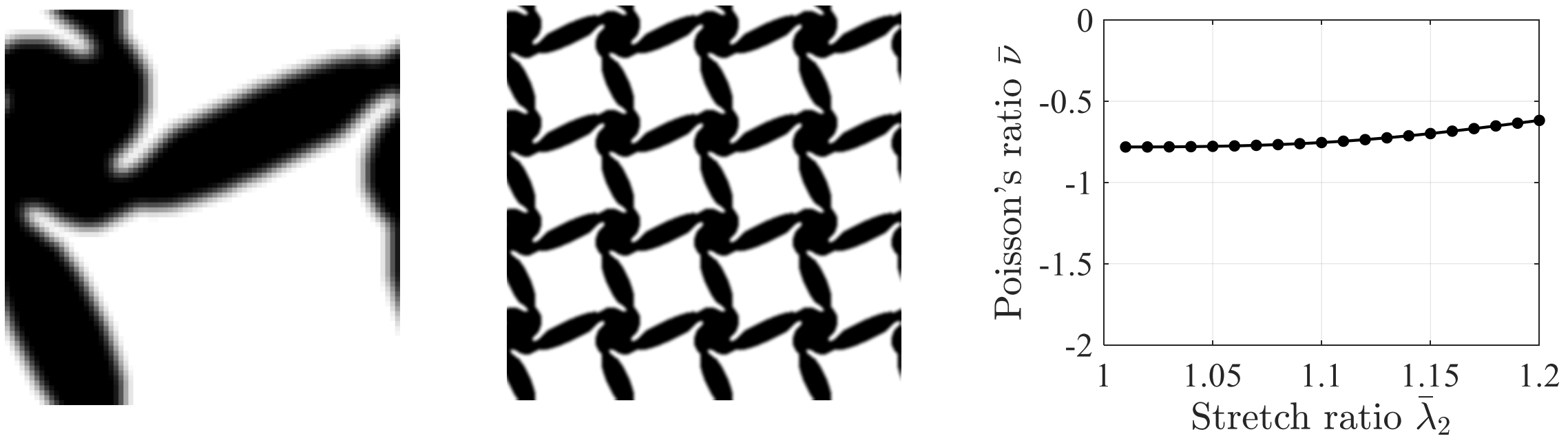

(c) $\bar{k}$ = 5 ($f_0$ = 2.92×10$^{-2}$, $V_f$ = 0.400)

Figure 5. Optimized designs with $V_T$ = 0.4 and different stiffness constraints $\bar{k}$ ($\alpha_1$ = 0).

This issue is remedied by incorporating a material usage term $\alpha_1 V_f(\boldsymbol{x}_1)$ in the optimization objective function, with $\alpha_1 > 0$ in Eq. (46). To remove the intermediate densities in Figure 5a, $\alpha_1$ = 0.01 is used after 200 iterations when an overall topology has already emerged, and the optimization history and results are shown in Figure 6. It can be seen in Figure 6 that compared to Figure 5a the optimality of the design is preserved with lower material usage (i.e. $V_f$ = 0.292) while still satisfying the stiffness constraint. It should be noted that the value of $\alpha_1$ should be carefully chosen, since a too small $\alpha_1$ is not effective towards removing intermediate densities, while a large $\alpha_1$ will shift the optimizer from minimizing $f_0$ to minimizing the material usage. Thus, $\alpha_1$ should be chosen depending on the relative values of $f_0$ and $V_f$ in Eq. (46). Figure 6d shows the deformed shape at $\bar{\lambda}_2$ = 1.2, where a clear negative Poisson's ratio effect can be seen. Note that only elements with density greater than or equal to 0.6 are plotted in Figure 6d.

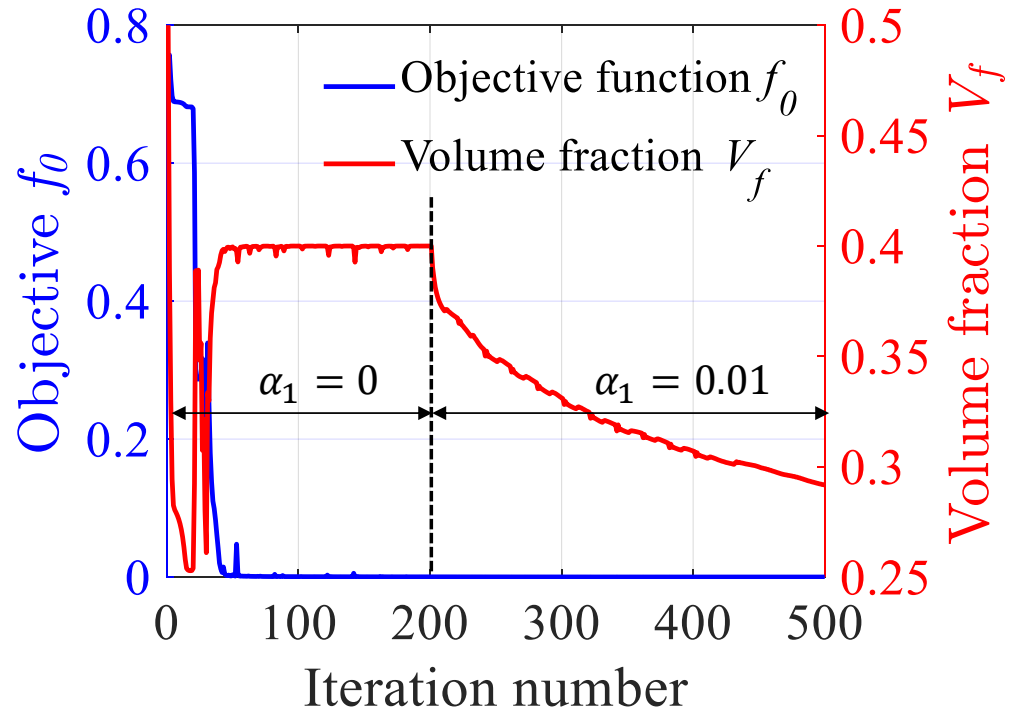

(a) Optimization history

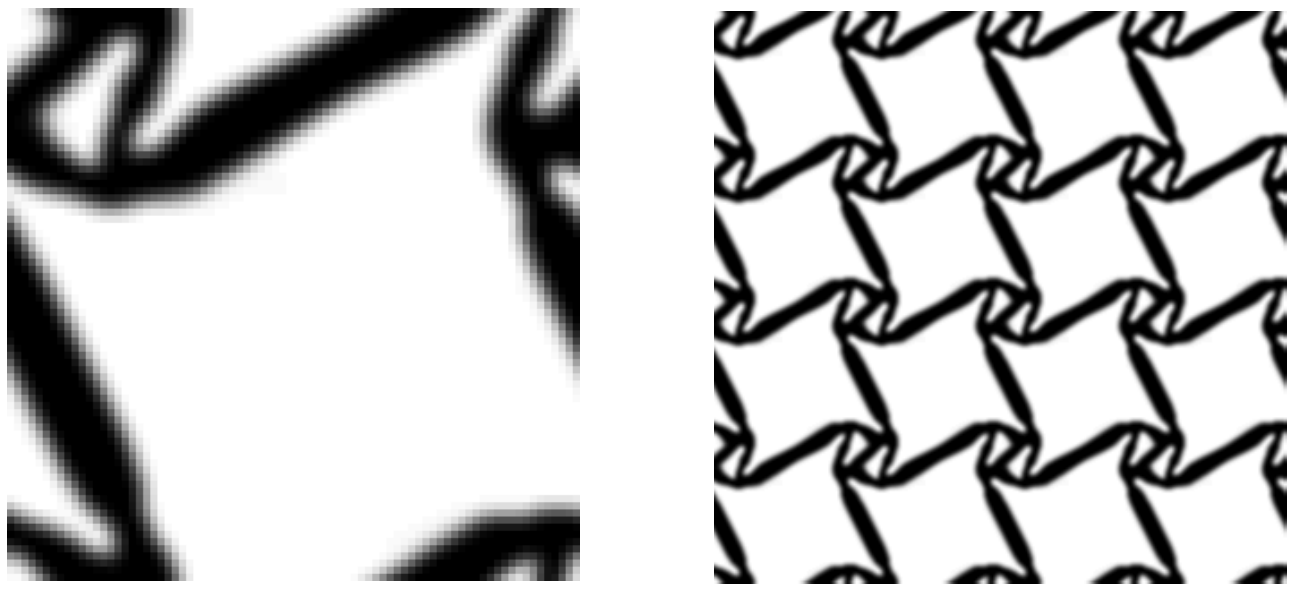

(b) Optimized unit cell

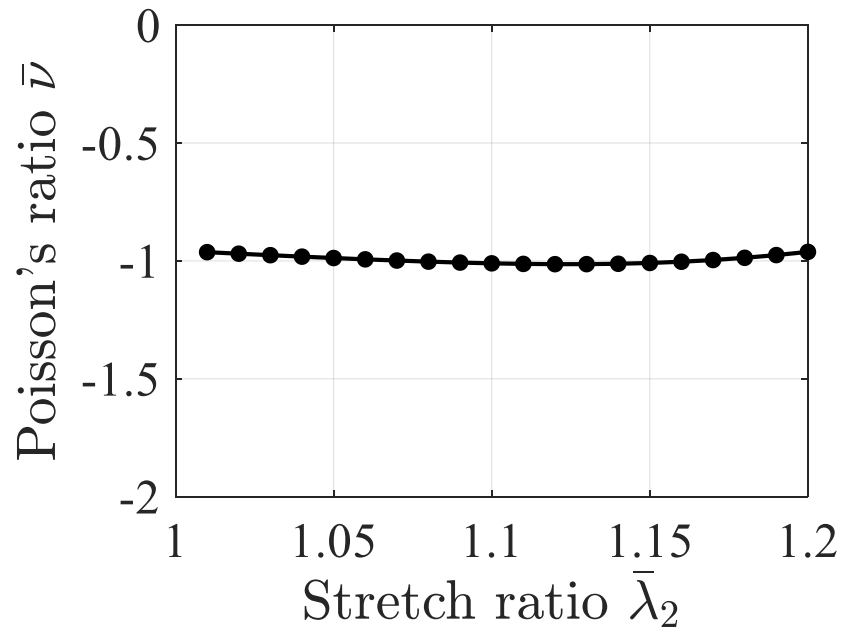

(c) Poisson's ratio over stretch

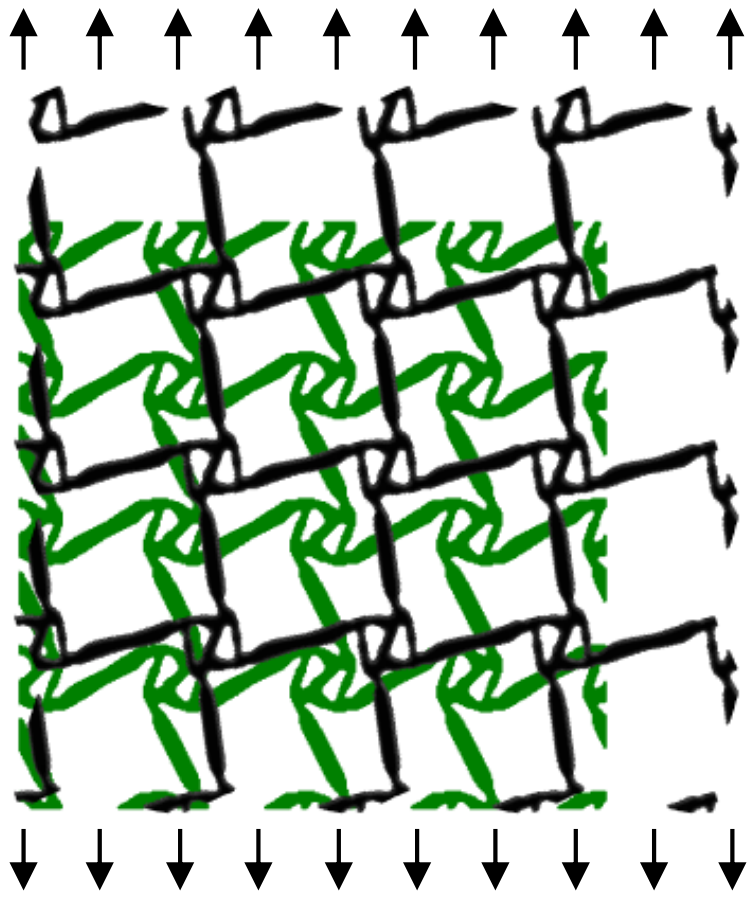

(d) Deformed shape at $\bar{\lambda}_2$ = 1.2 ($\bar{\nu} \approx$ -0.96)

Figure 6. Optimization results with $\bar{k}$ = 2 and $V_T$ = 0.4: $f_0$ = 1.03×10$^{-4}$, $V_f$ = 0.292 ($\alpha_1$ = 0.01 after 200 iter.).

### *5.2 NPR design under compression*

This example serves as a comparison study for demonstrating the differences in the optimized designs under tension and compression and for illustrating the difficulties in compression design.

With the same problem settings as in Figure 6, the NPR design is carried out under compression with macroscopic applied stretch ratio of $\bar{\lambda}_2 = 0.85$. The optimized result is shown in Figure 7, where obvious differences in the design can be seen when compared to the tension case in Figure 6. A notable difficulty in this compression case is the mesh distortion issue. Although the non-convergence of NR solver can be overcome by introducing linear energy interpolation as given in Section 3.4, the need for smaller step size in the adaptive NR solver and additional FEA due to the automatic updates of the parameter $c$ in the linear energy interpolation function (Eq. (53)) makes the optimization process slower when compared to the tension case. For example, the total number of FEA during the optimization process is 1125 (Figure 8), which far exceeds the number of optimization iterations. Due to the high computational cost associated with NPR designs under compression, the rest of the examples only consider NPR designs under tension.

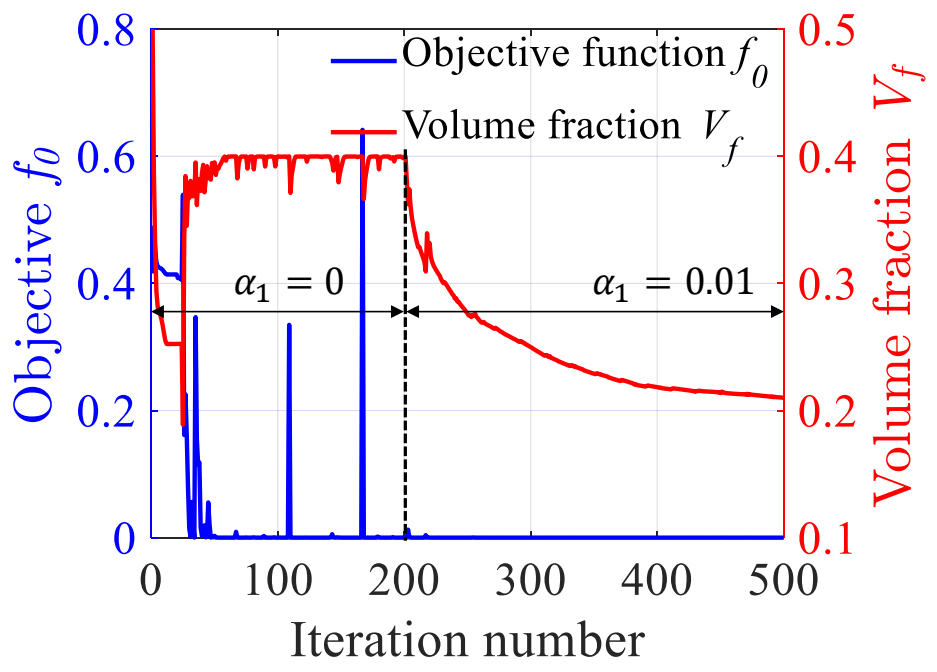

(a) Optimization history

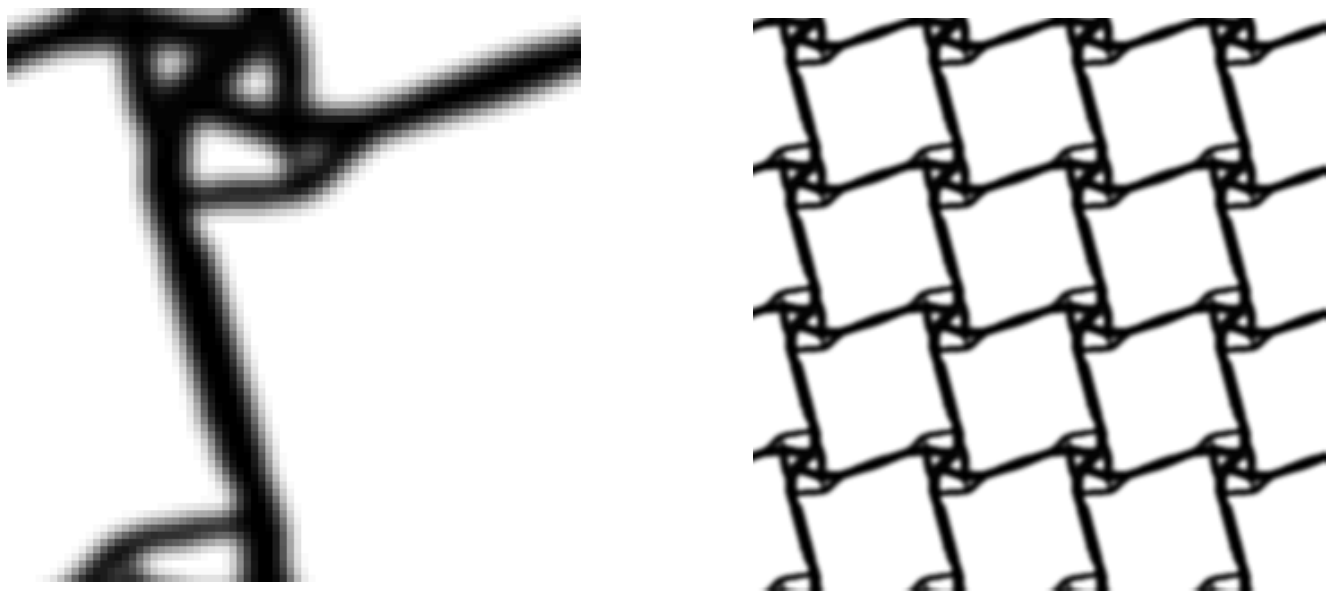

(b) Optimized unit cell

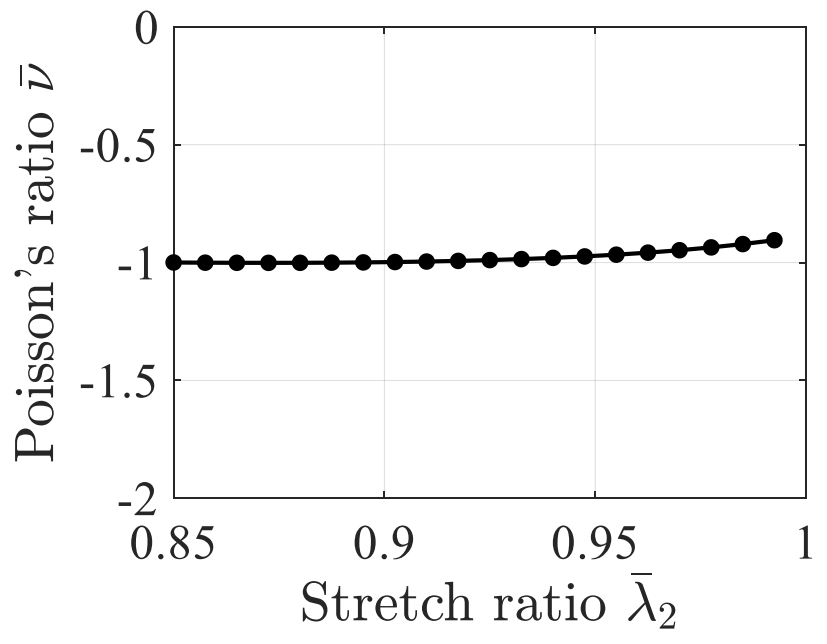

(c) Poisson's ratio over stretch

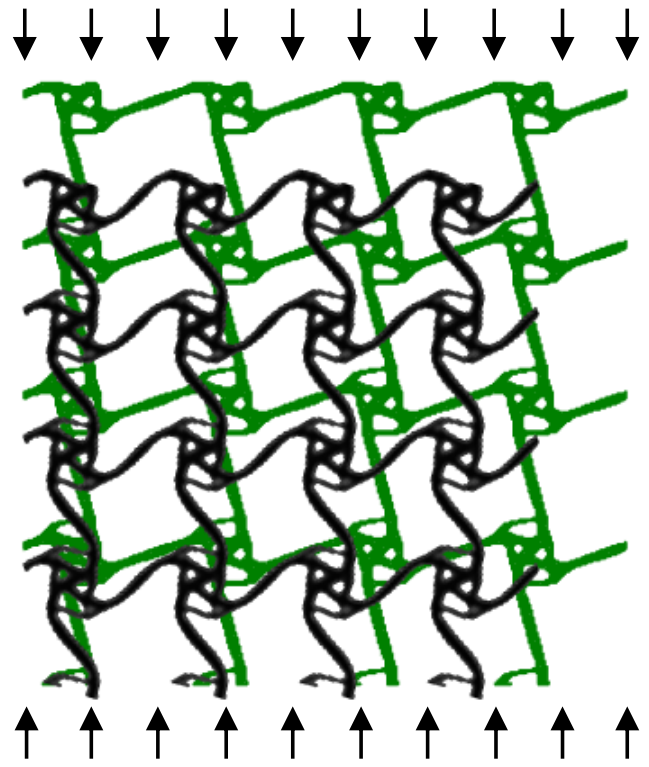

(d) Deformed shape at $\bar{\lambda}_2 = 0.85$ ($\bar{\nu} \approx -0.91$)

Figure 7. Optimization results with $\bar{k} = 2$ and $V_T = 0.4$: $f_0 = 1.67 \times 10^{-5}$, $V_f = 0.210$ ($\alpha_1 = 0.01$ after 200 iter.).

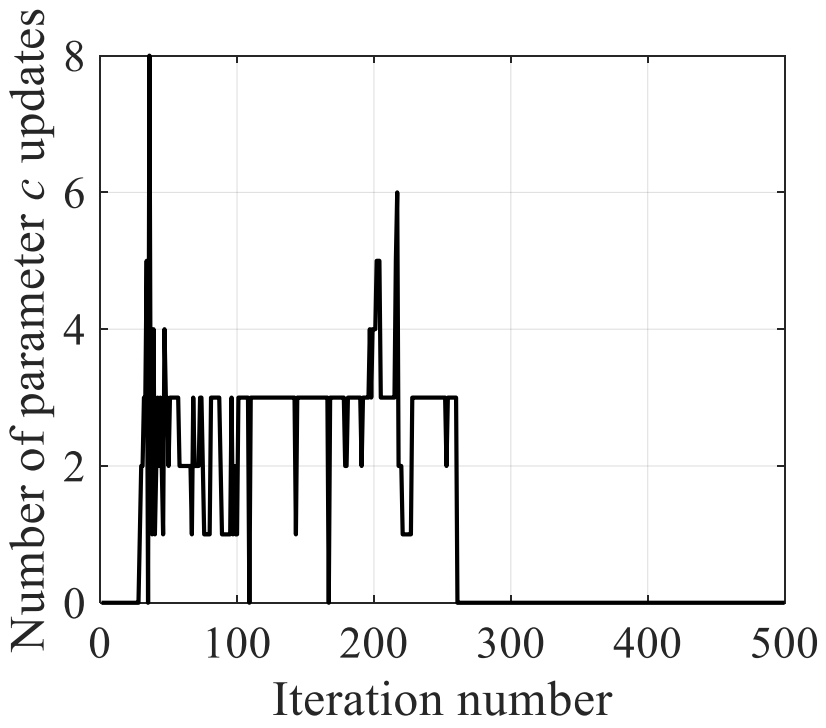

Figure 8. History of cutoff parameter $c$ updates.

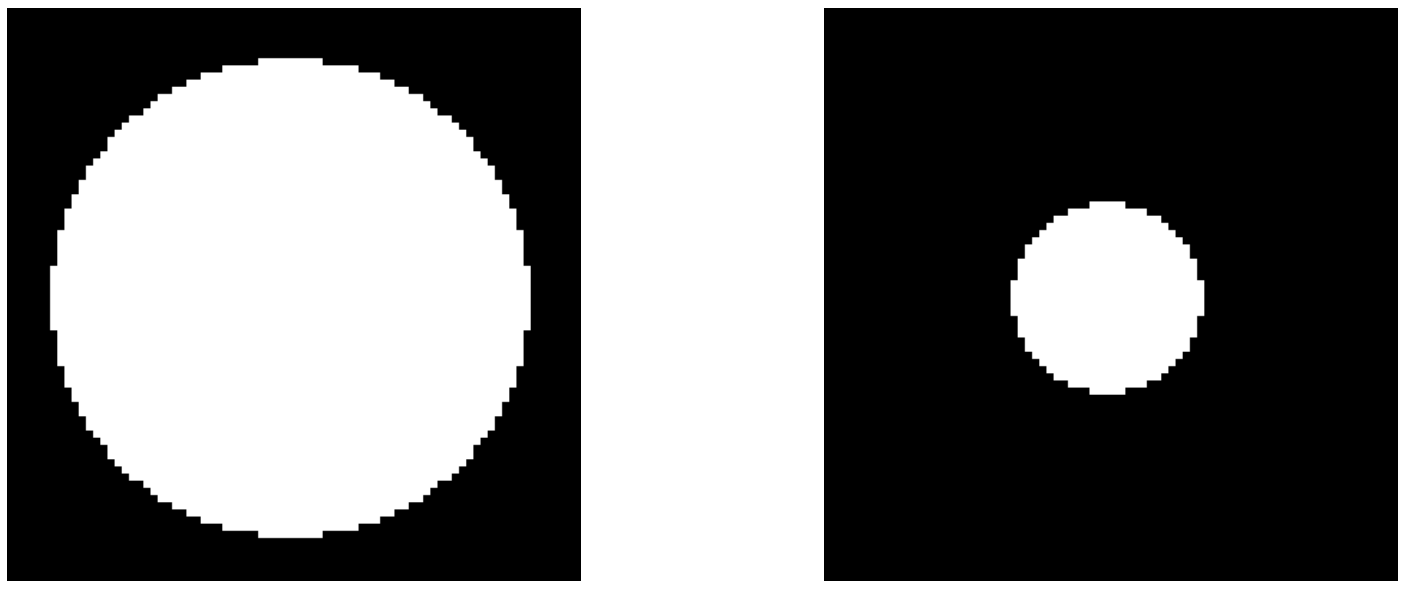

(a) Design-A (b) Design-B

Figure 9. Initial designs.

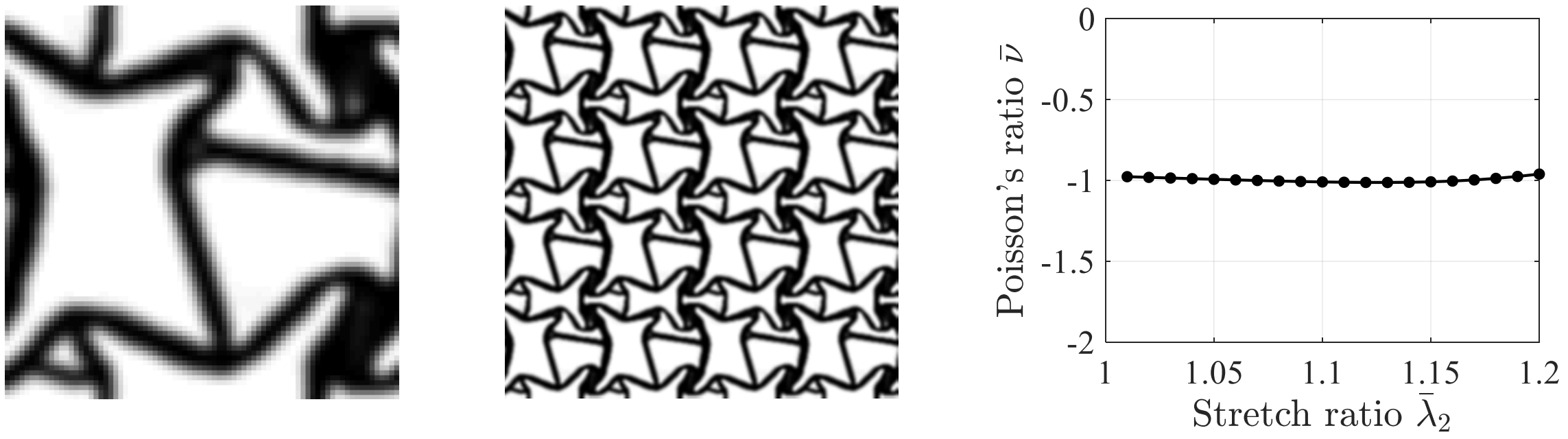

(a) Design-A ($f_0$ = 1.04×10$^{-4}$, $V_f$ = 0.388)

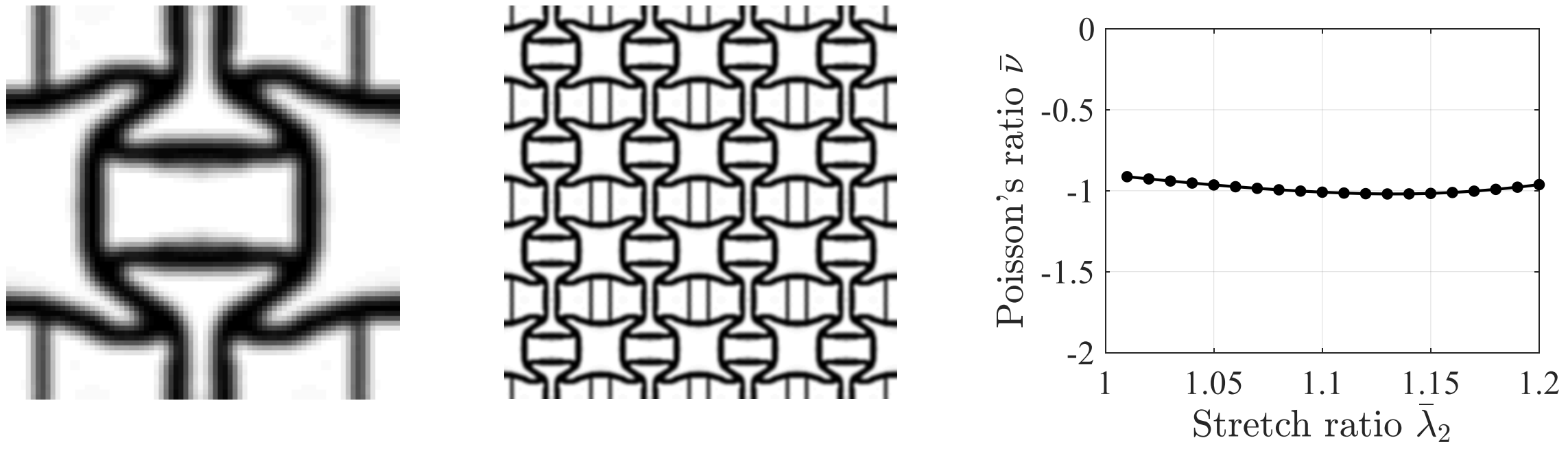

(b) Design-B ($f_0$ = 1.24×10$^{-4}$, $V_f$ = 0.331)

Figure 10. Optimized designs with $\bar{k}$ = 2 and $V_T$ = 0.4 ($\alpha_1$ = 0.01 after 200 iter.).

### 5.3 Influence of initial designs

In this test case, the influence of initial designs on the optimized NPR topologies is considered. With the same problem setting as in Figure 6, the optimization results corresponding to two different initial designs in Figure 9 are shown in Figure 10. Clearly, different optimized NPR metamaterial designs are obtained starting from different initial designs. This test case demonstrates a dependence of the optimized results on the initial design, suggesting that multiple initial designs can be examined in practice for auxetic metamaterial designs in order to achieve desirable results.

### 5.4 NPR designs under different loading magnitudes

It has been previously shown that for small and large deformations the NPR designs can be different [23]. In this subsection, the differences in NPR designs induced from the small and large macroscopic strains is explored together with the influence of the macroscopic strain magnitude on the selection of appropriate stiffness constraint $\bar{k}$. Note that the material usage term in Eq. (46) with $\alpha_1 > 0$ is only incorporated for cases where there are intermediate floating densities in the optimized topology and is activated only after 200 iterations when an overall design has already emerged. The initial design shown in Figure 4 is again used with the square unit cell design domain. Figure 11 shows the optimized topologies for the different macroscopic stretches $\bar{\lambda}_2 \in \{1.1, 1.4, 1.6\}$. As can be seen from these results, for a large stretch ratio, i.e. $\bar{\lambda}_2$ = 1.4 or 1.6, the stiffness constraint $\bar{k}$ is high enough to generate discrete topologies and there is no need for the material usage term, i.e. $\alpha_1 = 0$. However, as $\bar{\lambda}_2$ increases the objective $f_0$ is deteriorated, which is due to the high stiffness constraint. To verify this, $\bar{k}$ is decreased to 1 and the optimization results for $\bar{\lambda}_2$ = 1.4 and 1.6 are shown in Figure 12. As can be seen, with the decrease in $\bar{k}$ the material penalization term has to be activated in order to get a discrete topology for $\bar{\lambda}_2$ = 1.4 ($\alpha_1$ = 0.01) and the objective $f_0$ is obviously improved. For $\bar{\lambda}_2$ = 1.6, discrete topology can be obtained with $\alpha_1$ = 0, again indicating sufficient stiffness constraint. Moreover, the objective $f_0$ is still improved by decreasing $\bar{k}$. The deformed shapes of the topologies in Figure 12(a) and (b) are given in Figure 13, demonstrating the capability of maintaining target NPR over the specified strain range. This exercise demonstrates that the appropriate value of the stiffness constraint depends on the macroscopic strain range, and in general, $\bar{k}$ decreases as the target strain increases.

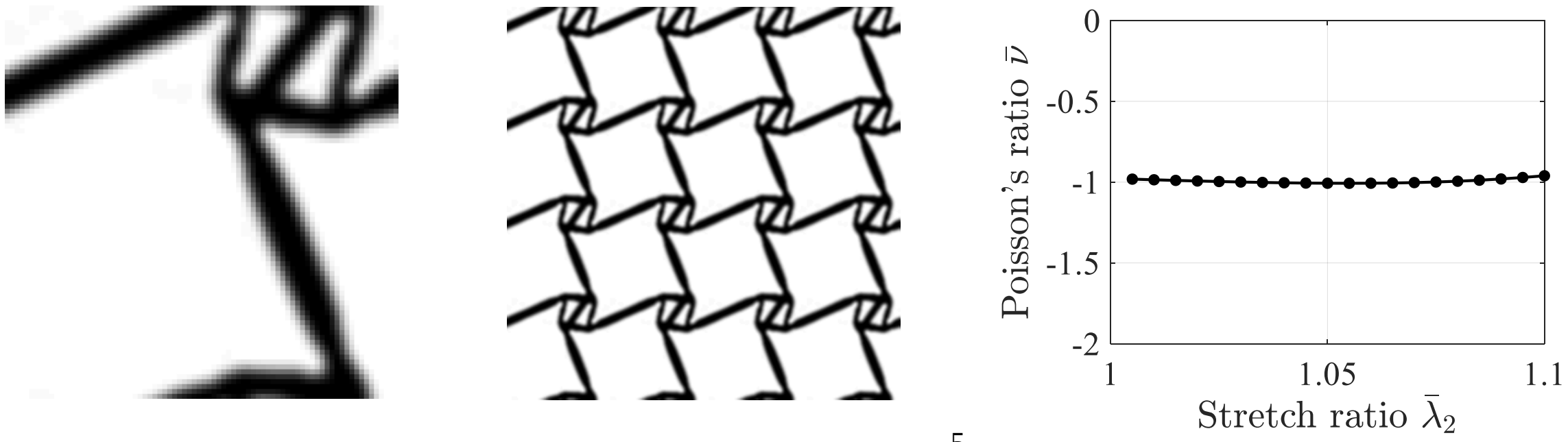

(a) $\bar{\lambda}_2$ = 1.1, $\alpha_1$ = 0.01 ($f_0$ = 2.77×10$^{-5}$, $V_f$ = 0.226)

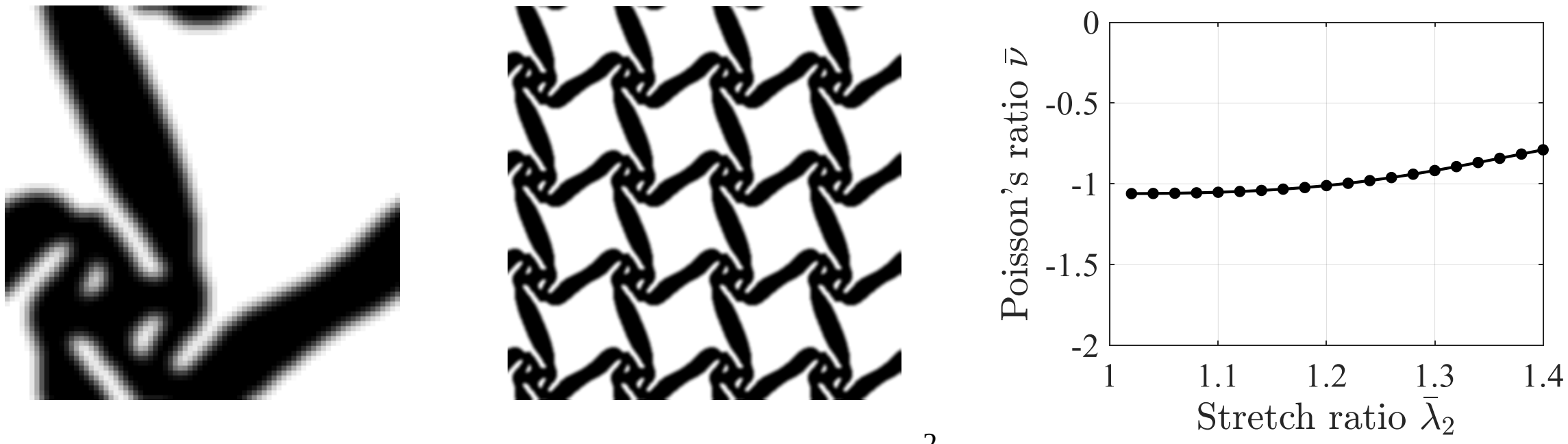

(b) $\bar{\lambda}_2$ = 1.4, $\alpha_1$ = 0 ($f_0$ = 1.96×10$^{-2}$, $V_f$ = 0.400)

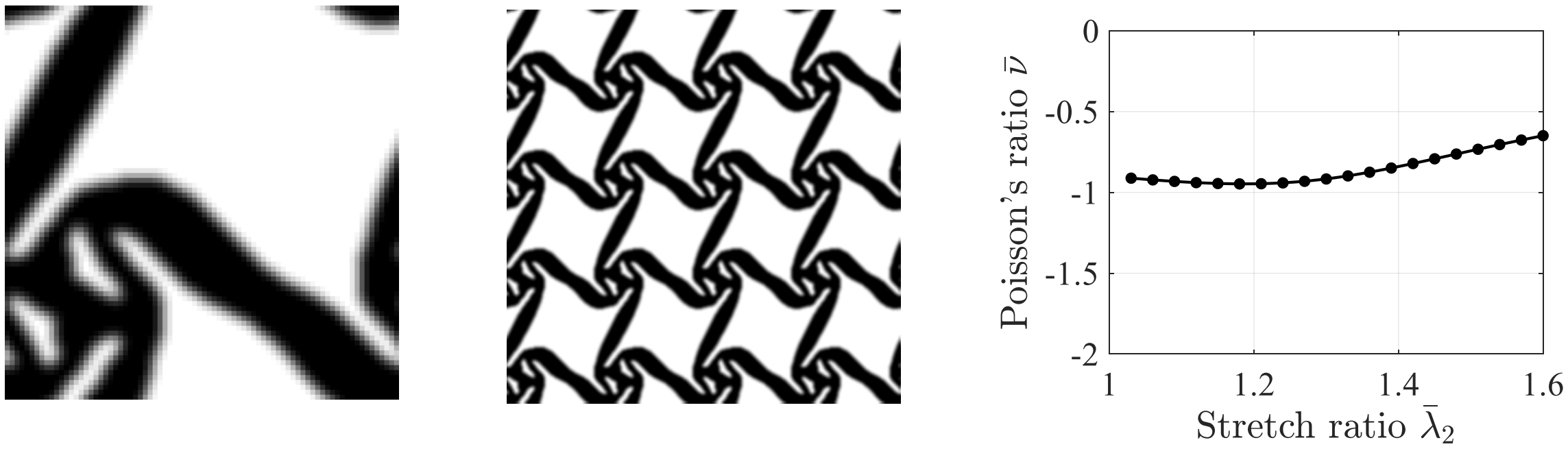

(c) $\bar{\lambda}_2$ = 1.6, $\alpha_1$ = 0 ($f_0$ = 1.58×10$^{-1}$, $V_f$ = 0.400)

Figure 11. Optimized designs with $\bar{k}$ = 2 and $V_T$ = 0.4 ($\alpha_1$ = 0 before 200 iter.).

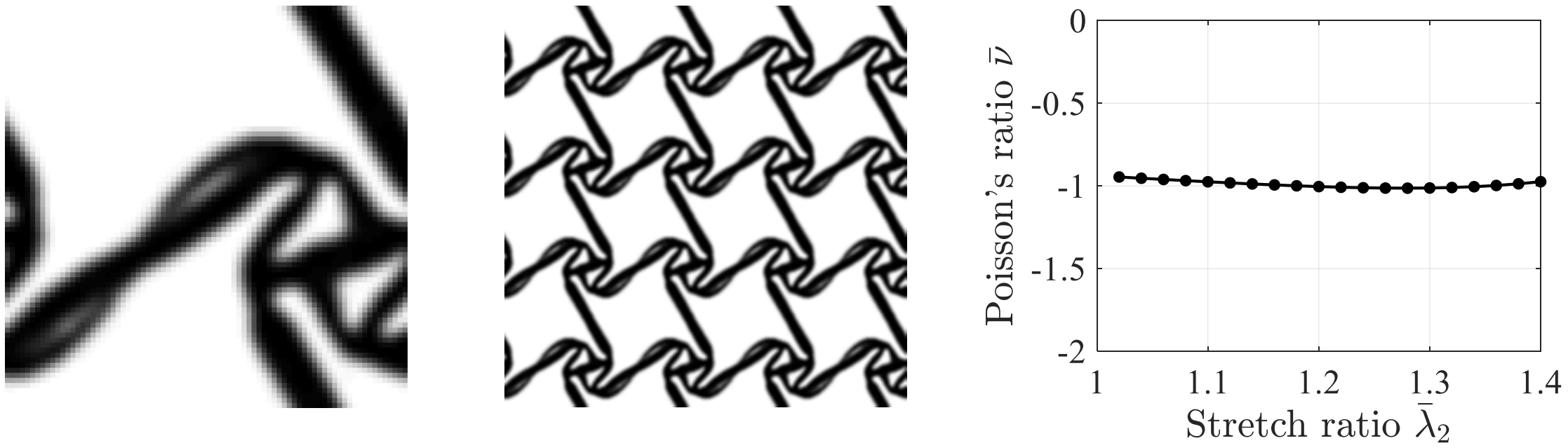

(a) $\bar{\lambda}_2$ = 1.4, $\alpha_1$ = 0.01 ($f_0$ = 2.22×10$^{-4}$, $V_f$ = 0.325)

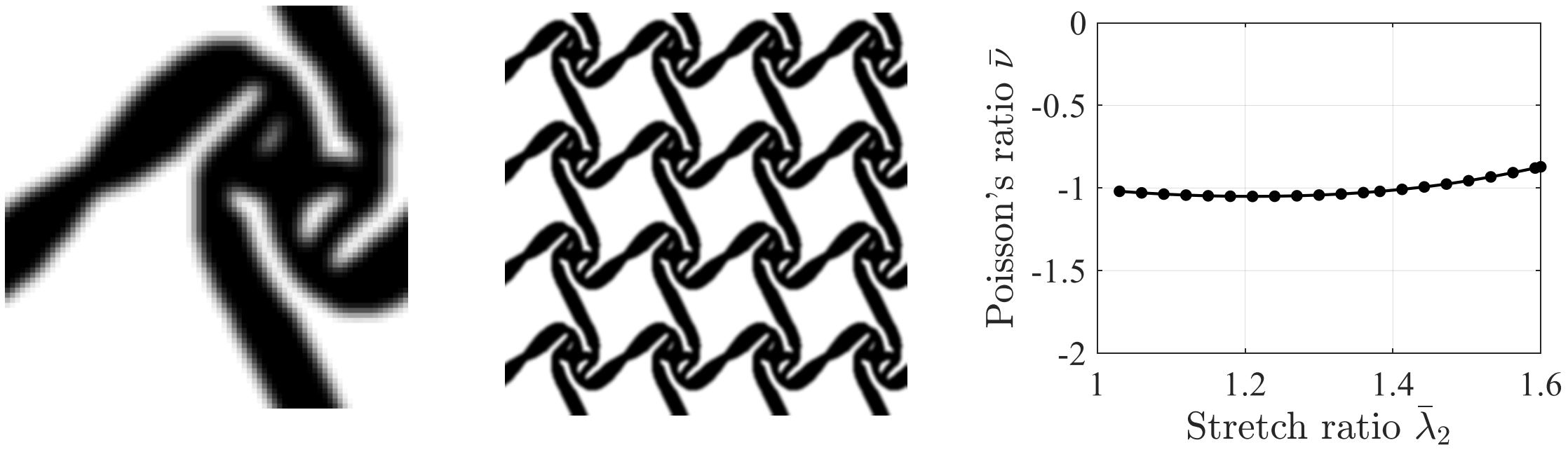

(b) $\bar{\lambda}_2$ = 1.6, $\alpha_1$ = 0 ($f_0$ = 1.68×10$^{-2}$, $V_f$ = 0.400)

Figure 12. Optimized designs with $\bar{k}$ = 1 and $V_T$ = 0.4 ($\alpha_1$ = 0 before 200 iter.).

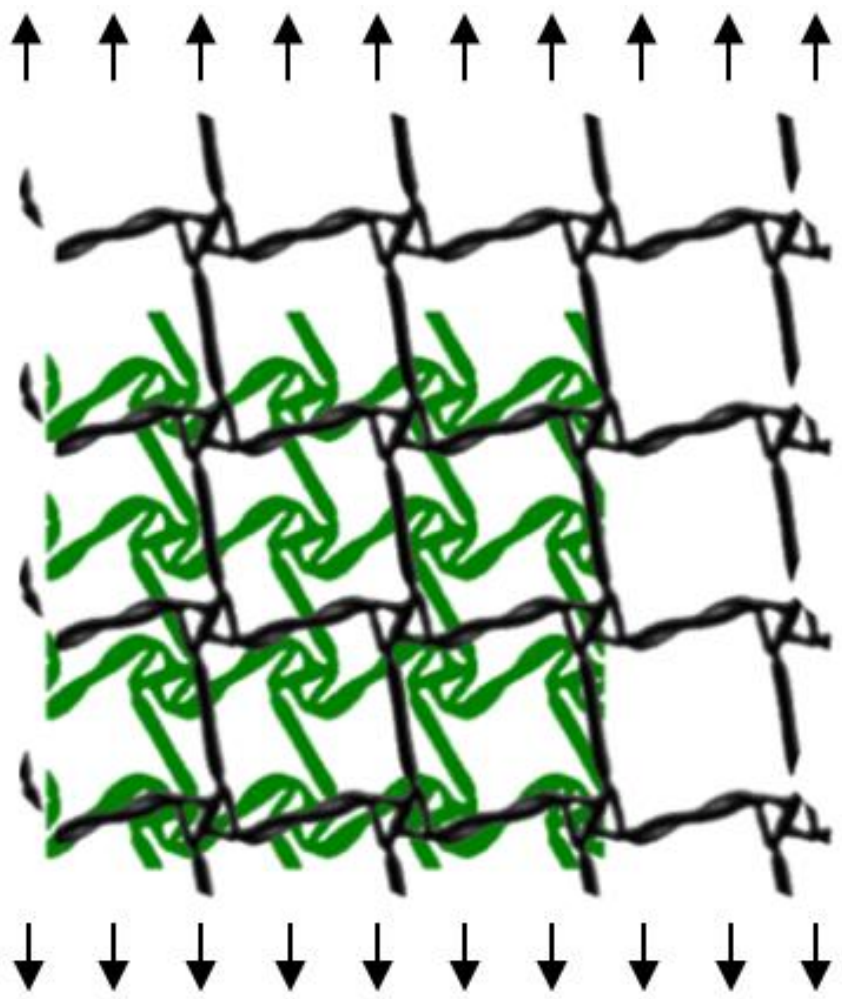

(a) $\bar{\lambda}_2$ = 1.4, $\bar{\nu} \approx$ -0.97

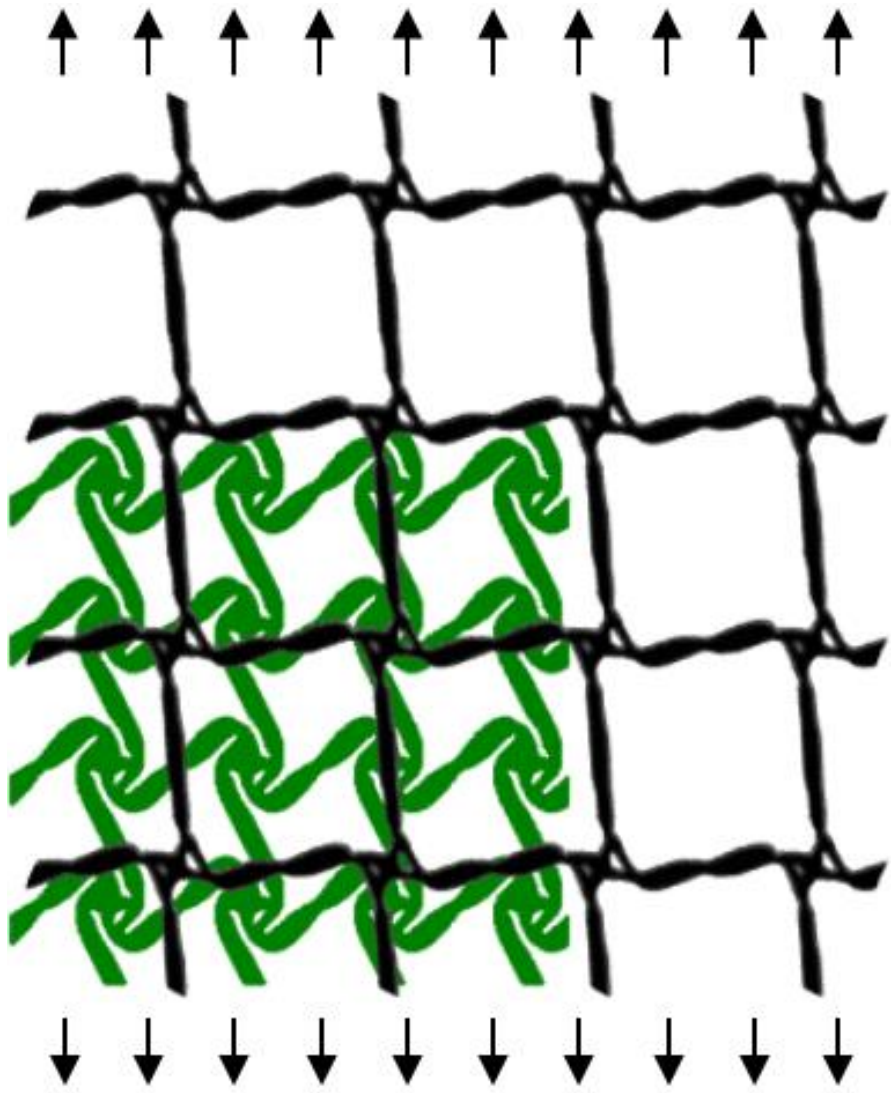

(b) $\bar{\lambda}_2$ = 1.6, $\bar{\nu} \approx$ -0.87

Figure 13. Deformed shape of the topologies in Figure 12 (a) and (b).

### 5.5 Designs with different target NPR

In this section, numerical studies are carried out to demonstrate the effectiveness of the proposed framework for designing metamaterials exhibiting various target NPRs, i.e. $\bar{\nu}_T$. Under the same material volume constraint $V_T$ = 0.4 and macroscopic stretch $\bar{\lambda}_2$ = 1.2, different values for stiffness constraint $\bar{k}$ and material penalization factor $\alpha_1$ have to be used in order to achieve desired Poisson's ratios. In general, following the reasons explained in Section 5.4, as the target NPR decreases, $\bar{k}$ has to be decreased in order to have a better design. Again $\alpha_1 > 0$ is only used in cases with intermediate densities and is only activated after 200 iterations. The optimization results are shown in Figure 14, where it can be seen that starting from the same initial design (Figure 4) different Poisson's ratios correspond to different topologies, as expected. Figure 15 shows the deformed shape of the designs in Figure 14(d) and (e), where the deformation mechanisms that lead to auxetic behavior can be clearly seen. It should be noted that designing auxetic material with target $\bar{\nu} = \bar{\nu}_T < 0$ along one loading direction is equivalent to designing material with target $\bar{\nu} = 1/\bar{\nu}_T$ along the orthogonal direction, see Figure 16 for an illustration. As a result, the difference between designing with $\bar{\nu}_T$ = -0.5 in Figure 14a and designing with $\bar{\nu}_T$ = -2 in Figure 14d can be understood as designing under different macroscopic stretch magnitudes. This also explains why a smaller value of $\bar{k}$ is sufficient for designing for smaller $\bar{\nu}_T$.

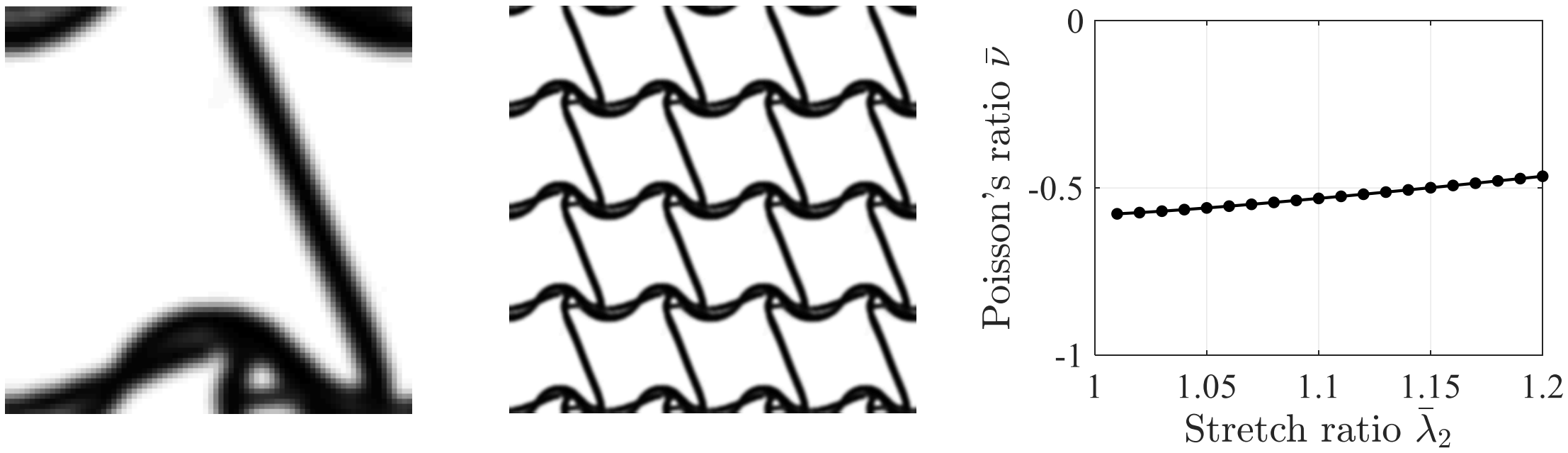

(a) Target $\bar{\nu}_T$ = -0.5, $\bar{k}$ = 2, $\alpha_1$ = 0.01 ($f_0$ = 1.93×10$^{-4}$, $V_f$ = 0.234)

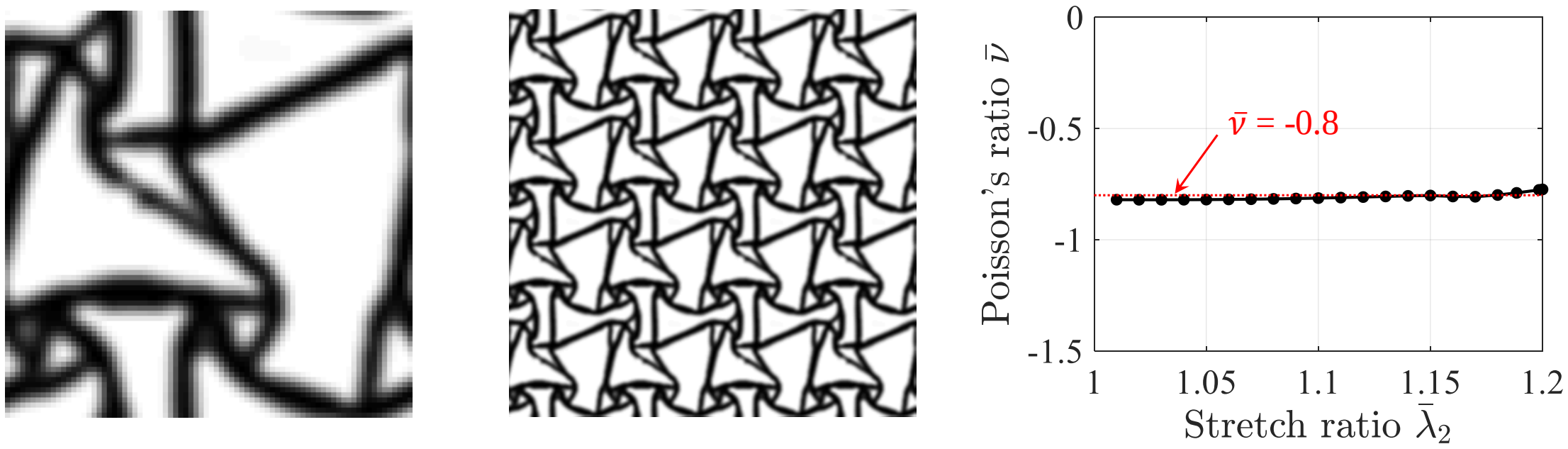

(b) Target $\bar{\nu}_T$ = -0.8, $\bar{k}$ = 2, $\alpha_1$ = 0.01 ($f_0$ = 6.95×10$^{-5}$, $V_f$ = 0.357)

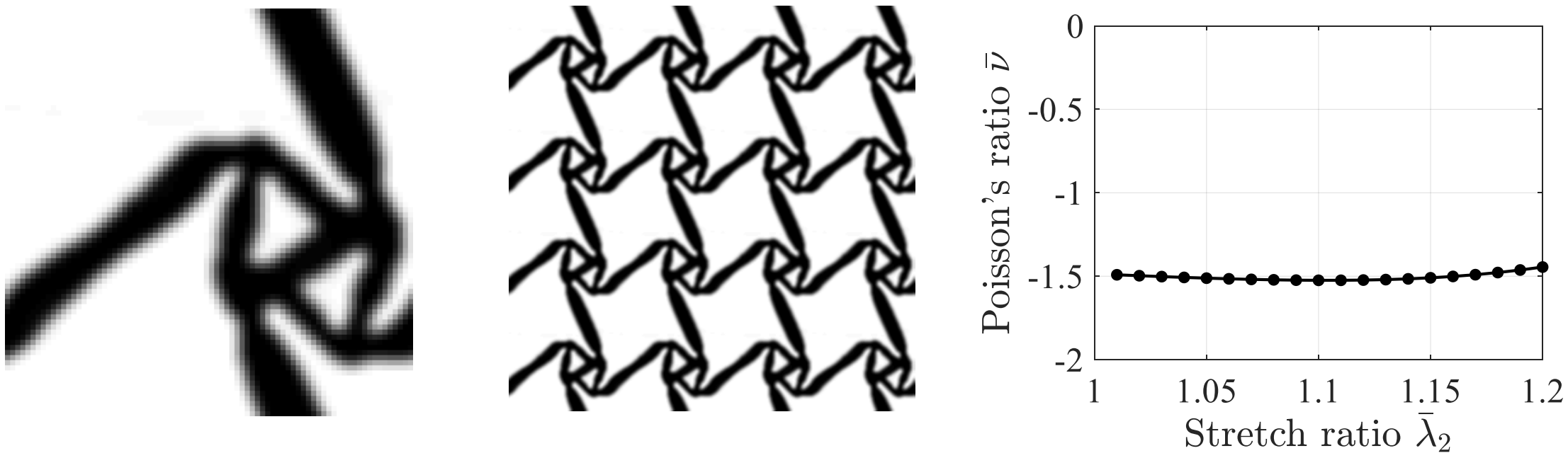

(c) Target $\bar{\nu}_T$ = -1.5, $\bar{k}$ = 1, $\alpha_1$ = 0.015 ($f_0$ = 2.34×10$^{-4}$, $V_f$ = 0.313)

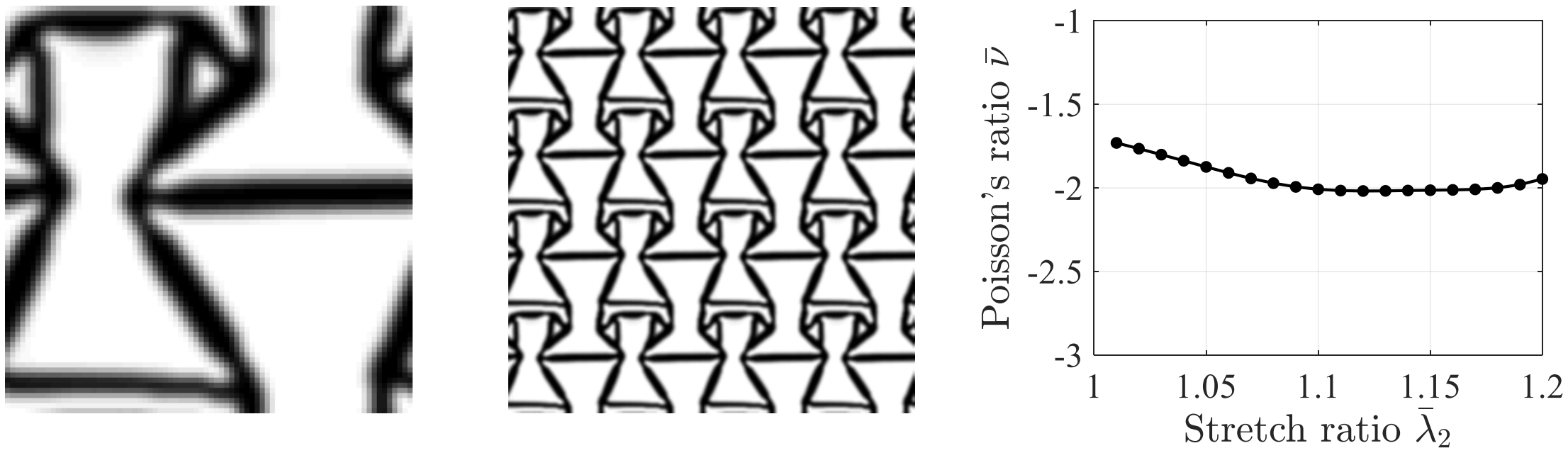

(d) Target $\bar{\nu}_T$ = -2, $\bar{k}$ = 1, $\alpha_1$ = 0.01 ($f_0$ = 3.54×10$^{-4}$, $V_f$ = 0.320)

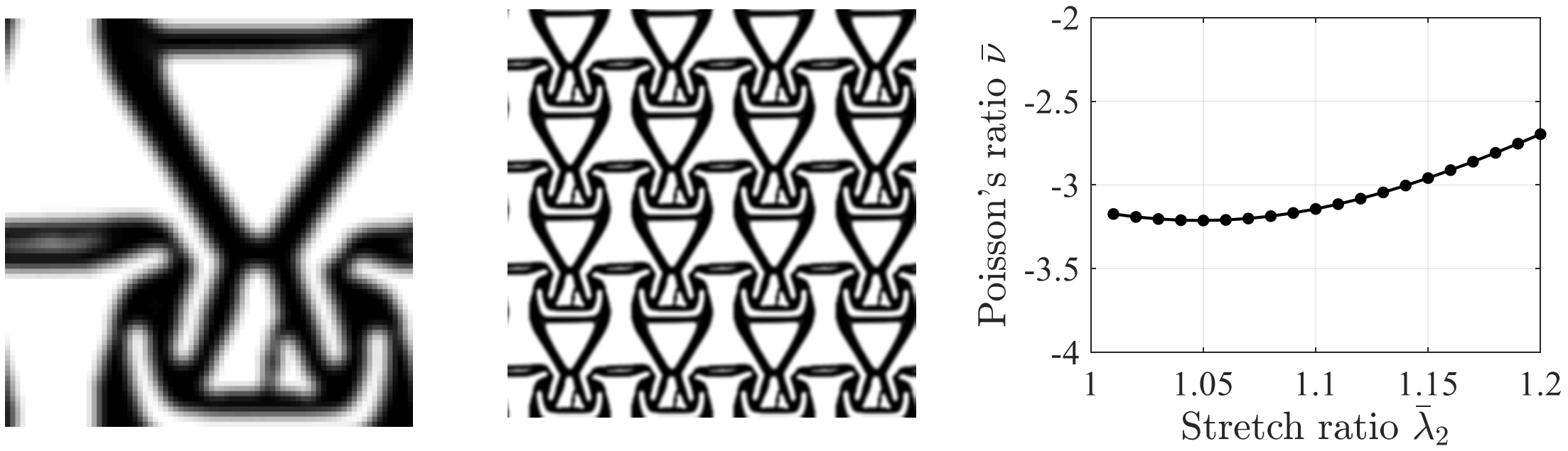

(e) Target $\bar{\nu}_T$ = -3, $\bar{k}$ = 0.2, $\alpha_1$ = 0 ($f_0$ = 9.41×10$^{-3}$, $V_f$ = 0.400)

Figure 14. Optimized designs for different target negative Poisson's ratios with $V_T$ = 0.4 ($\alpha_1$ = 0 before 200 iter.).

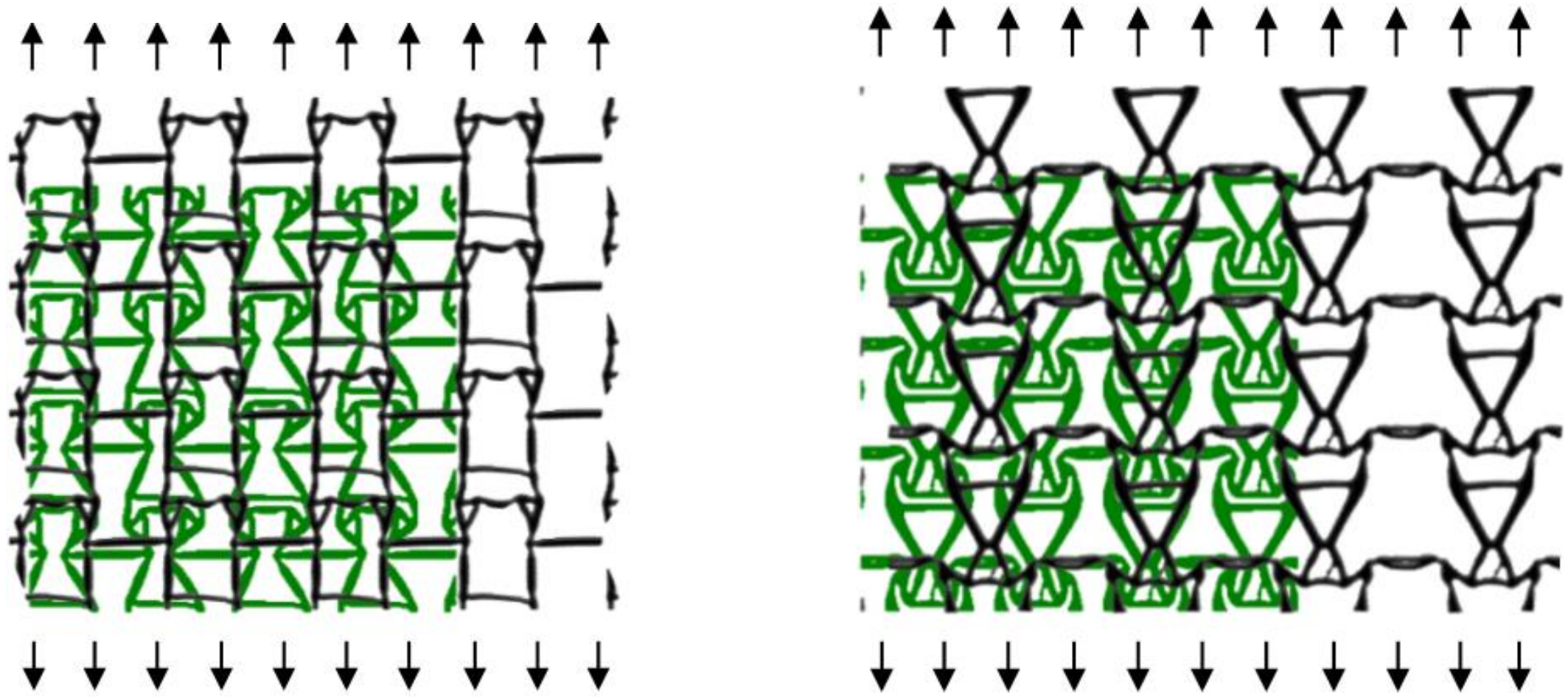

(a) $\bar{\lambda}_2$ = 1.2, $\bar{\nu} \approx$ -1.95 (b) $\bar{\lambda}_2$ = 1.2, $\bar{\nu} \approx$ -2.70

Figure 15. Deformed shape of the topologies in Figure 14 (d) and (e).

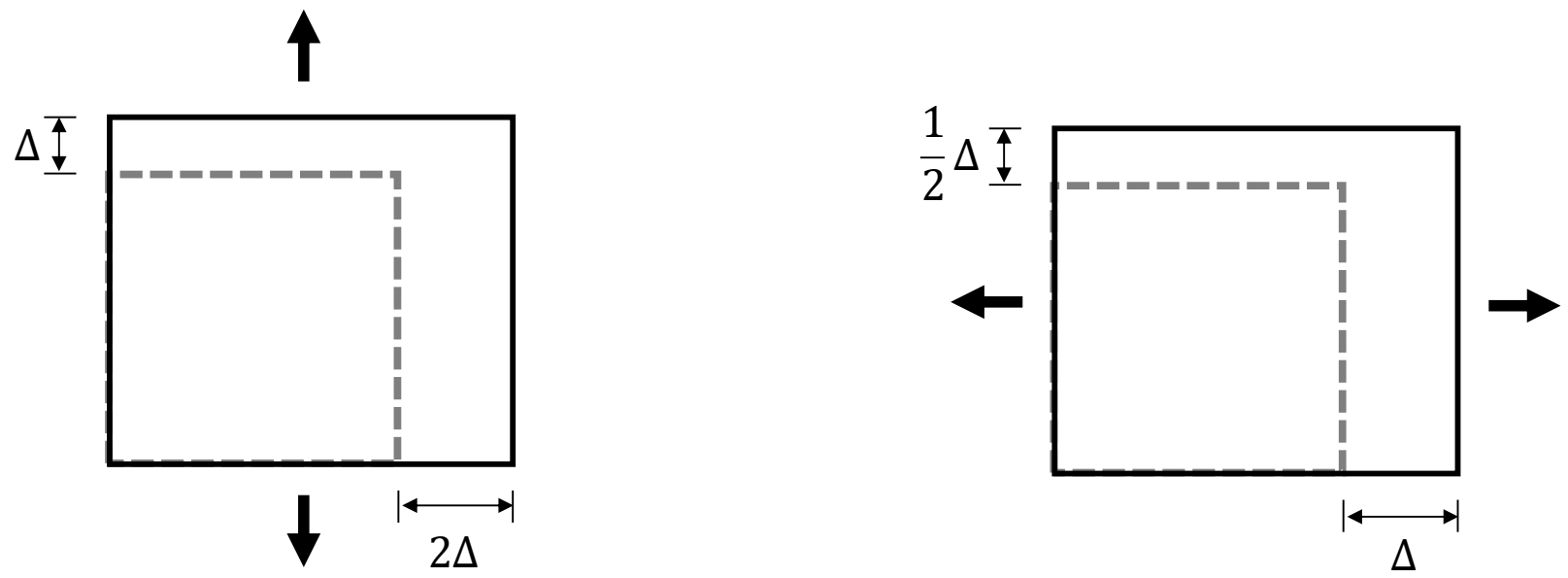

(a) Target $\bar{\nu}_T$ = -2 (b) Target $\bar{\nu}_T$ = -1/2

Figure 16. Relationship between designs with negative Poisson's ratio $\bar{\nu} = \bar{\nu}_T$ and $\bar{\nu} = 1/\bar{\nu}_T$: dashed grey line represents the initial square unit cell and black solid line represents the deformed shape.

### 5.6 Different unit cell geometries

For 2-D periodic metamaterials, the two periodic lattice vectors can be of any non-zero angle and of different lengths. Different set of periodic vectors constitute different shapes of unit cells. The most general shape of a unit cell is parallelogram, and the square and hexagonal unit cells can be seen as special cases of a parallelogram. Three different non-square unit cell design domains are investigated in this subsection, as shown in Figure 17. It should be noted that the regular hexagonal unit cell shown in Figure 17c has the same periodicity as the 60° parallelogram shown in Figure 17b (although their periodic vector lengths or equivalently unit cell sizes are different). Starting from the initial designs given in Figure 17 with $\bar{k}$ = 2 and $V_T$ = 0.4, the optimized designs for different unit cell shapes are shown in Figure 18, where appropriate $\alpha_1$ values are again used to achieve discrete topologies. As can be seen, different unit cell design domains can lead to different optimized topologies with similar auxetic behavior. It is worth mentioning that compared to the proposed framework in [23] for designing nonlinear auxetic materials, the framework in this study is based on nonlinear homogenization theory. As a result, different unit cell design domains can be consistently incorporated as shown above, which is not possible in the framework presented in [23].

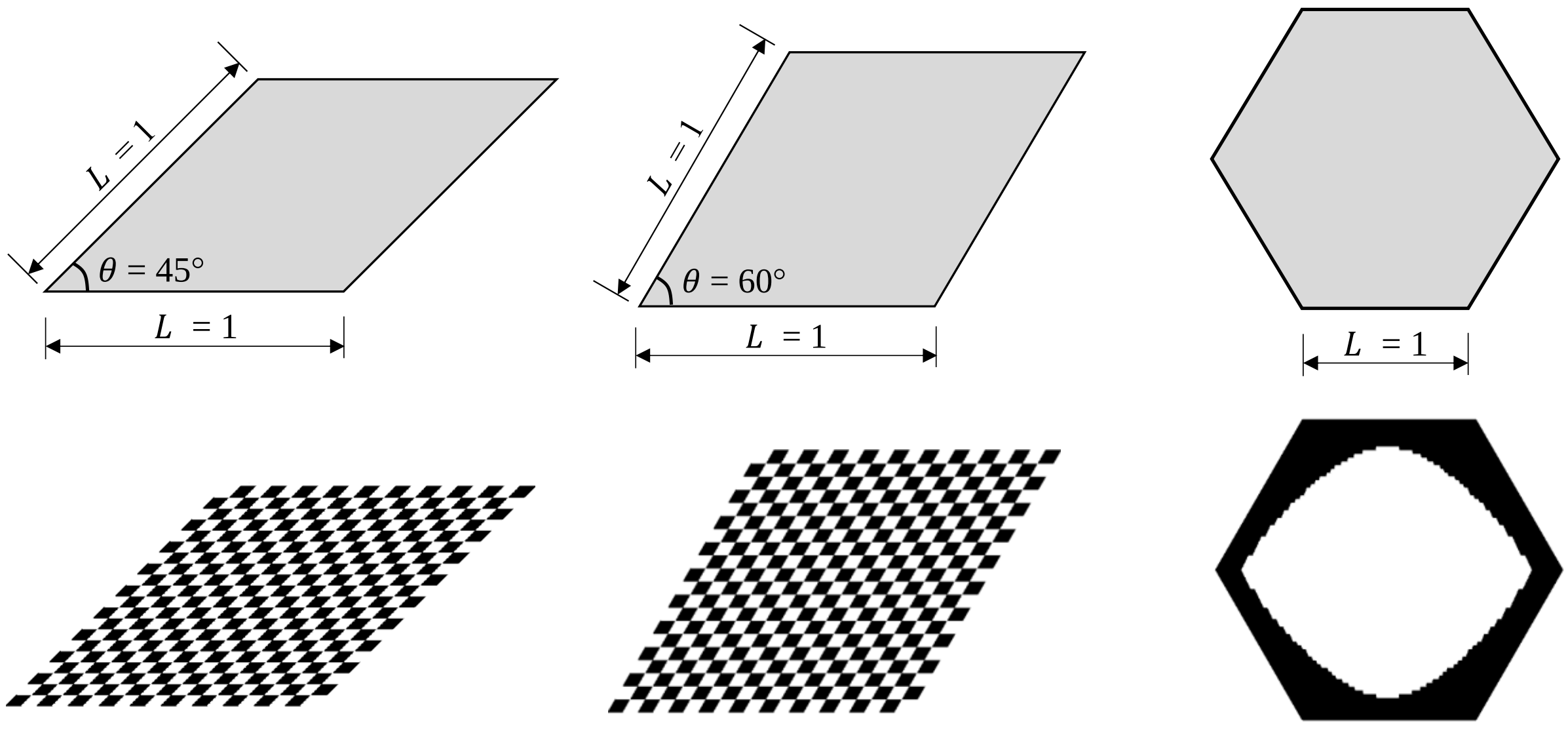

(a) 45° parallelogram (b) 60° parallelogram (c) Hexagon

Figure 17. Different unit cell shapes, first row: design domains, second row: initial designs.

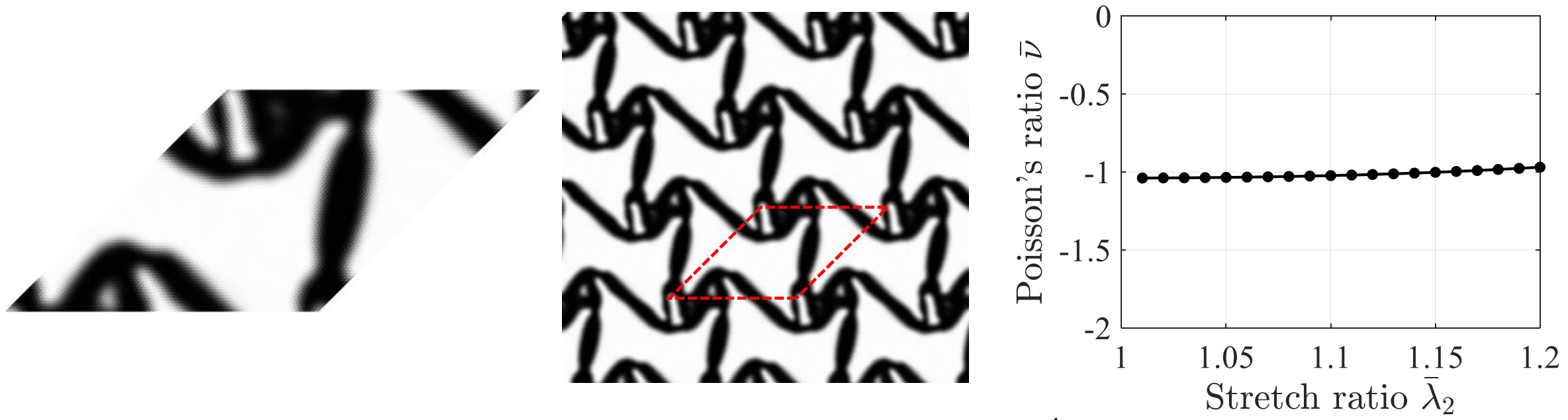

(a) 45° parallelogram, $\alpha_1$ = 0 ($f_0$ = 1.14×10$^{-4}$, $V_f$ = 0.400)

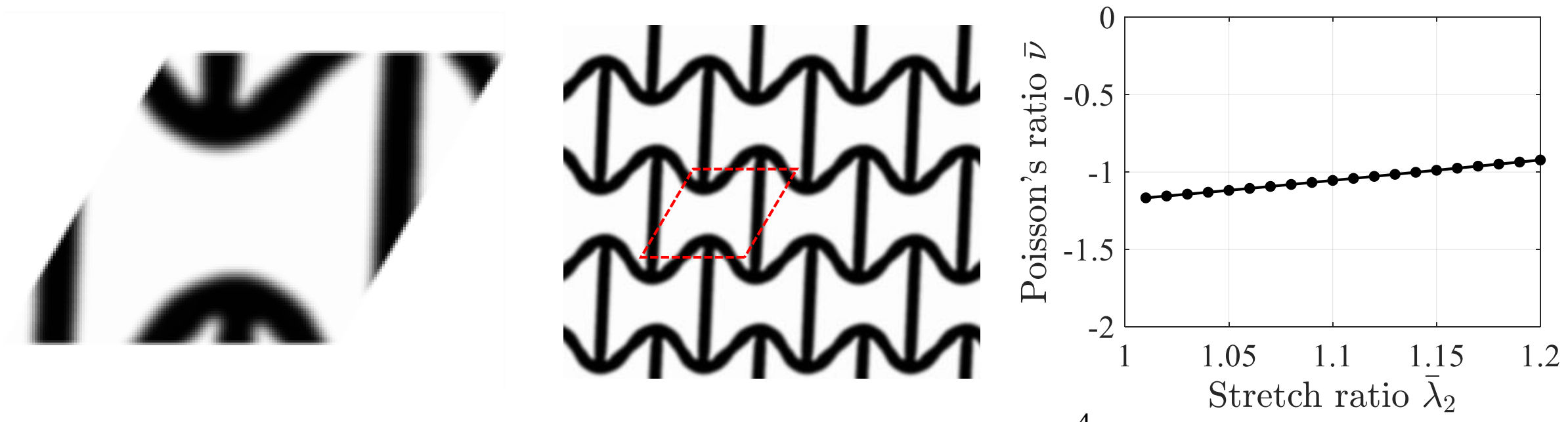

(b) 60° parallelogram, $\alpha_1$ = 0.02 ($f_0$ = 8.55×10$^{-4}$, $V_f$ = 0.333)

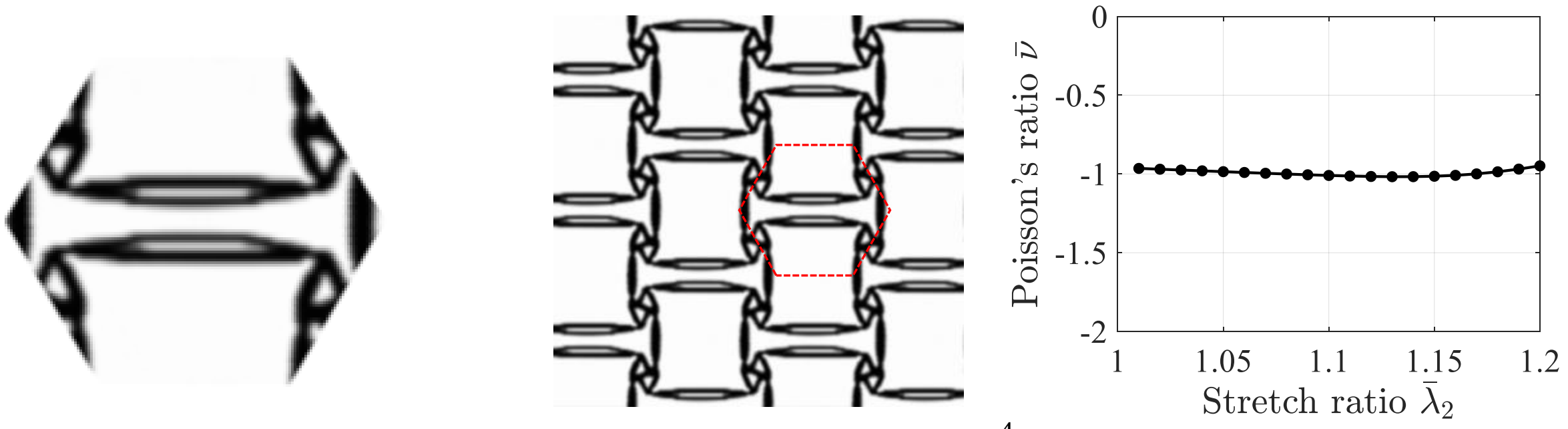

(c) Hexagon, $\alpha_1$ = 0.015 ($f_0$ = 1.74×10$^{-4}$, $V_f$ = 0.230)

Figure 18. Optimized designs for different unit cell shapes with $\bar{k}$ = 2 and $V_T$ = 0.4 ($\alpha_1$ = 0 before 200 iter.).

### *5.7 Loading axis*

Another merit of the presented framework based on a nonlinear homogenization formulation is that inclined loading scenario can be consistently considered. This is important in studying and controlling the auxetic behavior of material along different directions, which can also be useful when uncertainties in the loading direction are involved. To demonstrate this idea, this subsection considers square unit cell with three different loading axes $\theta$ = 15°, 30° and 45° (see Section 2.2). The material volume constraint and stiffness constraints are chosen as $V_T$ = 0.4 and $\bar{k}$ = 2. With the initial design shown in Figure 4, the optimized results for different loading axes are shown in Figure 19, where different topologies are achieved showing their dependencies on the loading directions. To further understand the auxetic behavior of the designed materials, their deformed shapes at $\bar{\lambda}_2$ = 1.2 under different loading scenarios are plotted in Figure 20.The results seem counter-intuitive since for usual material with positive Poisson's ratio, a uniaxial stretch along 45° direction should lead to mostly shear deformation. However, the resultant deformation shows a clear uniform expansion, i.e. volumetric expansion, see Figure 20c where the macroscopic strain $\bar{\boldsymbol{F}} = \begin{bmatrix} 1.198 & -0.00249 \\ -0.00249 & 1.198 \end{bmatrix}$ further establishes the observation. This auxetic behavior is clearly due to the negative Poisson's ratio effect and can be explained by Figure 21, where the role of Poisson's ratio in measuring the relative resistances to the volumetric and isochoric deformations can be seen.

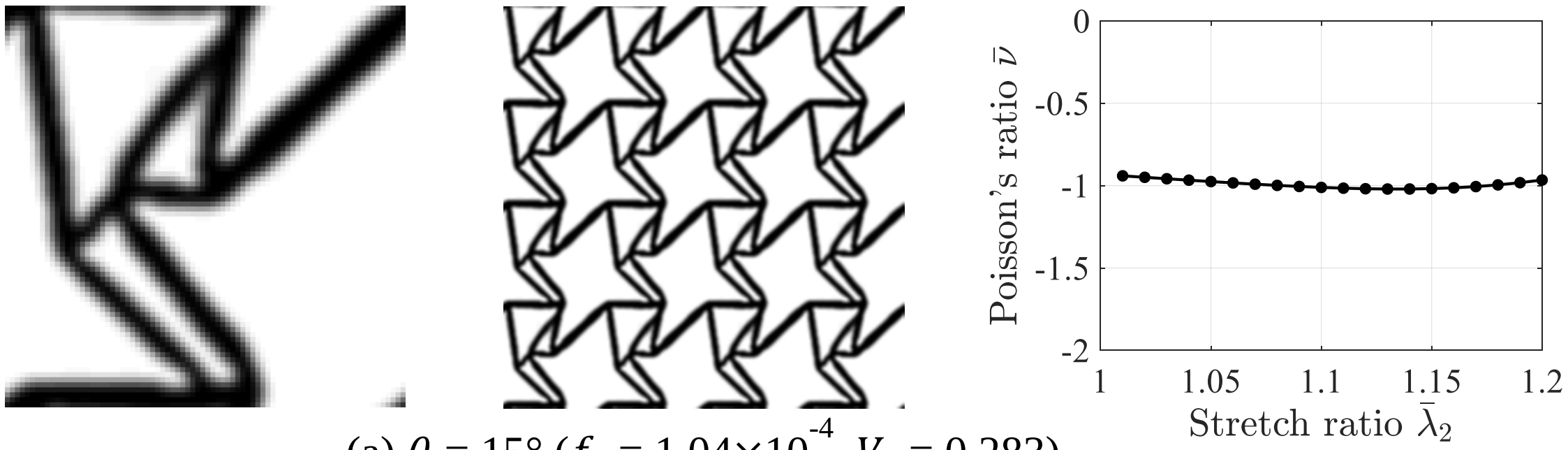

(a) $\theta$ = 15° ($f_0$ = 1.04×10$^{-4}$, $V_f$ = 0.283)

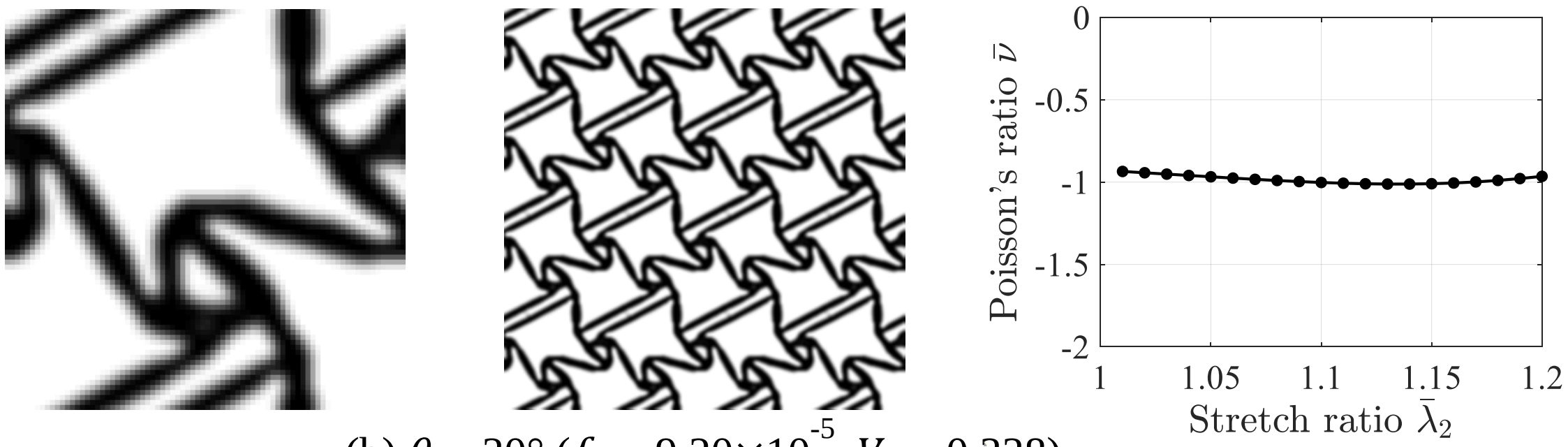

(b) $\theta$ = 30° ($f_0$ = 9.20×10$^{-5}$, $V_f$ = 0.338)

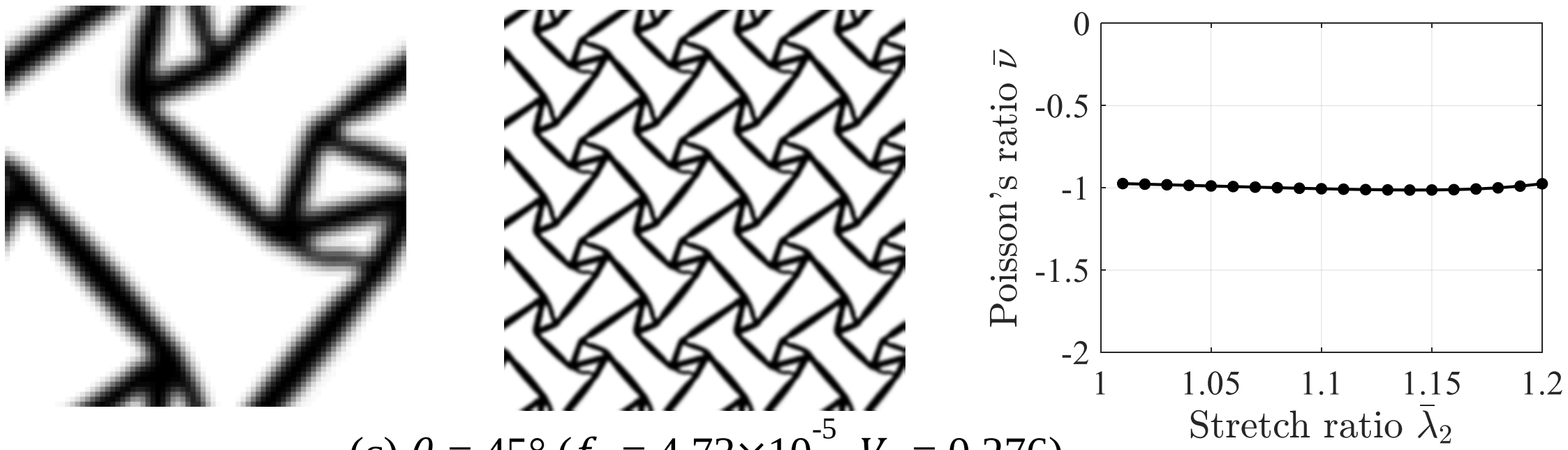

(c) $\theta$ = 45° ($f_0$ = 4.73×10$^{-5}$, $V_f$ = 0.276)

Figure 19. Optimized designs for different loading axes with $\bar{k}$ = 2 and $V_T$ = 0.4 ($\alpha_1$ = 0.01 after 200 iter.).

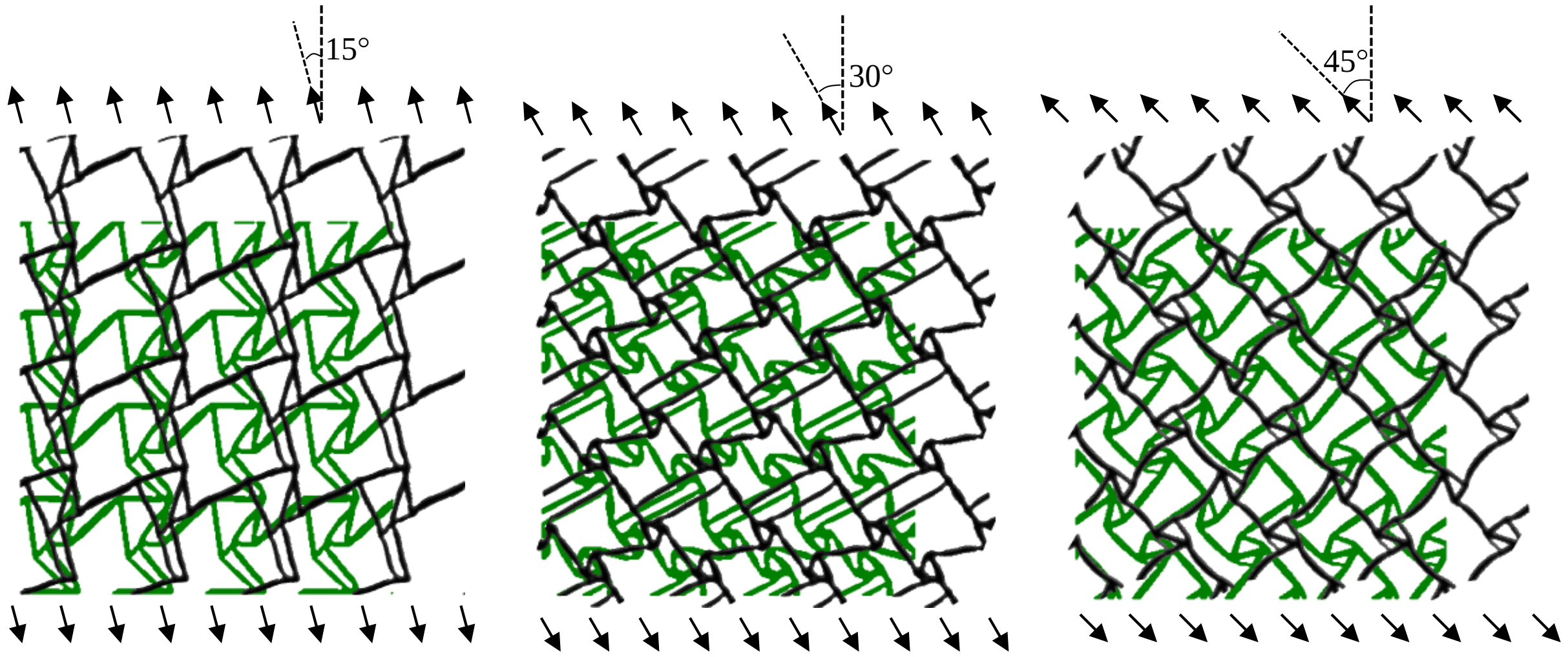

(a) $\theta$ = 15°, $\bar{\lambda}_1$ = 1.1932, $\bar{\boldsymbol{F}} = \begin{bmatrix} 1.194 & -0.001697 \\ -0.001697 & 1.200 \end{bmatrix}$

(b) $\theta$ = 30°, $\bar{\lambda}_1$ = 1.1930, $\bar{\boldsymbol{F}} = \begin{bmatrix} 1.195 & -0.003020 \\ -0.003020 & 1.198 \end{bmatrix}$

(c) $\theta$ = 45°, $\bar{\lambda}_1$ = 1.1950, $\bar{\boldsymbol{F}} = \begin{bmatrix} 1.198 & -0.00249 \\ -0.00249 & 1.198 \end{bmatrix}$

Figure 20. Deformed shape of the topologies in Figure 19 (a), (b) and (c) at $\bar{\lambda}_2$ = 1.2.

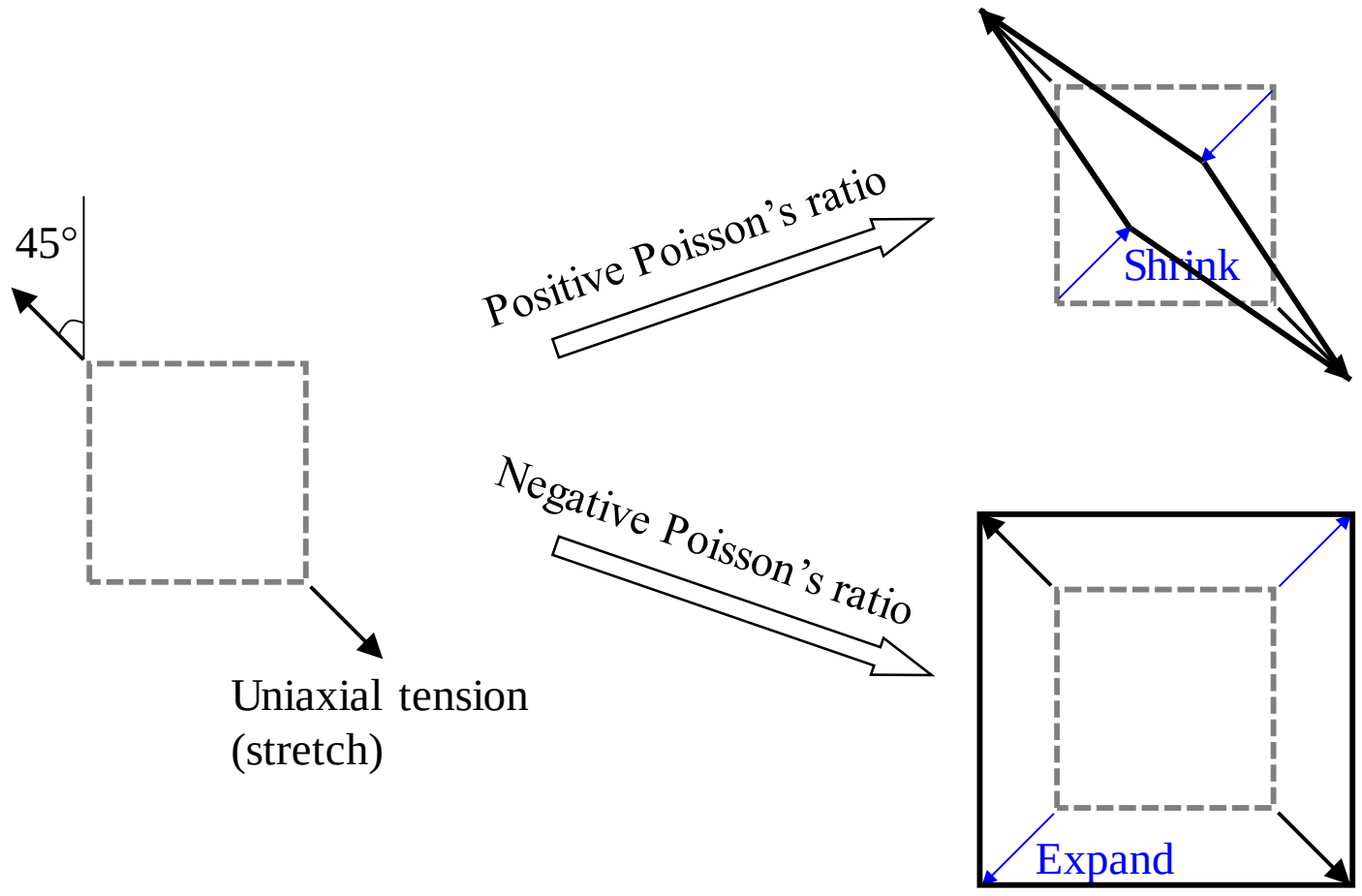

Figure 21. Illustration of Poisson's ratio in measuring the relative resistances to isochoric and volumetric deformations.

### *5.8 Multimaterial NPR design*

Lightweight auxetic material with prescribed stiffness might be desirable in engineering applications and this can be also achieved using the presented nonlinear homogenization framework by using the OF-2 (Eq. (47)), which considers multimaterial design with constraints on total mass. As illustrative examples, two hyperelastic materials with properties given in Table 1 are considered, where the softer and lighter material M2 has a higher stiffness to weight ratio as compared to the stiffer and heavier material M1. Using the optimization formulation in Eq. (47), the material weight constraint $M^*$ in Eq. (49) is chosen as $M^* = \omega^* \cdot V_D$ with $\omega^*$ = 500 (Figure 23) or 400 (Figure 24) and $V_D$ is the total volume of the unit cell design domain, i.e. $V_D$ = 1 for square unit cell, $V_D$ = 0.866 for 60° parallelogram unit cell and $V_D$ = 2.598 for hexagonal unit cell, respectively. The stiffness constraints are chosen as $\bar{k}$ = 4 in Eq. (47). Starting from the initial designs shown in Figure 22 for different unit cell shapes ($\rho_2$ = 0.5 while $\rho_1$ = 0 or 1 in the initial design), the corresponding optimized results for $\omega^*$ = 500 and $\omega^*$ = 400 are shown in Figure 23 and Figure 24, respectively. It can be seen that different constraints or initial designs or unit cell shapes greatly affect the optimized topologies, as expected. Besides, although not efficient in terms of stiffness/weight ratio, the material M1 is still needed in the optimized topologies due to the relatively high stiffness constraint. These examples also demonstrate the effectiveness of proposed multimaterial interpolation scheme in generating discrete and meaningful auxetic metamaterial designs.

Table 1. Material properties of M1 and M2 material phases.

| Material | Young's modulus $E$ | Density $\omega$ | $E/\omega$ |
|---|---|---|---|
| M1 (red) | 300 | 2100 | 0.143 |
| M2 (blue) | 100 | 500 | 0.200 |

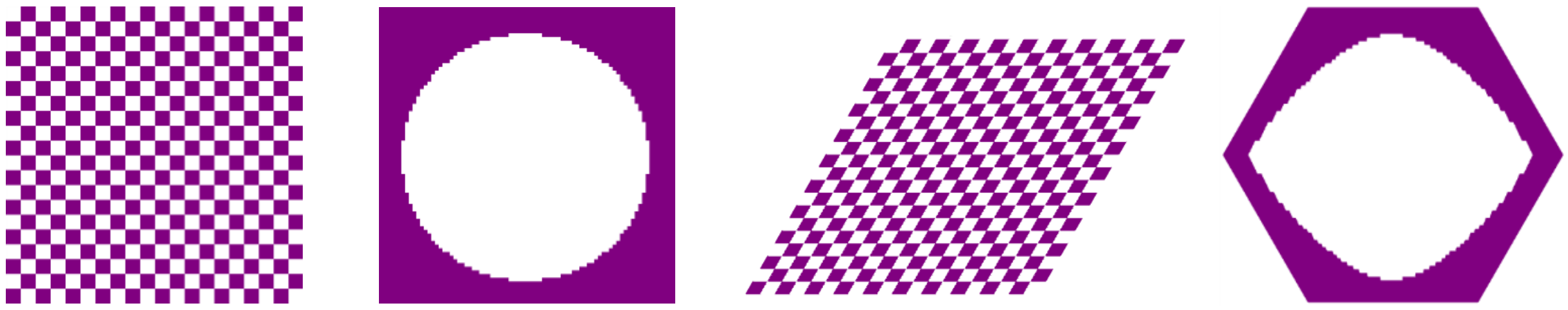

(a) (b) (c) (d)

Figure 22. Initial designs for: (a) and (b) square unit cell; (c) 60° parallelogram unit cell; and (d) hexagonal unit cell.

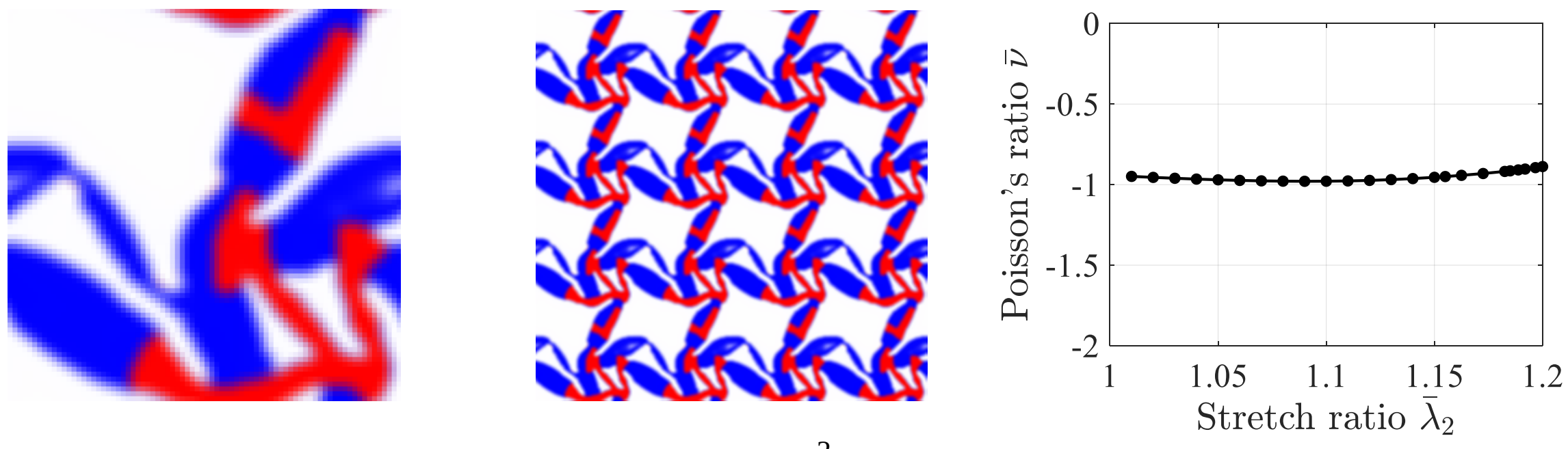

(a) $\alpha_2$ = 0 ($f_0$ = 2.46×10$^{-3}$, $M_f$ = 1)

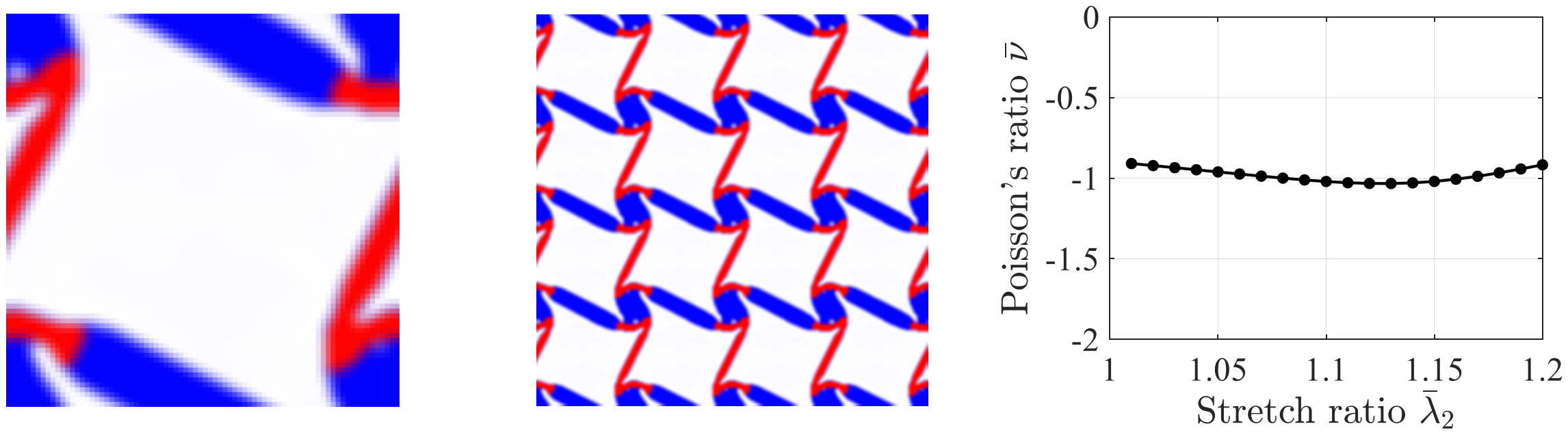

(b) $\alpha_2$ = 0.02 ($f_0$ = 5.26×10$^{-4}$, $M_f$ = 0.654)

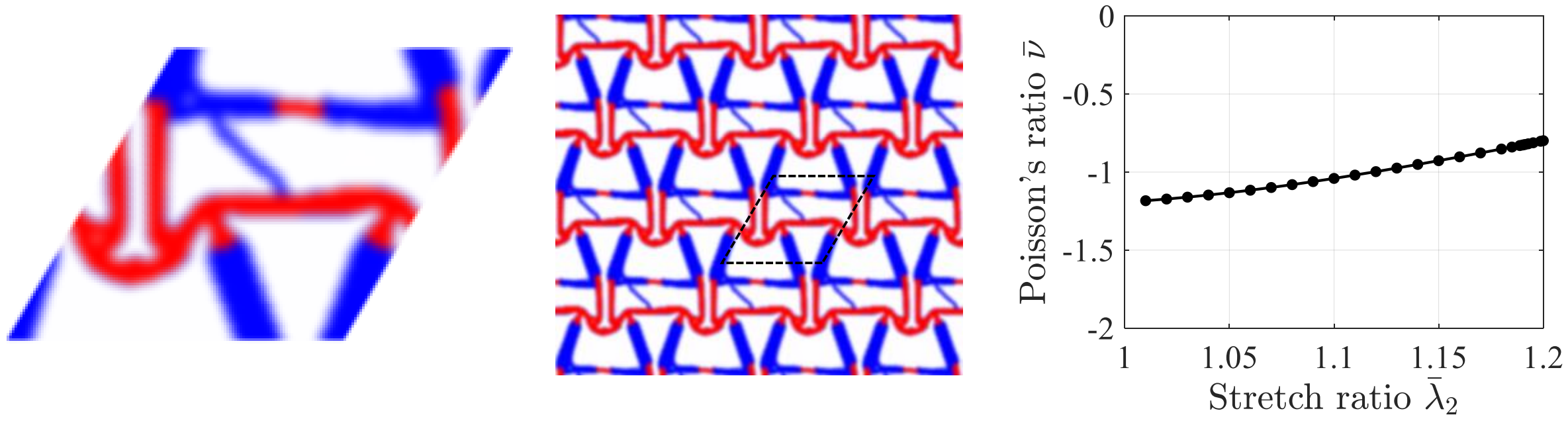

(c) $\alpha_2$ = 0.1 ($f_0$ = 1.17×10$^{-2}$, $M_f$ = 0.913)

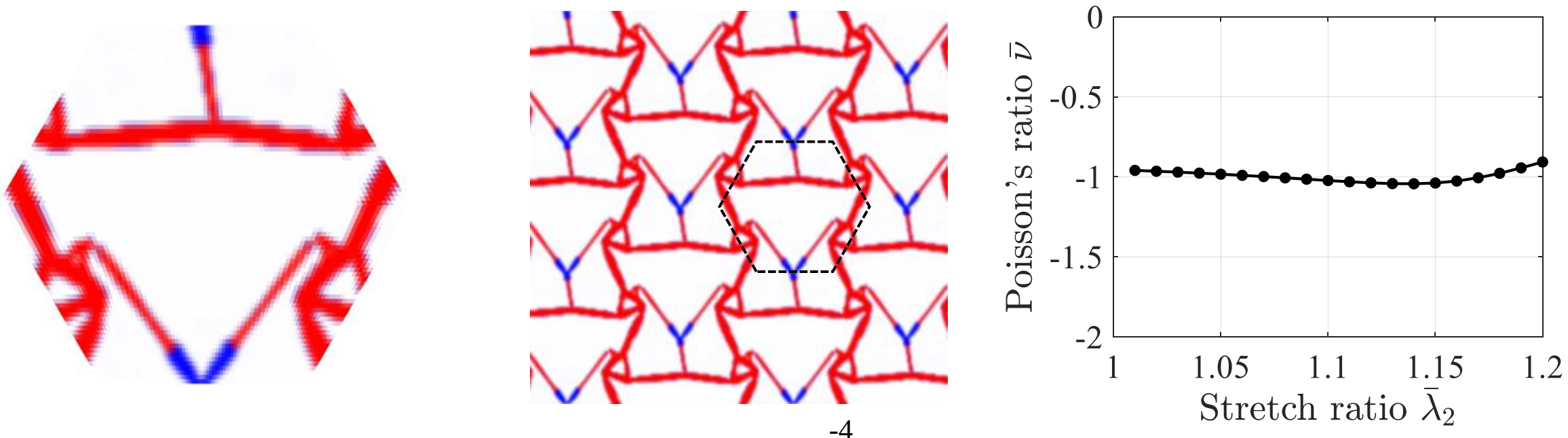

(d) $\alpha_2$ = 0.02 ($f_0$ = 6.34×10$^{-4}$, $M_f$ = 0.788)

Figure 23. Optimized multimaterial designs for different unit cell shapes and initial designs in Figure 22 with $\bar{k}$ = 4 and $\omega^*$ = 500 ($\alpha_2$ = 0 before 200 iter.).

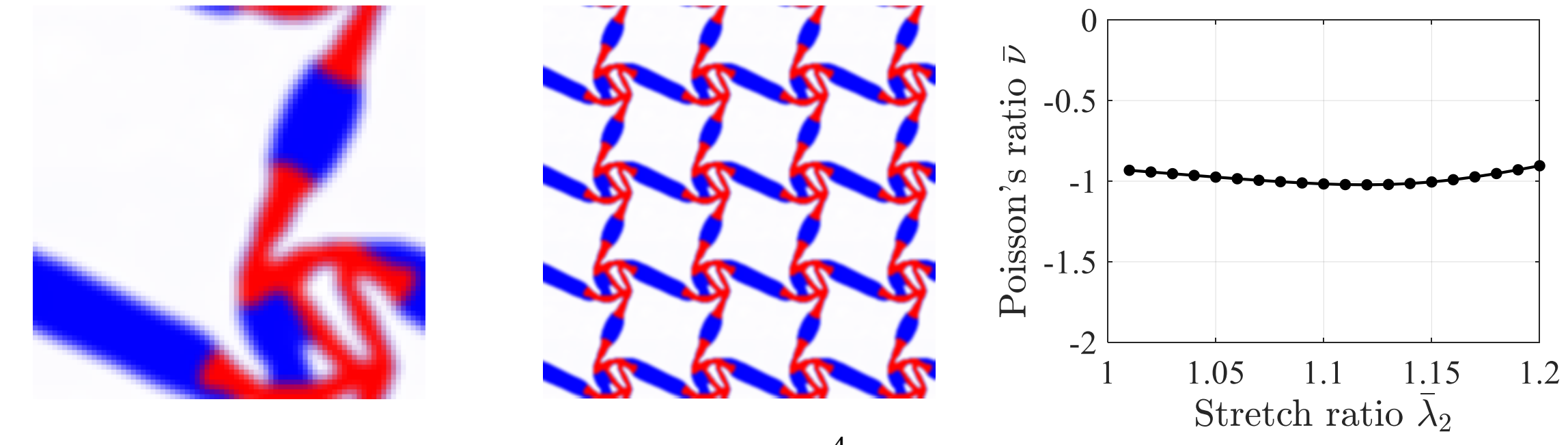

(a) $\alpha_2$ = 0.02 ($f_0$ = 6.78×10$^{-4}$, $M_f$ = 0.908)

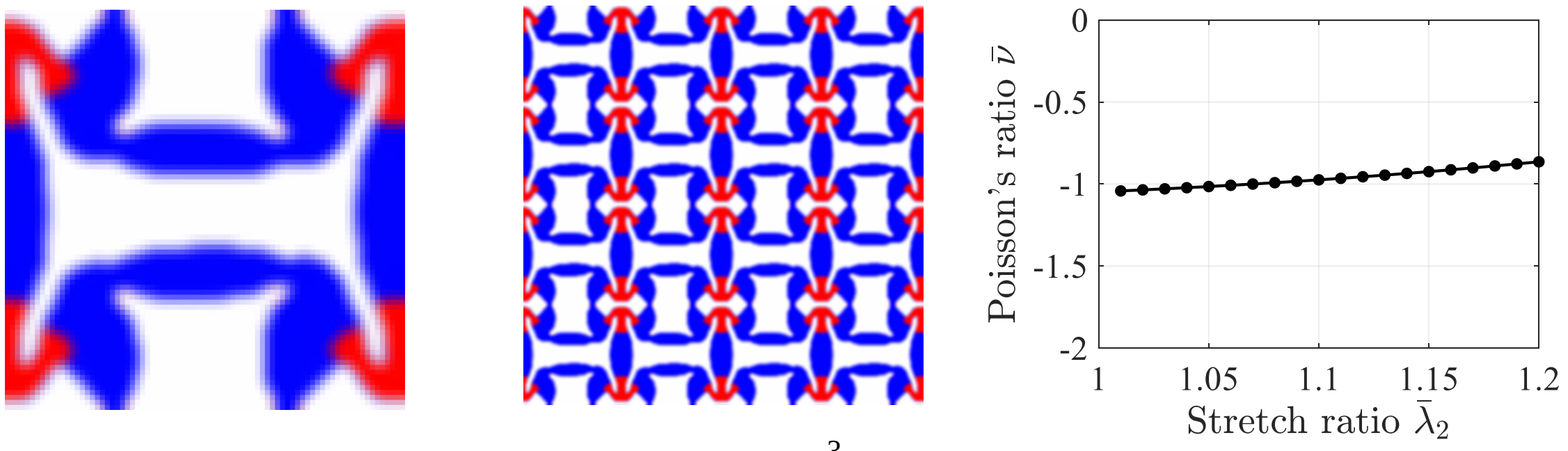

(b) $\alpha_2$ = 0 ($f_0$ = 2.46×10$^{-3}$, $M_f$ = 1)

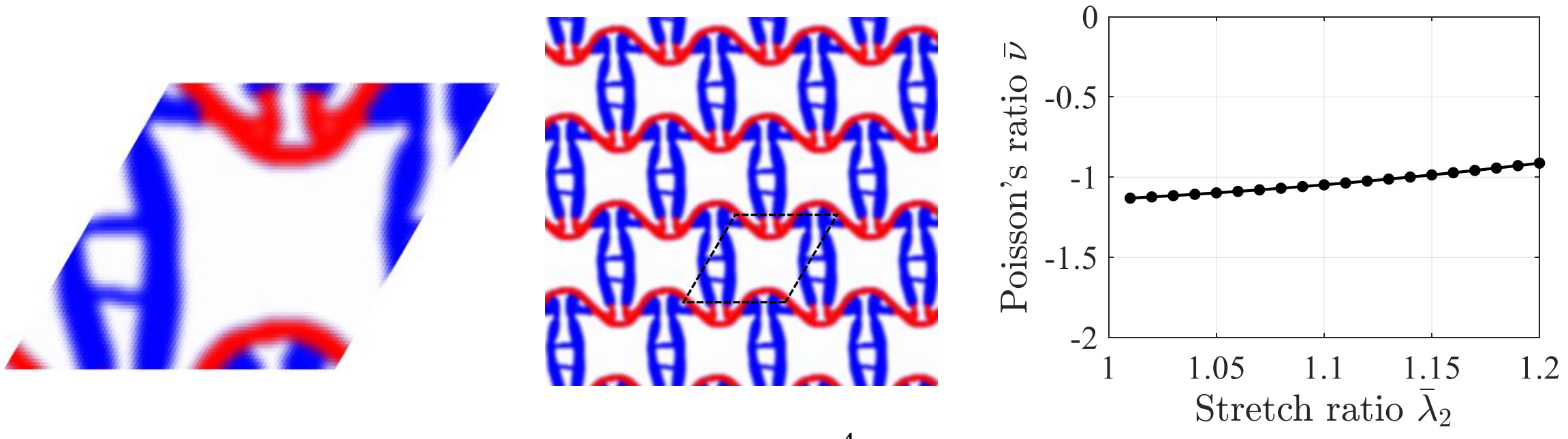

(c) $\alpha_2$ = 0 ($f_0$ = 9.00×10$^{-4}$, $M_f$ = 1)

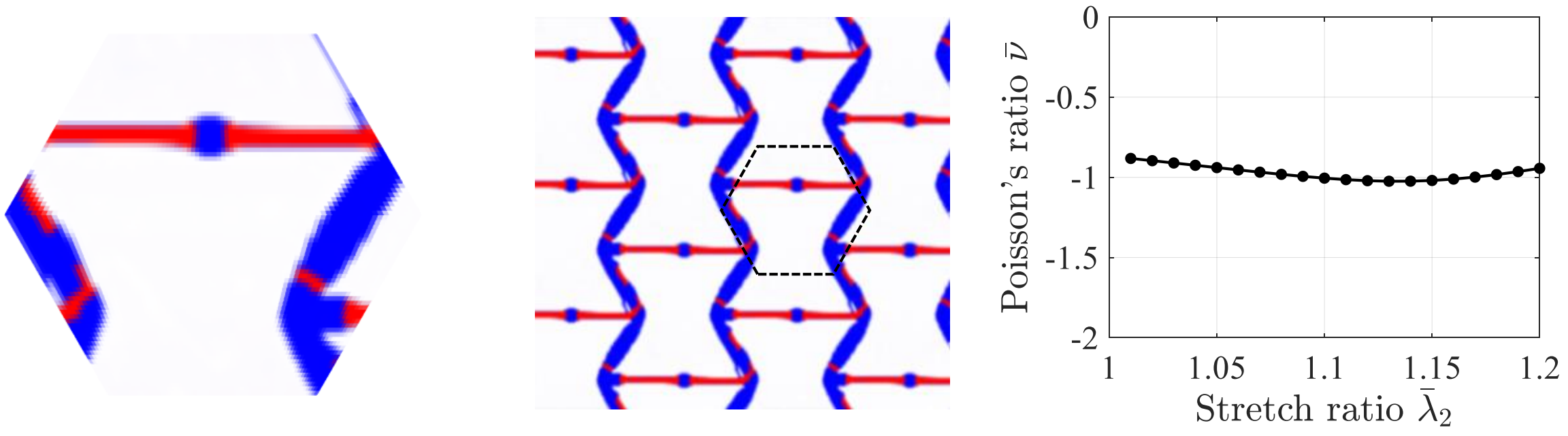

(d) $\alpha_2$ = 0.02 ($f_0$ = 2.69×10$^{-4}$, $M_f$ = 0.562)

Figure 24. Optimized multimaterial designs for different unit cell shapes and initial designs in Figure 22 with $\bar{k}$ = 4 and $\omega^*$ = 400 ($\alpha_2$ = 0 before 200 iter.).

## 6 Numerical testing of the optimized auxetic metamaterial designs

In this section, representative optimized topologies obtained in Section 5 are further validated using direct numerical simulations, where a bulk of material with 20×20 optimized unit cells is tested under uniaxial loading boundary conditions in ABAQUS [40] (Figure 26). The Poisson's ratio is measured by the negative ratio of engineering strains along the transverse and longitudinal directions of the central 2×2 unit cells. The geometries of the optimized unit cells are obtained by a B-spline curve fitting of the level set of the density value 0.5 (Figure 25). The optimized topologies in Figure 14e and Figure 24b are investigated and their corresponding fitted topologies (after FE discretization) are shown in Figure 26. Four node plane strain elements with reduced integration (CPE4R) are used, and the total number of elements for the 20×20 testing of Design-A in Figure 26a is 889200, while 2424000 elements are used in Design-B. Since these two topologies are optimized for stretch ratio $\bar{\lambda}_2$ = 1.2, the uniaxial loading condition is specified as $u_y$ = 4, which corresponds to 20% axial strain of the bulk of material. In the FEA, the actual stretches that the central 2×2 cells undergo are 0.3997 and 0.3984, respectively, for the Design-A and Design-B, which correspond to approximately 20% engineering strain. The analysis results are shown in Figure 27, where the deformation of the central 3×3 unit cells are plotted in Figure 27(a) and Figure 27(c) at different deformation stages. The measured Poisson's ratios for the two designs are calculated and compared with the results from the homogenization analysis of one unit cell in Figure 27(b), which show a close match between the two results. The small differences may be attributed to the fact that the boundary conditions are not perfectly matched with those assumed in the homogenization, and also that the scale separation assumption in homogenization is not completely fulfilled. Nevertheless, a good agreement further strengthen the basis for using the presented homogenization framework for designing nonlinear auxetic metamaterials. Moreover, the proposed measure for Poisson's ratio in Eq. (43) is also validated by the close matches in the two results. On the other hand, the resulted Poisson's ratios are not the same as the ones in the design phase (see Figure 14e and Figure 24b). This is due to the difference in the designs before and after B-spline fitting. It should be noted that these difference can be further mitigated by adopting projection techniques and by incorporating additional shape optimization based on the B-spline representation of the design. The investigation of these methods are out of this paper's scope, however, and the interested readers are referred to Refs [41-44].

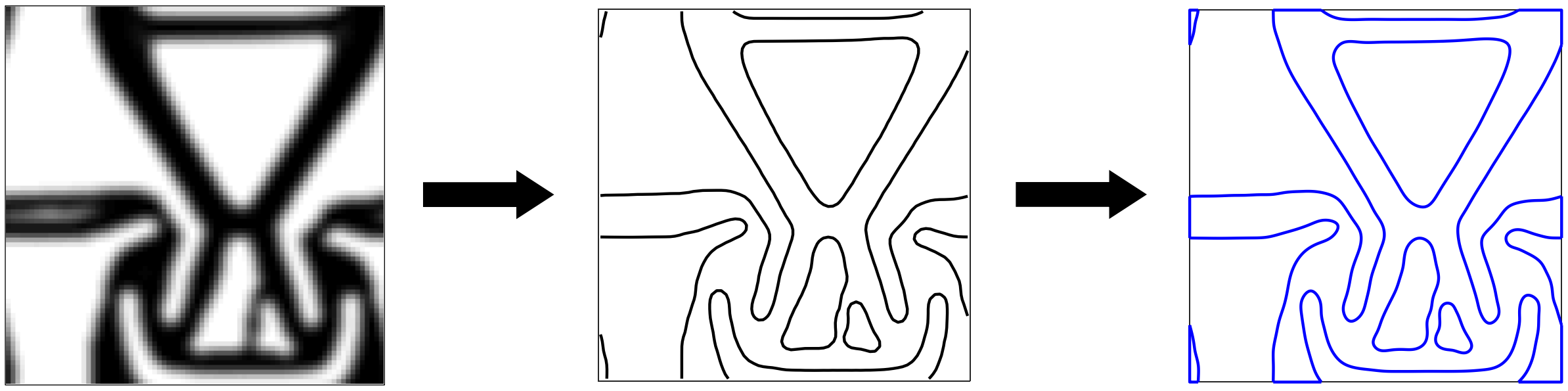

(a) Optimized topology (b) Contour at $\rho$ = 0.5 (c) B-spline fit

Figure 25. Contour plot and B-spline fit.

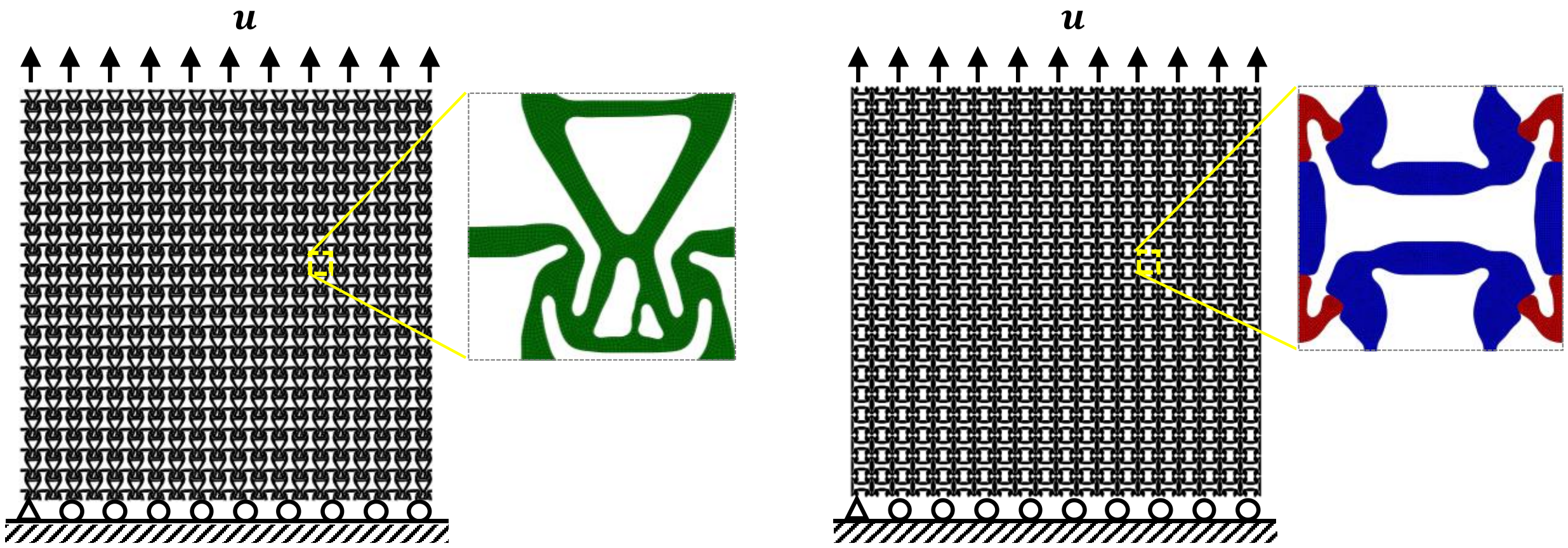

(a) Design-A (b) Design-B

Figure 26. Uniaxial test setups for two periodic solids including 20×20 unit cells.

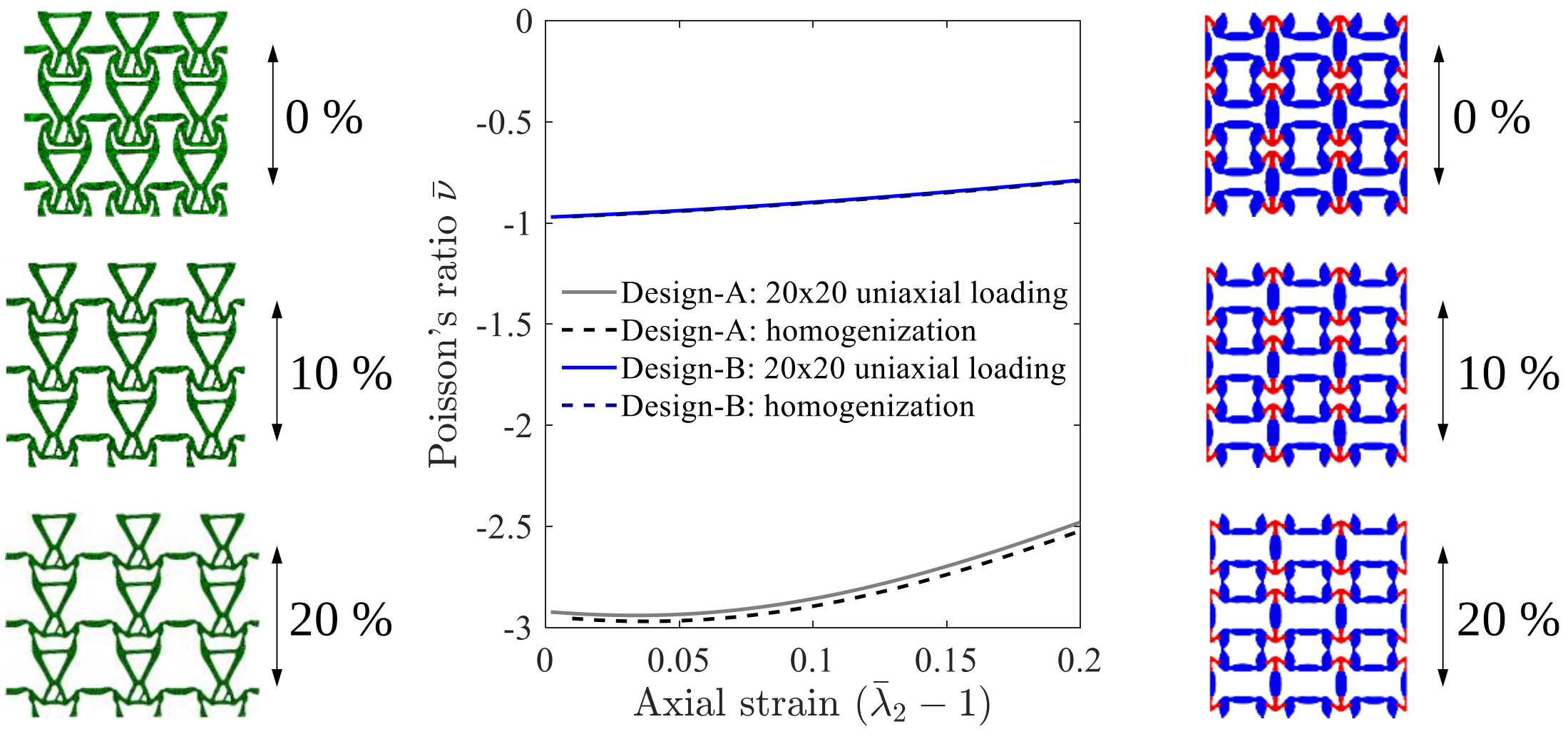

(a) (b) (c)

Figure 27. Uniaxial test results of the 20×20 periodic solids: Comparison between 20×20 uniaxial test results and the 1 unit cell homogenization results in (b); Deformation patterns of the central 3×3 unit cells at different loading strains for Design-A and Design-B, respectively, in (a) and (c).

## 7 Multiscale stability

A basic assumption in the homogenization analysis during optimization process is that one unit cell serves as the fundamental periodic cell during the entire loading process and can be taken as the RVE. For finitely strained nonlinearly elastic composites, this assumption, however, does not always hold. Upon loading, buckling can happen at the microscale at a wavelength possibly across arbitrary length, which leads to a change of the periodicity in the underlying microstructure. If the buckling mode is periodic, a fundamental periodic cell consists of more than one unit cell, while when the buckling mode is aperiodic, no fundamental periodic cell can be found [45]. From macroscopic viewpoint, though polyconvexity in the sense of Ball [46] of the underlying material phases is guaranteed, which ensures the strict rank-one convexity, the homogenized macroscopic metamaterial may still lose strict rank-one convexity [47]. As shown in the previous studies, there exists a close connection between microscale buckling and macroscale loss of strict rank-one convexity, i.e., the long wavelength buckling on the microscale corresponds to the loss of strict rank-one convexity (strong ellipticity) in the homogenized incremental moduli on the macroscale [27, 28]. Also noted is that the micro-instability occurs either before (short wavelength buckling) or simultaneously (long wavelength buckling) with the macro-instability [27, 28]. Compared to the macro-stability check where a rank-one convexity examination of the homogenized incremental moduli is only needed, the micro-stability check is much more computationally expensive since the length scale of the buckling mode is not a priori known.

### *7.1 Microscale stability*

For rate-independent solids, the stability is governed by the Hill's stability criterion [48]. The principal solution branch ceases to be stable if the functional $\beta(\lambda)$ defined by

$$\beta(\lambda) = \inf_{\boldsymbol{v}} Q(\boldsymbol{v}; \mathbb{R}^d) \qquad \text{with} \qquad Q(\boldsymbol{v}; \Omega) \overset{\text{def}}{=} \frac{\int_\Omega \boldsymbol{\nabla v}^* : \mathbb{A} : \boldsymbol{\nabla v} \, dV}{\int_\Omega \boldsymbol{\nabla v}^* : \boldsymbol{\nabla v} \, dV} \tag{65}$$

loses positive definiteness, where $\boldsymbol{v}$ is taken from the kinematically admissible displacement variation space $H_0^1(\Omega)$ for the corresponding macroscale boundary value problem. For periodic solids of infinite extent ($\Omega \to \mathbb{R}^d$), $\boldsymbol{v}$ is taken from locally integrable, bounded functions that ensures the finiteness of the ratio $Q$ [28]. The symbol $*$ denotes the complex conjugate and $\lambda$ stands for the load parameter, which in this study can be taken as identical to the macroscopic stretch ratio $\bar{\lambda}_2$ (see Section 2.2). The tensor $\mathbb{A}$ represents the tangent moduli under the loading parameter $\lambda$ with the same periodicity as one unit cell. It was shown in [27] that this infimum $\beta(\lambda)$ can be computed through Bloch wave analysis, where the calculation is carried out within one unit cell $\Omega_0^\mu$ and is expressed as

$$\beta(\lambda) = \inf_{\boldsymbol{k}} \inf_{\boldsymbol{u}} Q\left(\boldsymbol{v}_B(\boldsymbol{k}, \boldsymbol{u}); \Omega_0^\mu\right) \quad \text{with} \quad \boldsymbol{v}_B(\boldsymbol{k}, \boldsymbol{u}) = e^{i\boldsymbol{k}.\boldsymbol{X}}\boldsymbol{u} \tag{66}$$

where $\boldsymbol{v}_B$ is the Bloch wave function representing the eigenmode in which $\boldsymbol{u}$ is periodic functions with the same periodicity as one unit cell, i.e. $\boldsymbol{u}(\boldsymbol{X} + c_i\boldsymbol{a}_i) = \boldsymbol{u}(\boldsymbol{X})$ with $c_i$ arbitrary integers and $\boldsymbol{a}_i$ the $i^{th}$ periodic lattice vector ($i$ = 1, …, $d$), while the wavevector $\boldsymbol{k}$ is chosen in the 1st Brillouin zone (BZ) in the reciprocal space spanned by the reciprocal bases $\boldsymbol{b}_i$ ($i$ = 1, …, $d$) defined by $\boldsymbol{a}_i\boldsymbol{b}_j = 2\pi\delta_{ij}$ [49]. For the square unit cell, which is the case of interest in this section, the 1st BZ can be simply chosen as $k_i \in [0,1)$, $i = 1, \dots, d$ with $\boldsymbol{k} = \sum_i k_i\boldsymbol{b}_i$. It is worth noting that two physically different types of buckling modes exist in the neighborhood of $\boldsymbol{k} = \boldsymbol{0}$, i.e., the long wavelength instability with $\boldsymbol{k} \to \boldsymbol{0}$ that leads to the loss of rank-one convexity of the homogenized tangent moduli at the macroscale, and the highly localized buckling mode with $\boldsymbol{k} = \boldsymbol{0}$ which has the same periodicity as one unit cell. The following relationship was also established in [27]

$$\beta(\lambda) \leq \beta(\lambda; \boldsymbol{k} \to \boldsymbol{0}) \leq \beta(\lambda; \boldsymbol{k} = \boldsymbol{0}) \tag{67}$$

which states that the short wavelength instabilities always precede the long wavelength instabilities which always precede the highly localized ones. Interested readers are referred to Refs [27, 28, 50] for further theoretical details and numerical implementations.

### *7.2 Macroscale stability*

As a measure of the macroscopic stability, the strict rank-one convexity of the homogenized tangent moduli ensures the absence of discontinuities in the deformation gradient field on the macroscale. It can be assessed by examining the positive definiteness of the ellipticity indicator $B(\lambda)$ defined by

$$B(\lambda) = \min_{\bar{\boldsymbol{m}}, \bar{\boldsymbol{M}}} \; (\bar{\boldsymbol{m}} \otimes \bar{\boldsymbol{M}}) : \bar{\mathbb{A}} : (\bar{\boldsymbol{m}} \otimes \bar{\boldsymbol{M}}) \tag{68}$$

where $\bar{\boldsymbol{m}}$ and $\bar{\boldsymbol{M}}$ span over all possible directions with $\|\bar{\boldsymbol{m}}\| = \|\bar{\boldsymbol{M}}\| = 1$. A recent study have also shown that upon the loss of strict rank-one convexity there is not always a discontinuous/localized deformation pattern on the bifurcated branch [51]. The presence or absence of the localized deformation depends on the stability of the bifurcated branch [51]. When there is a discontinuous deformation corresponds to the loss of strict rank-one convexity, i.e. $B(\lambda) = 0$, the corresponding minimizing vector $\bar{\boldsymbol{M}}$ represents the normal to the curves across which the jump discontinuities appear and $\bar{\boldsymbol{m}}$ determines the nature of the discontinuous mode (simple shear if $\bar{\boldsymbol{m}}$ is orthogonal

to $\bar{\boldsymbol{M}}$ or pure splitting if $\bar{\boldsymbol{m}}$ is parallel to $\bar{\boldsymbol{M}}$ or mixture otherwise) [52]. Also, the loss of strict rank-one convexity corresponds to a long wavelength microscale buckling [27], i.e.

$$B(\lambda) = 0 \quad \text{if} \quad \beta(\lambda; \boldsymbol{k} \to \boldsymbol{0}) = 0 \tag{69}$$

### *7.3 Examples – Stability investigation of the optimized topologies*

Since in the Bloch wave analysis, the wave vector $\boldsymbol{k}$ needs to scan for the infimum over the whole 1st BZ which contains infinite points, the computational cost is extremely high. As a result, a direct incorporation of the Bloch wave stability calculation in the topology optimization process is not feasible in the presented topology optimization framework. However, the stability of the optimized topologies can be checked to further validate their performance. To this end, in this subsection, some of the optimized designs in Section 5 fitted using B-splines are examined for both micro and macro stabilities.

#### *7.3.1 Design for compression*

Figure 28a shows the FE mesh of the optimized auxetic design under compression (Section 5.2) fitted using B-spline. The macroscale rank-one convexity is first examined by slowly increasing the loading factor $\lambda$ ($\equiv \bar{\lambda}_2$) until $B(\lambda) \leq 0$ at every $\pi/720$ radian increment in both $\bar{\boldsymbol{m}}$ and $\bar{\boldsymbol{M}}$ space. The smallest load for the loss of rank-one convexity is detected as $\bar{\lambda}_2$ = 0.972175. Figure 28c shows the $B_\alpha(\lambda) = \min_{\bar{\boldsymbol{m}}}(\bar{\boldsymbol{m}} \otimes \bar{\boldsymbol{M}}) : \mathbb{A} : (\bar{\boldsymbol{m}} \otimes \bar{\boldsymbol{M}})$ versus the angle $\alpha \in [0, \pi)$ of the normal of the singular surface with respect to the horizontal axis, i.e. $\bar{\boldsymbol{M}} = \begin{bmatrix} \cos\alpha \\ \sin\alpha \end{bmatrix}$. By definition, the microscale stability must have been lost at this step as well, either with short or long wavelength buckling mode. The microscopic stability is investigated using the Bloch wave analysis where the 1st BZ, i.e., $(k_1, k_2) \in [0,1) \times [0,1)$, is discretized with a 400×400 uniform mesh together with $100 \times 100$ uniform meshes in three refined zones $(0,0.0025) \times (0.0025,1]$, $(0.0025,1] \times (0,0.0025)$ and $(0,0.0025) \times (0,0.0025)$. The stability indicator $\beta_{\boldsymbol{k}}(\lambda)$, which is defined as $\beta_{\boldsymbol{k}}(\lambda) = \inf_{\boldsymbol{u}} Q\left(\boldsymbol{v}_B(\boldsymbol{k}, \boldsymbol{u}); \Omega_0^\mu\right)$, is computed at each discretized point $\boldsymbol{k}$ in the 1st BZ and the results are shown in Figure 28b. Two wave vectors that lead to a change of sign of $\beta_{\boldsymbol{k}}(\lambda)$ were detected at the same $\bar{\lambda}_2$ = 0.972175 load step, which are $(k_1, k_2)$ = (1.9307×10$^{-3}$ , 0.99003) and $(k_1, k_2)$ = (1.9554×10$^{-3}$, 0.99003). With the same $\boldsymbol{k}$-mesh in the Bloch analysis, no microscale instabilities were found when increasing $\bar{\lambda}_2$ by $\Delta\bar{\lambda}_2$ = 7.5×10$^{-5}$. Moreover, it can be observed that the origin $(k_1, k_2) = (0, 0)$ is a singular point, which shows that the highly localized buckling mode is not occurring simultaneously with the long wavelength buckling and $\beta(\lambda) < \beta(\lambda, \boldsymbol{k} = \mathbf{0})$. Thus, within the precisions of the underlying computational study, it can be concluded that the optimized design loses both micro and macro stability on the principal branch at the stretch ratio $\bar{\lambda}_2$ = 0.972175, which is far from the design target $\bar{\lambda}_2$ = 0.85 for which this topology is designed. Thus, the homogenization results cease to be valid once $\bar{\lambda}_2 < 0.972175$, and the optimality of the optimized design in Figure 7 cannot be justified.

### *7.3.2 Design for tension*

Two representative topologies (Figure 12b and Figure 24b) optimized under tension are investigated for their multiscale stabilities. As opposed to the compression case, both micro and macro stabilities are maintained during the loading process even with large stretch ratios ($\bar{\lambda}_2$ = 1.6 in Figure 29a). The macro rank-one convexity curves are shown in Figure 29(a) and Figure 29(b), respectively, for the two designs at different loading steps. The Bloch wave analysis is carried out for both designs using the same $\boldsymbol{k}$-mesh as that in Section 7.3.1 in the 1st BZ and no microscopic instability was detected.

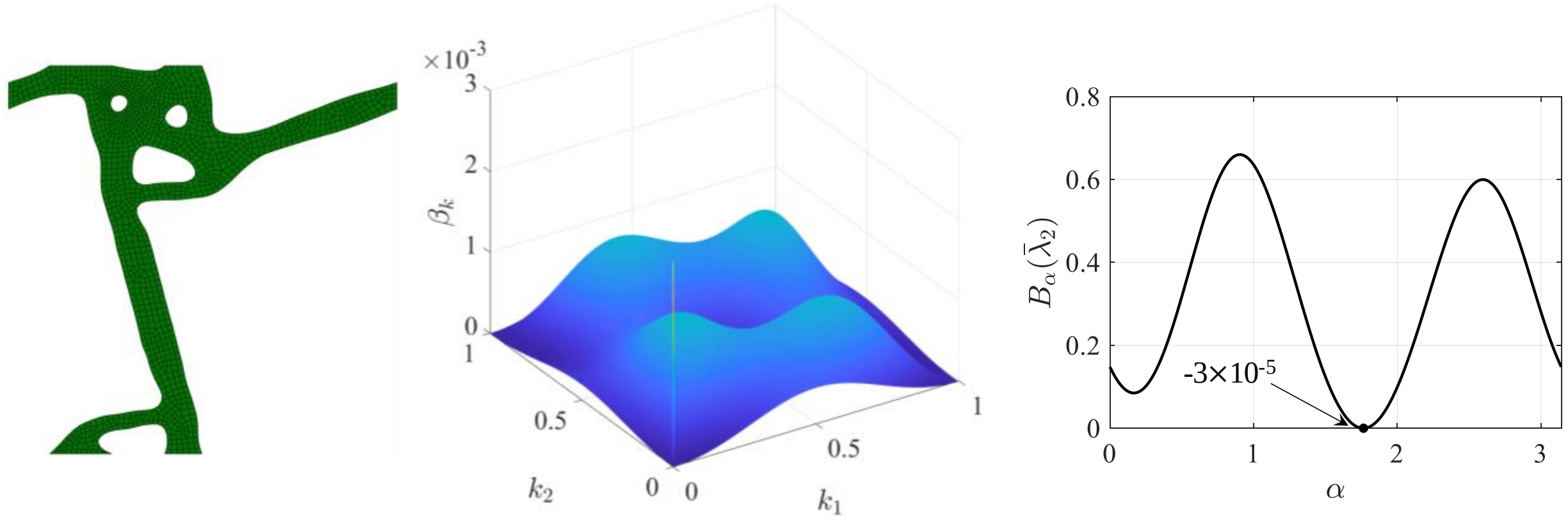

(a) Unit cell FE mesh (b) Microscale stability surface (c) Macroscale stability

Figure 28. Loss of micro and macro stabilities of the topology in Figure 7b at $\bar{\lambda}_2$ = 0.972175.

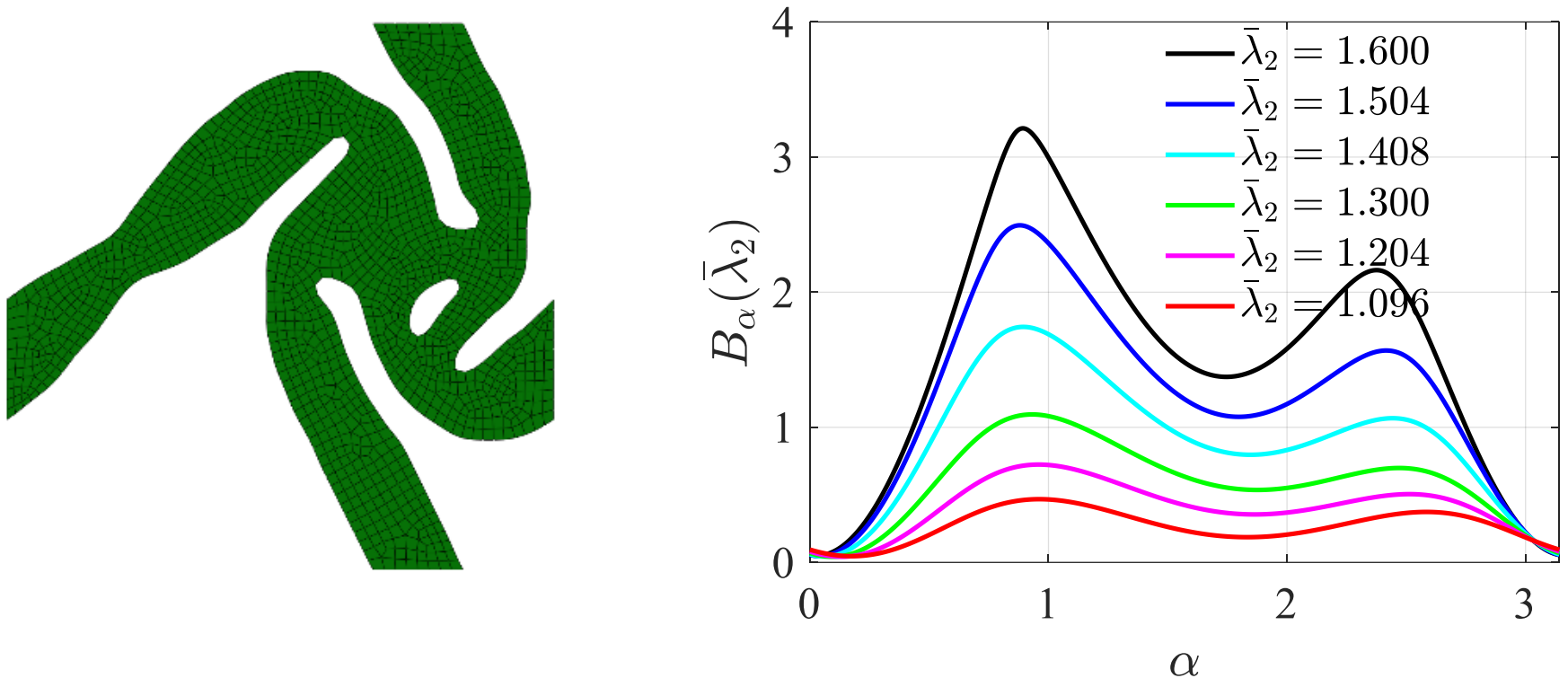

Unit cell FE mesh Rank-one convexity curves

(a) Topology in Figure 12b under uniaxial tension $\bar{\lambda}_2 = 1.6$

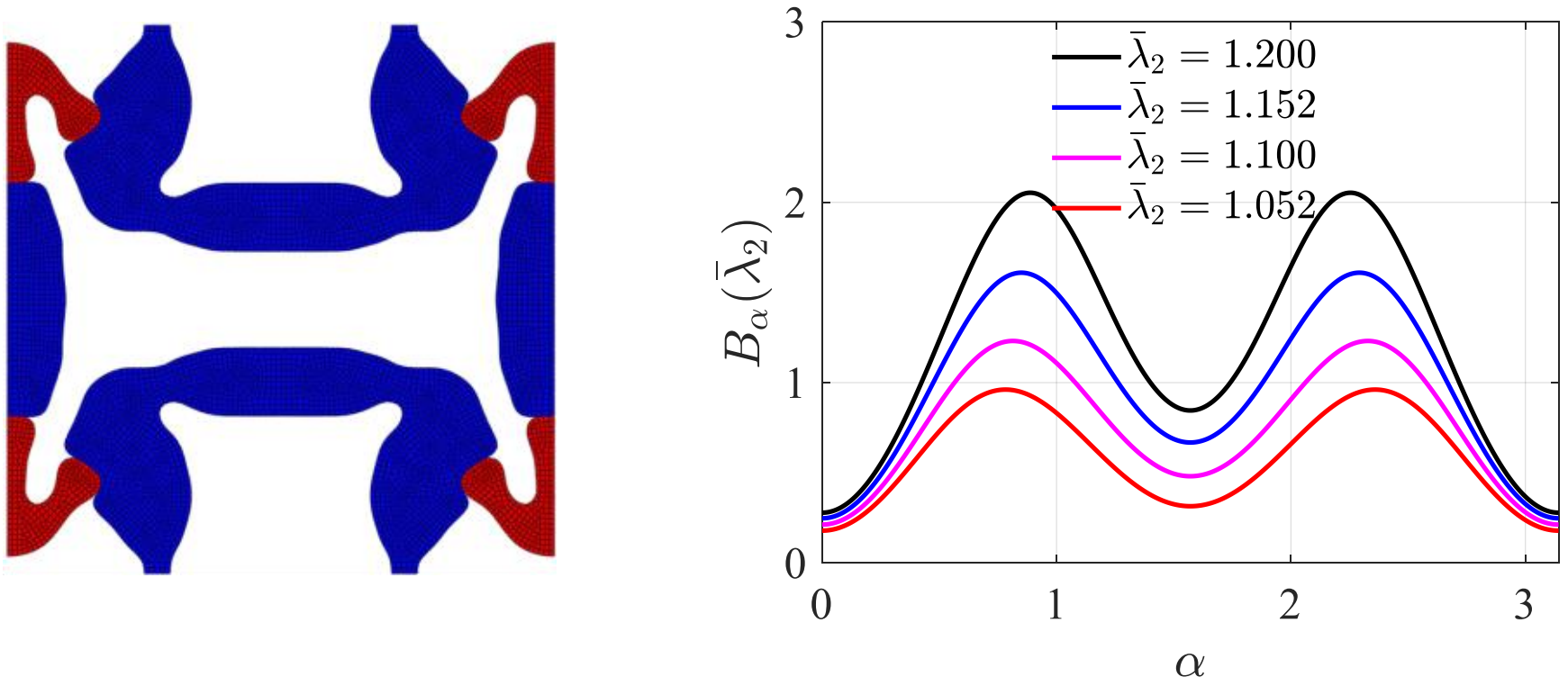

Unit cell FE mesh Rank-one convexity curves

(b) Topology in Figure 24b under uniaxial tension $\bar{\lambda}_2 = 1.2$

Figure 29. Macroscale rank-one convexity investigations of different topologies under tension.

## 8 Conclusion

In this study, a computational framework is proposed for designing nonlinear auxetic metamaterials based on the nonlinear mixed stress/deformation driven homogenization and density-based multimaterial topology optimization. Optimization formulations with multimaterial hyperelastic phases are considered, and challenges associated with obtaining discrete topologies are addressed by multiobjective formulations, wherein material usage term is considered in the overall objective function. Compared to the previous studies on auxetic metamaterial designs, the merits of the proposed framework are:

- A consistent nonlinear homogenization method is incorporated which enables clear representation of the homogenized material properties of periodic solids under finite strains.

- In the presented homogenization framework, the design space is greatly expanded. For example, designs with different unit cell shapes such as square, parallelogram and hexagon can be consistently explored and different loading orientations can be considered.
- Design of auxetic metamaterials is extended from single to multiple material phases.
- In the presented homogenization-based formulation, the homogenized tangent moduli can be evaluated, which allows for the multiscale stability examination of the optimized metamaterials by Bloch wave analyses and rank-one convexity checks.

Many novel single and multimaterial auxetic metamaterial designs are created using the proposed framework. Performance of the optimized designs as well as the validity of the proposed homogenization method is successfully demonstrated by the direct numerical simulations on the bulk of periodic solids with optimized microstructures. Multiscale stability investigation using Bloch wave analyses and rank-one convexity checks show that the optimized metamaterial designs might lose both micro and macro stabilities during the loading process. Thus, the optimized topology should always be examined for their micro and macro stabilities in order to validate the structural performance under applied loads. Although only 2D problems are considered in this study, the proposed framework can be canonically extended to 3D cases. Finally, the consideration of multiscale stability in the optimization phase is challenging, and further investigations are needed to directly consider the multiscale stability constraints in the design phase.

## Acknowledgements

The presented work is supported in part by the US National Science Foundation through Grant CMMI-1762277. Any opinions, findings, conclusions, and recommendations expressed in this paper are those of the authors and do not necessarily reflect the views of the sponsors.

## Appendix A. Explicit derivatives required for the adjoint sensitivity analysis

This appendix gives the derivatives that are used in the sensitivity calculation of $f_0$ (see Section 4.1). In the following derivations, the tensor forms and matrix-vector forms are both utilized for notational simplicity and the appropriate form should be clear by the context.

### A.1 Derivative $\partial f_0/\partial v^k$

The calculation of the derivative $\partial f_0/\partial v^k$ is straightforward

$$\frac{\partial f_0}{\partial v^k} \equiv \frac{\partial f_0}{\partial \bar{\lambda}_1^k} = 2\left(\bar{\lambda}_1^k + a\bar{\lambda}_2^k - \bar{v}_T - 1\right) \tag{A 1}$$

### A.2 Derivatives $\partial \boldsymbol{R}^k/\partial \boldsymbol{\rho}_A$ $(A = 1,2)$

Taking the F-bar formulation (Eqns. (55) and (56)) and linear energy interpolation (Eqns. (53) and (54)) into account, some useful derivatives are first calculated.

The 1st PK stress $\boldsymbol{P}$ is computed as

$$\boldsymbol{P} = r^{-1/2}\boldsymbol{P}^b \quad \text{with} \quad \boldsymbol{P}^b = \boldsymbol{P}_v^b + \boldsymbol{P}_1^b + \boldsymbol{P}_2^b \tag{A 2}$$

where $r = r(\rho_1, \boldsymbol{u})$ is a function of $\rho_1$ and $\boldsymbol{u}$, and $\boldsymbol{P}_v^b$, $\boldsymbol{P}_1^b$ and $\boldsymbol{P}_2^b$ are 1st PK stresses contributed from different material phases. Using this information, the following derivatives are derived

$$\begin{aligned}&\frac{\partial r}{\partial \rho_1} = \frac{\partial r}{\partial \boldsymbol{F}}:\frac{\partial \boldsymbol{F}}{\partial \rho_1} + \frac{\partial r}{\partial \boldsymbol{F}_0}:\frac{\partial \boldsymbol{F}_0}{\partial \rho_1} \quad \text{with} \quad \frac{\partial r}{\partial \boldsymbol{F}} = -r\boldsymbol{F}^{-T} \text{ and } \frac{\partial r}{\partial \boldsymbol{F}_0} = r\boldsymbol{F}_0^{-T} \\ &\text{and} \quad \frac{\partial \boldsymbol{F}}{\partial \rho_1} = \frac{\partial \gamma}{\partial \rho_1}\boldsymbol{\nabla}_{\boldsymbol{X}}\boldsymbol{u} \quad \text{and} \quad \frac{\partial \boldsymbol{F}_0}{\partial \rho_1} = \frac{\partial \gamma}{\partial \rho_1}\boldsymbol{\nabla}_{\boldsymbol{X}}^0\boldsymbol{u}\end{aligned} \tag{A 3}$$

where $\boldsymbol{\nabla}_{\boldsymbol{X}}^0$ denotes the gradient operator evaluated at the centroid of the element. Also,

$$\frac{\partial r}{\partial \boldsymbol{u}} = \frac{\partial r}{\partial \boldsymbol{F}}:\frac{\partial \boldsymbol{F}}{\partial \boldsymbol{u}} + \frac{\partial r}{\partial \boldsymbol{F}_0}:\frac{\partial \boldsymbol{F}_0}{\partial \boldsymbol{u}} \quad \text{with} \quad \frac{\partial \boldsymbol{F}}{\partial \boldsymbol{u}} = \gamma\boldsymbol{B} \text{ and } \frac{\partial \boldsymbol{F}_0}{\partial \boldsymbol{u}} = \gamma\boldsymbol{B}_0 \tag{A 4}$$

where $\boldsymbol{B}$ and $\boldsymbol{B}_0$ are the shape functions derivative matrices evaluated at the integration points and the centroid, respectively. Besides, due to the dependence of $\boldsymbol{F}^b$ on $\rho_1$ and $\boldsymbol{u}$ the following derivatives are obtained

$$\frac{\partial \boldsymbol{F}^b}{\partial \rho_1} = \frac{1}{2}r^{-1/2}\frac{\partial r}{\partial \rho_1}\boldsymbol{F} + r^{1/2}\frac{\partial \boldsymbol{F}}{\partial \rho_1} \quad \text{and} \quad \frac{\partial \boldsymbol{F}^b}{\partial \boldsymbol{u}} = \frac{1}{2}r^{-1/2}\boldsymbol{F} \otimes \frac{\partial r}{\partial \boldsymbol{u}} + r^{1/2}\frac{\partial \boldsymbol{F}}{\partial \boldsymbol{u}} \tag{A 5}$$

With all the above derivatives and noting that the explicit dependence of $\boldsymbol{R}^k$ on $\boldsymbol{\rho}_1$ comes from the linear energy interpolation parameter $\gamma(\rho_1)$ as well as the interpolated constitutive model parameters, the derivative $\partial \boldsymbol{R}^k/\partial \boldsymbol{\rho}_1$ is computed as

$$\frac{\partial \boldsymbol{R}^k}{\partial \boldsymbol{\rho}_1} = \begin{bmatrix} \dfrac{\partial \boldsymbol{F}_{int}^k}{\partial \boldsymbol{\rho}_1} \\ \boldsymbol{0} \\ \boldsymbol{0} \end{bmatrix} \quad \text{with} \quad \frac{\partial \boldsymbol{F}_{int}^k}{\partial \boldsymbol{\rho}_1} = \mathop{\mathcal{A}}_{e=1}^{n_{ele}} \frac{\partial \boldsymbol{F}_{int,e}^k}{\partial \boldsymbol{\rho}_1} \quad \text{and} \quad \frac{\partial \boldsymbol{F}_{int,e}^k}{\partial \boldsymbol{\rho}_1} = \begin{bmatrix} \dfrac{\partial \boldsymbol{F}_{int,e}^k}{\partial \rho_1^1} & \cdots & \dfrac{\partial \boldsymbol{F}_{int,e}^k}{\partial \rho_1^{n_{ele}}} \end{bmatrix}$$

where $\dfrac{\partial \boldsymbol{F}_{int,e}^k}{\partial \rho_1^j} = \boldsymbol{0}$ if $j \neq e$ and

$$\frac{\partial \boldsymbol{F}_{int,e}^k}{\partial \rho_1^e} = \sum_{s=1}^{n_{ipt}} \boldsymbol{B}_{e_s}^T \left( \frac{\partial \gamma}{\partial \rho_1^e} \boldsymbol{P}_{e_s}^k + \gamma \frac{\partial \boldsymbol{P}_{e_s}^k}{\partial \rho_1^e} \right) w_{e_s} \tag{A 6}$$

$$+ \sum_{s=1}^{n_{ipt}} \boldsymbol{B}_{L,e_s}^T \left[ -2\gamma \frac{\partial \gamma}{\partial \rho_1^e} \left( \mathbb{C} : \boldsymbol{\varepsilon}_{e_s} \right) + (1-\gamma^2) \left( \frac{\partial \mathbb{C}}{\partial \rho_1^e} : \boldsymbol{\varepsilon}_{e_s} \right) \right] w_{e_s} \quad \text{with}$$

$$\frac{\partial \boldsymbol{P}}{\partial \rho_1} = -\frac{1}{2} r^{-\frac{3}{2}} \frac{\partial r}{\partial \rho_1} \boldsymbol{P}^b + r^{-\frac{1}{2}} \left( \left. \frac{\partial \boldsymbol{P}^b}{\partial \rho_1} \right|_{\boldsymbol{F}^b \text{ fixed}} + \frac{\partial \boldsymbol{P}^b}{\partial \boldsymbol{F}^b} : \frac{\partial \boldsymbol{F}^b}{\partial \rho_1} \right)$$

$$\text{where} \quad \frac{\partial \boldsymbol{P}^b}{\partial \boldsymbol{F}^b} = \mathbb{A}_v^b + \mathbb{A}_1^b + \mathbb{A}_2^b \tag{A 7}$$

and $\mathbb{A}_v^b$, $\mathbb{A}_1^b$ and $\mathbb{A}_2^b$ are the tangent moduli evaluated from each constitutive model with material interpolation, i.e. $\mathbb{A}_v^b \stackrel{\text{def}}{=} \partial \boldsymbol{P}_v^b / \partial \boldsymbol{F}^b$ , $\mathbb{A}_1^b \stackrel{\text{def}}{=} \partial \boldsymbol{P}_1^b / \partial \boldsymbol{F}^b$ and $\mathbb{A}_2^b \stackrel{\text{def}}{=} \partial \boldsymbol{P}_2^b / \partial \boldsymbol{F}^b$ . The subscript "$s$" denotes the $s^{th}$ quadrature point and in total $n_{ipt}$ quadrature points are used in each element. The subscript "$e_s$" is used to denote the quantity evaluated at the $s^{th}$ quadrature point in $e^{th}$ element and $w$ represents the quadrature weight. The term $\left. \frac{\partial \boldsymbol{P}^b}{\partial \rho_1} \right|_{\boldsymbol{F}^b \text{ fixed}}$ is computed by

$$\left. \frac{\partial \boldsymbol{P}^b}{\partial \rho_1} \right|_{\boldsymbol{F}^b \text{ fixed}} = \left. \frac{\partial \boldsymbol{P}_v^b}{\partial \rho_1} \right|_{\boldsymbol{F}^b \text{ fixed}} + \left. \frac{\partial \boldsymbol{P}_1^b}{\partial \rho_1} \right|_{\boldsymbol{F}^b \text{ fixed}} + \left. \frac{\partial \boldsymbol{P}_2^b}{\partial \rho_1} \right|_{\boldsymbol{F}^b \text{ fixed}} \tag{A 8}$$

$$\text{where} \quad \left. \frac{\partial \boldsymbol{P}_v^b}{\partial \rho_1} \right|_{\boldsymbol{F}^b \text{ fixed}} = -p_e \rho_1^{p_e - 1} \left( \widehat{\boldsymbol{P}}_{v,0}^b + \widetilde{\boldsymbol{P}}_{v,0}^b \right)$$

$$\left. \frac{\partial \boldsymbol{P}_1^b}{\partial \rho_1} \right|_{\boldsymbol{F}^b \text{ fixed}} = \rho_2^p \left( \frac{\partial \zeta_1^\kappa}{\partial \rho_1} \widehat{\boldsymbol{P}}_{1,0}^b + \frac{\partial \zeta_1^\mu}{\partial \rho_1} \widetilde{\boldsymbol{P}}_{1,0}^b \right) \tag{A 9}$$

$$\left. \frac{\partial \boldsymbol{P}_2^b}{\partial \rho_1} \right|_{\boldsymbol{F}^b \text{ fixed}} = (1 - \rho_2)^p \left( \frac{\partial \zeta_2^\kappa}{\partial \rho_1} \widehat{\boldsymbol{P}}_{2,0}^b + \frac{\partial \zeta_2^\mu}{\partial \rho_1} \widetilde{\boldsymbol{P}}_{2,0}^b \right)$$

where "0" in the subscript denotes that the term is evaluated with the non-interpolated solid material parameters. Again, the upper hat denotes the volumetric part while upper tilde denotes the isochoric part, e.g. $\widehat{\boldsymbol{P}}_{1,0}^b = \partial \widehat{\psi}_1 / \partial \boldsymbol{F}^b$ where $\widehat{\psi}_1$ is evaluated with solid phase parameters. On the other hand, the dependence of $\boldsymbol{R}^k$ on $\boldsymbol{\rho}_2$ comes from the constitutive model parameters. As a result, the derivative $\partial\boldsymbol{R}^k/\partial\boldsymbol{\rho}_2$ is computed as

$$\frac{\partial\boldsymbol{R}^k}{\partial\boldsymbol{\rho}_2}=\begin{bmatrix}\frac{\partial\boldsymbol{F}_{int}^k}{\partial\boldsymbol{\rho}_2}\\ \boldsymbol{0}\\ \boldsymbol{0}\end{bmatrix}\quad\text{with}\quad\frac{\partial\boldsymbol{F}_{int}^k}{\partial\boldsymbol{\rho}_2}=\underset{e=1}{\overset{n_{ele}}{\mathcal{A}}}\frac{\partial\boldsymbol{F}_{int,e}^k}{\partial\boldsymbol{\rho}_2}\quad\text{and}\quad\frac{\partial\boldsymbol{F}_{int,e}^k}{\partial\boldsymbol{\rho}_2}=\begin{bmatrix}\frac{\partial\boldsymbol{F}_{int,e}^k}{\partial\rho_2^1} & \cdots & \frac{\partial\boldsymbol{F}_{int,e}^k}{\partial\rho_2^{n_{ele}}}\end{bmatrix}$$

where $\frac{\partial\boldsymbol{F}_{int,e}^k}{\partial\rho_2^j}=\boldsymbol{0}$ if $j\neq e$ and

$$\frac{\partial\boldsymbol{F}_{int,e}^k}{\partial\rho_2^e}=\sum_{s=1}^{n_{ipt}}\gamma\boldsymbol{B}_{e_s}^T\frac{\partial\boldsymbol{P}_{e_s}^k}{\partial\rho_2^e}w_{e_s}\quad\text{with}\quad\frac{\partial\boldsymbol{P}}{\partial\rho_2}=r^{-\frac{1}{2}}\frac{\partial\boldsymbol{P}^b}{\partial\rho_2}\tag{A 10}$$

where $\frac{\partial\boldsymbol{P}^b}{\partial\rho_2}=\frac{\partial\boldsymbol{P}_1^b}{\partial\rho_2}+\frac{\partial\boldsymbol{P}_2^b}{\partial\rho_2}$ with

$$\frac{\partial\boldsymbol{P}_1^b}{\partial\rho_2}=p\rho_2^{p-1}\left(\zeta_1^\kappa\widehat{\boldsymbol{P}}_{1,0}^b+\zeta_1^\mu\widetilde{\boldsymbol{P}}_{1,0}^b\right)\quad\text{and}\quad\frac{\partial\boldsymbol{P}_2^b}{\partial\rho_2}=-p(1-\rho_2)^{p-1}\left(\zeta_2^\kappa\widehat{\boldsymbol{P}}_{2,0}^b+\zeta_2^\mu\widetilde{\boldsymbol{P}}_{2,0}^b\right)$$

### A.3 Derivative $\partial\boldsymbol{R}^k/\partial\widehat{\boldsymbol{u}}^k$

According to the expression of $\boldsymbol{R}^k$ given in Eq. (18), its derivative w.r.t. $\widehat{\boldsymbol{u}}^k$ can be derived as

$$\frac{\partial\boldsymbol{R}^k}{\partial\widehat{\boldsymbol{u}}^k}=\begin{bmatrix}\frac{\partial\boldsymbol{F}_{int}^k}{\partial\boldsymbol{u}^k} & -\boldsymbol{M}_1^T & -\boldsymbol{M}_2^T\\ -\boldsymbol{M}_1 & \boldsymbol{0} & \boldsymbol{0}\\ -\boldsymbol{M}_2 & \boldsymbol{0} & \boldsymbol{0}\end{bmatrix}\quad\text{with}\quad\frac{\partial\boldsymbol{F}_{int}^k}{\partial\boldsymbol{u}^k}=\underset{e=1}{\overset{n_{ele}}{\mathcal{A}}}\frac{\partial\boldsymbol{F}_{int,e}^k}{\partial\boldsymbol{u}_e^k}\tag{A 11}$$

$$\begin{aligned}\text{where }\frac{\partial\boldsymbol{F}_{int,e}^k}{\partial\boldsymbol{u}_e^k}=&\sum_{s=1}^{n_{ipt}}\gamma\boldsymbol{B}_{e_s}^T\left(r^{-1/2}\frac{\partial\boldsymbol{P}^b}{\partial\boldsymbol{F}^b}:\frac{\partial\boldsymbol{F}^b}{\partial\boldsymbol{u}}-\frac{1}{2}r^{-\frac{3}{2}}\boldsymbol{P}^b\otimes\frac{\partial r}{\partial\boldsymbol{u}}\right)w_{e_s}\\&+\sum_{s=1}^{n_{ipt}}(1-\gamma^2)\boldsymbol{B}_{L,e_s}^T[\mathbb{C}]\boldsymbol{B}_{L,e_s}w_{e_s}\end{aligned}\tag{A 12}$$

### A.4 Derivative $\partial\boldsymbol{R}^k/\partial v^k$

The derivative $\partial\boldsymbol{R}^k/\partial v^k$ can be calculated as

$$\frac{\partial \boldsymbol{R}^k}{\partial v^k} \equiv \frac{\partial \boldsymbol{R}^k}{\partial \bar{\lambda}_1^k} = \begin{bmatrix} \mathbf{0} \\ \mathbf{0} \\ \dfrac{\partial \boldsymbol{b}}{\partial \bar{\lambda}_1^k} \end{bmatrix} \quad \text{with} \quad \frac{\partial \boldsymbol{b}}{\partial \bar{\lambda}_1^k} = [\boldsymbol{L}_M] \begin{bmatrix} \dfrac{\partial \bar{F}_{11}}{\partial \bar{\lambda}_1^k} \\ \dfrac{\partial \bar{F}_{21}}{\partial \bar{\lambda}_1^k} \\ \dfrac{\partial \bar{F}_{12}}{\partial \bar{\lambda}_1^k} \\ \dfrac{\partial \bar{F}_{22}}{\partial \bar{\lambda}_1^k} \end{bmatrix} \tag{A 13}$$

where the derivatives $\partial \bar{F}_{ij} / \partial \bar{\lambda}_1^k$ are given in Eq. (35).

### A.5 Derivative $\partial H^k / \partial \hat{\boldsymbol{u}}^k$

The derivative $\partial H^k / \partial \hat{\boldsymbol{u}}^k$ is calculated as

$$\frac{\partial H^k}{\partial \hat{\boldsymbol{u}}^k} = \begin{bmatrix} \dfrac{\partial H^k}{\partial \boldsymbol{u}^k} & \dfrac{\partial H^k}{\partial \boldsymbol{\lambda}^k} & \dfrac{\partial H^k}{\partial \boldsymbol{\mu}^k} \end{bmatrix} = \begin{bmatrix} \mathbf{0} & \mathbf{0} & \dfrac{\partial H^k}{\partial \boldsymbol{\mu}^k} \end{bmatrix} \quad \text{with} \quad \frac{\partial H^k}{\partial \boldsymbol{\mu}^k} \equiv \frac{\partial g\left(\bar{P}_{ij}^k\right)}{\partial \boldsymbol{\mu}^k} \tag{A 14}$$

where according to Eqns. (32) and (33) the only terms that need to be calculated are $\partial \bar{P}_{ij}^k / \partial \boldsymbol{\mu}^k$, which can be obtained by

$$\frac{\partial [\bar{\boldsymbol{P}}]}{\partial \boldsymbol{\mu}^k} = \frac{1}{V} [\boldsymbol{L}_M]^T \tag{A 15}$$

## References

[1] N.A. Fleck, V.S. Deshpande, M.F. Ashby, Micro-architectured materials: past, present and future, Proceedings of the Royal Society A: Mathematical, Physical and Engineering Sciences, 466 (2010) 2495-2516. doi:10.1098/rspa.2010.0215
[2] X. Yu, J. Zhou, H. Liang, Z. Jiang, L. Wu, Mechanical metamaterials associated with stiffness, rigidity and compressibility: A brief review, Progress in Materials Science, 94 (2018) 114-173. doi:10.1016/j.pmatsci.2017.12.003
[3] J. Bauer, L.R. Meza, T.A. Schaedler, R. Schwaiger, X. Zheng, L. Valdevit, Nanolattices: An Emerging Class of Mechanical Metamaterials, Advanced Materials, 29 (2017) 1701850. doi:10.1002/adma.201701850
[4] K.E. Evans, M.A. Nkansah, I.J. Hutchinson, S.C. Rogers, Molecular network design, Nature, 353 (1991) 124-124. doi:10.1038/353124a0
[5] F.A. Robert, An isotropic three-dimensional structure with Poisson's ratio =−1, Journal of Elasticity, 15 (1985) 427-430. doi:10.1007/bf00042531
[6] R. Lakes, Foam Structures with a Negative Poisson's Ratio, Science, 235 (1987) 1038-1040. doi:10.1126/science.235.4792.1038
[7] K.K. Saxena, R. Das, E.P. Calius, Three Decades of Auxetics Research − Materials with Negative Poisson's Ratio: A Review, Advanced Engineering Materials, 18 (2016) 1847-1870. doi:10.1002/adem.201600053
[8] H. Wan, H. Ohtaki, S. Kotosaka, G. Hu, A study of negative Poisson's ratios in auxetic honeycombs based on a large deflection model, European Journal of Mechanics - A/Solids, 23 (2004) 95-106. doi:10.1016/j.euromechsol.2003.10.006
[9] A. Alderson, K.L. Alderson, D. Attard, K.E. Evans, R. Gatt, J.N. Grima, W. Miller, N. Ravirala, C.W. Smith, K. Zied, Elastic constants of 3-, 4- and 6-connected chiral and anti-chiral honeycombs subject to uniaxial in-plane loading, Composites Science and Technology, 70 (2010) 1042-1048. doi:10.1016/j.compscitech.2009.07.009
[10] J.N. Grima, A. Alderson, K.E. Evans, Auxetic behaviour from rotating rigid units, physica status solidi (b), 242 (2005) 561-575. doi:10.1002/pssb.200460376
[11] R.S. Lakes, Negative-Poisson's-Ratio Materials: Auxetic Solids, Annual Review of Materials Research, 47 (2017) 63-81. doi:10.1146/annurev-matsci-070616-124118
[12] M.P. Bendsøe, N. Kikuchi, Generating optimal topologies in structural design using a homogenization method, Computer Methods in Applied Mechanics and Engineering, 71 (1988) 197-224. doi:10.1016/0045-7825(88)90086-2
[13] O. Sigmund, Materials with prescribed constitutive parameters: An inverse homogenization problem, International Journal of Solids and Structures, 31 (1994) 2313-2329. doi:10.1016/0020-7683(94)90154-6
[14] O. Sigmund, Design of material structures using topology optimization (PhD Thesis). Technical University of Denmark Denmark, 1994
[15] O. Sigmund, S. Torquato, Design of materials with extreme thermal expansion using a three-phase topology optimization method, Journal of the Mechanics and Physics of Solids, 45 (1997) 1037-1067. doi:10.1016/S0022-5096(96)00114-7
[16] L.V. Gibiansky, O. Sigmund, Multiphase composites with extremal bulk modulus, Journal of the Mechanics and Physics of Solids, 48 (2000) 461-498. doi:10.1016/S0022-5096(99)00043-5

[17] X. Huang, A. Radman, Y.M. Xie, Topological design of microstructures of cellular materials for maximum bulk or shear modulus, Computational Materials Science, 50 (2011) 1861-1870. doi:10.1016/j.commatsci.2011.01.030
[18] O. Sigmund, J.S. Jensen, Systematic design of phononic band-gap materials and structures by topology optimization, Philosophical Transactions of the Royal Society of London. Series A: Mathematical, Physical and Engineering Sciences, 361 (2003) 1001-1019. doi:doi:10.1098/rsta.2003.1177
[19] Y.-M. Yi, S.-H. Park, S.-K. Youn, Design of microstructures of viscoelastic composites for optimal damping characteristics, International Journal of Solids and Structures, 37 (2000) 4791-4810. doi:10.1016/S0020-7683(99)00181-X
[20] R. Alberdi, K. Khandelwal, Design of periodic elastoplastic energy dissipating microstructures, Structural and Multidisciplinary Optimization, 59 (2019) 461-483. doi:10.1007/s00158-018-2076-2
[21] P.B. Nakshatrala, D.A. Tortorelli, K.B. Nakshatrala, Nonlinear structural design using multiscale topology optimization. Part I: Static formulation, Computer Methods in Applied Mechanics and Engineering, 261-262 (2013) 167-176. doi:10.1016/j.cma.2012.12.018
[22] J. Kato, D. Yachi, T. Kyoya, K. Terada, Micro-macro concurrent topology optimization for nonlinear solids with a decoupling multiscale analysis, International Journal for Numerical Methods in Engineering, 113 (2018) 1189-1213. doi:10.1002/nme.5571
[23] F. Wang, O. Sigmund, J.S. Jensen, Design of materials with prescribed nonlinear properties, Journal of the Mechanics and Physics of Solids, 69 (2014) 156-174. doi:10.1016/j.jmps.2014.05.003
[24] R. Hill, On constitutive macro-variables for heterogeneous solids at finite strain, Proceedings of the Royal Society of London. A. Mathematical and Physical Sciences, 326 (1972) 131-147. doi:10.1098/rspa.1972.0001
[25] P.P. Castañeda, Exact second-order estimates for the effective mechanical properties of nonlinear composite materials, Journal of the Mechanics and Physics of Solids, 44 (1996) 827-862. doi:10.1016/0022-5096(96)00015-4
[26] S. Saeb, P. Steinmann, A. Javili, Aspects of Computational Homogenization at Finite Deformations: A Unifying Review From Reuss' to Voigt's Bound, Applied Mechanics Reviews, 68 (2016) 050801-050801-050833. doi:10.1115/1.4034024
[27] G. Geymonat, S. Müller, N. Triantafyllidis, Homogenization of nonlinearly elastic materials, microscopic bifurcation and macroscopic loss of rank-one convexity, Archive for Rational Mechanics and Analysis, 122 (1993) 231-290. doi:10.1007/bf00380256
[28] N. Triantafyllidis, M.D. Nestorović, M.W. Schraad, Failure Surfaces for Finitely Strained Two-Phase Periodic Solids Under General In-Plane Loading, Journal of Applied Mechanics, 73 (2005) 505-515. doi:10.1115/1.2126695
[29] E.A. de Souza Neto, P.J. Blanco, P.J. Sánchez, R.A. Feijóo, An RVE-based multiscale theory of solids with micro-scale inertia and body force effects, Mechanics of Materials, 80 (2015) 136-144. doi:10.1016/j.mechmat.2014.10.007
[30] C. Miehe, Strain-driven homogenization of inelastic microstructures and composites based on an incremental variational formulation, International Journal for Numerical Methods in Engineering, 55 (2002) 1285-1322. doi:10.1002/nme.515
[31] J. Mandel, Plasticité classique et Viscoplasticité, volume 97 of CISM Lecture Notes, in, Springer-Verlag, Wien, 1972.

[32] K. Pham, V.G. Kouznetsova, M.G.D. Geers, Transient computational homogenization for heterogeneous materials under dynamic excitation, Journal of the Mechanics and Physics of Solids, 61 (2013) 2125-2146. doi:10.1016/j.jmps.2013.07.005
[33] G. Zhang, R. Alberdi, K. Khandelwal, Topology optimization with incompressible materials under small and finite deformations using mixed u/p elements, International Journal for Numerical Methods in Engineering, 115 (2018) 1015-1052. doi:10.1002/nme.5834
[34] B. Bourdin, Filters in topology optimization, International Journal for Numerical Methods in Engineering, 50 (2001) 2143-2158. doi:10.1002/nme.116
[35] T.E. Bruns, D.A. Tortorelli, Topology optimization of non-linear elastic structures and compliant mechanisms, Computer Methods in Applied Mechanics and Engineering, 190 (2001) 3443-3459. doi:10.1016/S0045-7825(00)00278-4
[36] F. Wang, B.S. Lazarov, O. Sigmund, J.S. Jensen, Interpolation scheme for fictitious domain techniques and topology optimization of finite strain elastic problems, Computer Methods in Applied Mechanics and Engineering, 276 (2014) 453-472. doi:10.1016/j.cma.2014.03.021
[37] E.A. de Souza Neto, D. Perić, M. Dutko, D.R.J. Owen, Design of simple low order finite elements for large strain analysis of nearly incompressible solids, International Journal of Solids and Structures, 33 (1996) 3277-3296. doi:10.1016/0020-7683(95)00259-6
[38] R. Alberdi, G. Zhang, L. Li, K. Khandelwal, A unified framework for nonlinear path-dependent sensitivity analysis in topology optimization, International Journal for Numerical Methods in Engineering, 115 (2018) 1-56. doi:10.1002/nme.5794
[39] K. Svanberg, The method of moving asymptotes—a new method for structural optimization, International Journal for Numerical Methods in Engineering, 24 (1987) 359-373. doi:10.1002/nme.1620240207
[40] Dassault-Systèmes, Abaqus User Manual (Ver 6.14-2), (2014)
[41] J.K. Guest, J.H. Prévost, T. Belytschko, Achieving minimum length scale in topology optimization using nodal design variables and projection functions, International Journal for Numerical Methods in Engineering, 61 (2004) 238-254. doi:10.1002/nme.1064
[42] V. Braibant, C. Fleury, Shape optimal design using B-splines, Computer Methods in Applied Mechanics and Engineering, 44 (1984) 247-267. doi:10.1016/0045-7825(84)90132-4
[43] S. Cho, S.-H. Ha, Isogeometric shape design optimization: exact geometry and enhanced sensitivity, Structural and Multidisciplinary Optimization, 38 (2008) 53. doi:10.1007/s00158-008-0266-z
[44] W.A. Wall, M.A. Frenzel, C. Cyron, Isogeometric structural shape optimization, Computer Methods in Applied Mechanics and Engineering, 197 (2008) 2976-2988. doi:10.1016/j.cma.2008.01.025
[45] N. Triantafyllidis, M.W. Schraad, Onset of failure in aluminum honeycombs under general in-plane loading, Journal of the Mechanics and Physics of Solids, 46 (1998) 1089-1124. doi:10.1016/S0022-5096(97)00060-4
[46] J.M. Ball, Convexity conditions and existence theorems in nonlinear elasticity, Archive for Rational Mechanics and Analysis, 63 (1976) 337-403. doi:10.1007/bf00279992
[47] N. Triantafyllidis, B.N. Maker, On the Comparison Between Microscopic and Macroscopic Instability Mechanisms in a Class of Fiber-Reinforced Composites, Journal of Applied Mechanics, 52 (1985) 794-800. doi:10.1115/1.3169148
[48] R. Hill, A general theory of uniqueness and stability in elastic-plastic solids, Journal of the Mechanics and Physics of Solids, 6 (1958) 236-249. doi:10.1016/0022-5096(58)90029-2
[49] C. Kittel, P. McEuen, Introduction to solid state physics, Wiley New York, 1996.

[50] R. Alberdi, G. Zhang, K. Khandelwal, A framework for implementation of RVE-based multiscale models in computational homogenization using isogeometric analysis, International Journal for Numerical Methods in Engineering, 114 (2018) 1018-1051. doi:10.1002/nme.5775
[51] M.P. Santisi d'Avila, N. Triantafyllidis, G. Wen, Localization of deformation and loss of macroscopic ellipticity in microstructured solids, Journal of the Mechanics and Physics of Solids, 97 (2016) 275-298. doi:10.1016/j.jmps.2016.07.009
[52] M. Ortiz, Y. Leroy, A. Needleman, A finite element method for localized failure analysis, Computer Methods in Applied Mechanics and Engineering, 61 (1987) 189-214. doi:10.1016/0045-7825(87)90004-1